\documentclass[
	aps, prd, reprint,
	10pt, notitlepage, a4paper,
        floats, floatfix,
	amsmath, amssymb, amsfonts, eqsecnum,
	superscriptaddress,
	showpacs, showkeys,
	nofootinbib,
 	longbibliography,
]{revtex4-1}

\usepackage{graphicx} % include figures
\usepackage[usenames,dvipsnames]{xcolor}
\usepackage{xspace} % Sensible space treatment at end of simple macros
\usepackage{bm} % bold math
\usepackage{amsmath,amsfonts,amssymb,amsthm}
\usepackage[utf8]{inputenc} % for some references from inspirehep.net

\xdefinecolor{mylinkcolor}{rgb}{0,0,0.5}
\usepackage[
	bookmarksnumbered, bookmarksopen, bookmarksopenlevel=2,
	breaklinks=true,
	colorlinks=true, filecolor=mylinkcolor, citecolor=mylinkcolor,
	linkcolor=mylinkcolor, urlcolor=mylinkcolor, menucolor=mylinkcolor,
]{hyperref}

\usepackage{verbatim}
\usepackage{mathrsfs}
\usepackage{color}
\usepackage{enumitem}
\usepackage[dvipsnames]{xcolor}
\usepackage{comment}

\newcommand{\nin}{\noindent}
\newcommand{\nnm}{\nonumber}
\newcommand{\doe}{\partial}
\newcommand{\be}{\begin{equation}}
\newcommand{\ee}{\end{equation}}
\newcommand{\bse}{\begin{subequations}}
\newcommand{\ese}{\end{subequations}}

\newcommand{\mr}{\mathrm}
\newcommand{\tr}{\textrm}
\newcommand{\mc}{\mathcal}

\newcommand{\bs}{\boldsymbol}

\newcommand{\bpm}{\begin{pmatrix}}
\newcommand{\epm}{\end{pmatrix}}

\newcommand{\AEI}{\affiliation{Max Planck Institute for Gravitational Physics (Albert Einstein Institute), Am M\"uhlenberg 1, Potsdam 14476, Germany}}
\newcommand{\Maryland}{\affiliation{Department of Physics, University of Maryland, College Park, MD 20742, USA}}

\begin{document}

\title{Spinning-black-hole scattering and the test-black-hole limit \\ at second post-Minkowskian order}

\author{Justin Vines}\email{justin.vines@aei.mpg.de}\AEI
\author{Jan Steinhoff}\email{jan.steinhoff@aei.mpg.de}\AEI
\author{Alessandra Buonanno}\email{alessandra.buonanno@aei.mpg.de}\AEI\Maryland

\date{\today}

\begin{abstract}

  Recently, the gravitational scattering of two black holes (BHs)
  treated at the leading order in the weak-field, or post-Minkowskian
  (PM), approximation to general relativity has been shown to map
  bijectively onto a simpler effectively one-body process: the
  scattering of a test BH in a stationary BH spacetime.  Here, for BH
  spins aligned with the orbital angular momentum, we propose a simple
  extension of that mapping to 2PM order.  We provide evidence for the validity
  and utility of this 2PM mapping by demonstrating its compatibility
  with all known analytical results for the conservative local-in-time dynamics of
  binary BHs in the post-Newtonian (weak-field and slow-motion)
  approximation and, separately, in the test-BH limit. 
 Our result could be employed in the construction of improved effective-one-body models for the conservative dynamics of inspiraling spinning binary BHs.

\end{abstract}

%\pacs{}

\maketitle

\section{Introduction}

Sometimes it is easier to solve complicated problems by looking at
seemingly more complicated ones.  For example, new insights into the
study of the (classical) relativistic gravitational interaction of
massive bodies might be gained by considering the even more formidable
problem of their quantum gravitational interaction.\footnote{As 
often emphasized by J.~A.~Wheeler (see, e.g., Box 25.3 in Ref.~\cite{Misner:1974qy}),   
quantum mechanics can help us elucidate the essence of classical mechanics.}

Attempts to understand surprisingly simple results from computations
of quantum scattering amplitudes, at higher orders in perturbation
theory, for gauge theories as well as gravity theories, have led in
recent years to an \textit{amplitudes revolution} of physical insights 
breeding new more efficient computational techniques
\cite{Parke:1986gb,Bern:1994zx,Bern:1994cg,Britto:2004nc,Britto:2004ap,Britto:2005fq,Bern:2008qj,Bern:2010yg,Bern:2010ue}.
A central theme has been that 
%gauge-invariant 
amplitudes or S-matrices
are determined to a surprising extent by general principles such as
symmetries, unitarity, and locality.  Very recent progress along these
lines has included analyses of tree and loop amplitudes involving
quantum particles with arbitrary masses and spins
\cite{Guevara:2017csg,Arkani-Hamed:2017jhn,Ochirov:2018uyq}.

Directly connecting such advances to the classical dynamics of
spinning black holes (BHs) would be highly valuable, particularly for
the study of binary BHs and their gravitational-wave (GW) emissions, 
with important applications to the new field of GW astrophysics~\cite{Abbott:2016blz}.  In
spite of some progress along these lines \cite{Vaidya:2014kza}, it
remains unclear to what extent the scattering of (minimally
coupled) quantum particles might correspond to scattering of
classical BHs, especially when the particles and BHs are spinning, and
when we consider their complete multipole series.  It is hence
important to approach such questions from both the quantum and
classical sides.  The present paper is concerned with classical
scattering of spinning BHs, but we will make tangential contact with (and draw inspiration from) aspects of amplitudes approaches.

As another example of this section's opening maxim (being particularly
relevant for gravitational scattering), in an analytic treatment of
the binary BH problem, we can trade the more easily handled
post-Newtonian (PN) approximation
\cite{Blanchet:2013haa,Schafer:2018kuf,Futamase:2007zz,Poisson:2014,Goldberger:2007hy,Foffa:2013qca,Porto:2016pyg,Rothstein:2014sra,Levi:2018nxp}
for the 
post-Minkowskian (PM) approximation \cite{Blanchet:2013haa,Poisson:2014,Bertotti:1956,Bertotti:1960,Rosenblum:1978zr,Bel:1981be,Damour:1981bh,Portilla:1979xx,Portilla:1980uz,Westpfahl:1979gu,Westpfahl:1985,Schafer:1986,Westpfahl:1987}:
weak-field perturbation theory on a background flat Minkowski
spacetime, without the further assumption of nonrelativistic speeds
which would lead to the PN approximation.  The PM approximation has
recently been a subject of renewed interest concerning its
applications to classical and quantum gravitational scattering of
massive bodies and to the dynamics of bound binary systems
\cite{Ledvinka:2008tk,Foffa:2013gja,Damour:2016gwp,Bini:2017xzy,Vines:2017hyw,Damour:2017zjx,Bini:2018ywr,Blanchet:2018yvb,Bjerrum-Bohr:2013bxa,Holstein:2004dn,Bjerrum-Bohr:2018xdl,Neill:2013wsa,Vaidya:2014kza,Cheung:2018wkq} 
(see also Refs.~\cite{Duff:1973zz, Iwasaki:1971vb,Kosower:2018adc}). A related and very active line of
research aims at deriving predictions in classical gravity from
double-copy constructions, or color-kinematics dualities
\cite{Bern:2008qj,Bern:2010yg,Bern:2010ue,Bern:2018jmv}, between
scattering amplitudes for gauge theories and gravity theories
\cite{Monteiro:2014cda,Luna:2015paa,Luna:2016hge,Luna:2016due,Luna:2017dtq,Goldberger:2016iau,Goldberger:2017vcg,Goldberger:2017ogt,Goldberger:2017frp,Chester:2017vcz,Shen:2018ebu,Plefka:2018dpa}.

References \cite{Damour:2016gwp,Bini:2017xzy,Vines:2017hyw,Damour:2017zjx,Bini:2018ywr}
in particular have considered both PM two-body scattering and its
relationship to effective-one-body (EOB) models for binary dynamics
\cite{Buonanno:1998gg,Buonanno:2000ef,Damour:2001tu,Damour:2008qf,Barausse:2009xi}.
This was initiated in Ref.~\cite{Damour:2016gwp} with an analysis at
1PM order (at linear order in the gravitational coupling $G$, or in
linearized gravity) of a system of two pure-monopole/point-mass
bodies.  This was followed by a treatment of
dipole/linear-in-spin/spin-orbit effects at 1PM order in
Ref.~\cite{Bini:2017xzy}.  The point-mass case was considered at 2PM
order (quadratic order in $G$) in Ref.~\cite{Damour:2017zjx}, and 2PM
spin-orbit effects were treated in Ref.~\cite{Bini:2018ywr}.

The pole-dipole (point-mass and spin-orbit) contributions to the
classical gravitational dynamics of a system of massive bodies (in
vacuum) are universal, i.e., they are independent of the nature of the
bodies \cite{Mathisson:1937,Mathisson:2010,Dixon:1979,Dixon:2015vxa}.
This reflects local conservation of linear and angular momenta, due to
local Poincar\'e invariance. A body's internal structure influences its orbital dynamics through
its intrinsic quadrupole and higher multipole moments.  The leading
contributions to the $2^l$-pole moments $M_l$ of a BH are fully
determined by its mass $m$ and spin (intrinsic angular momentum) $S$,
its monopole and dipole, according to Hansen's formula
\cite{Hansen:1974zz} for a stationary Kerr BH, $M_l\sim m(i a)^l$,
with the rescaled spin $a=S/mc$ being the radius of the BH's ring singularity.  For the
specific case of a two-spinning-BH system, the analysis of PM
scattering was extended to treat all-multipole/all-orders-in-spin
effects at 1PM order in Ref.~\cite{Vines:2017hyw}.

\begin{figure*}

\begin{tabular}{cp{5.5cm}}
\includegraphics[scale=.5]{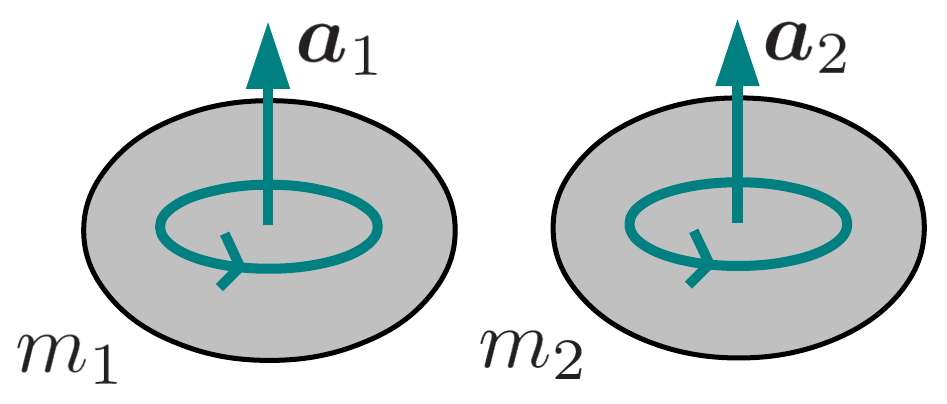}$\qquad$
& \vspace{-1.5cm}
Two-BH system with arbitrary masses 

$m_1$ and $m_2$ and rescaled spins 

$\bs a_1=\bs S_1/m_1c$ and $\bs a_2=\bs S_2/m_2c$

\\
\includegraphics[scale=.5]{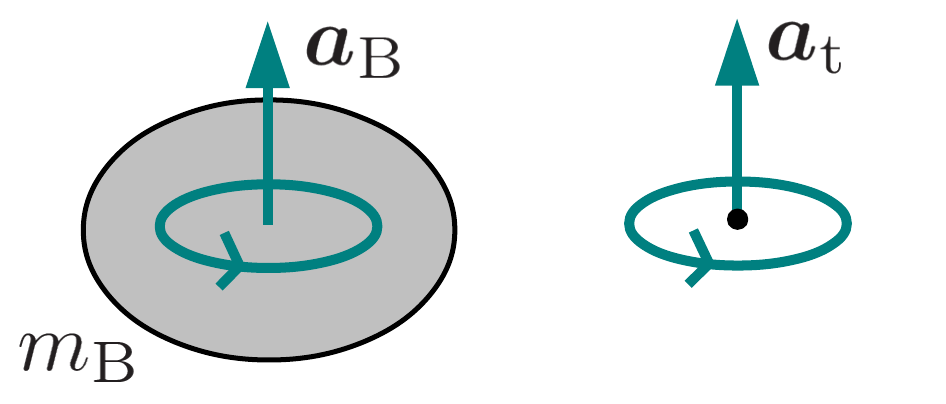}$\qquad$
& \vspace{-1.7cm} 
Spinning test BH with negligible mass 

and finite rescaled spin $\bs a_\mr t$

in a background Kerr spacetime 

with mass $m_\mr B$ and rescaled spin $\bs a_\mr B$
\\
\includegraphics[scale=.5]{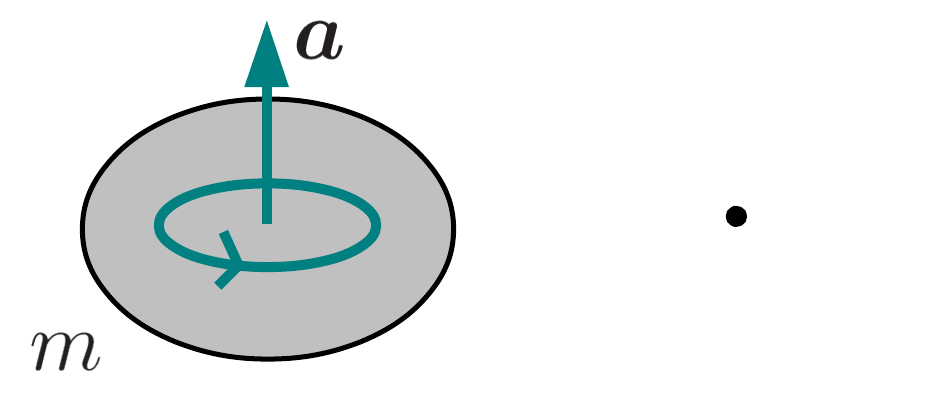}$\qquad$
& \vspace{-1.7cm}
Monopolar test point mass with 

negligible mass and spin following

a geodesic in a Kerr spacetime

with mass $m$ and rescaled spin $\bs a$
\\
\end{tabular}

\caption{A schematic diagram of an aligned-spin two-BH system and its
  limits as discussed in Sec.~\ref{sec:SEOB}.  We depict in green the spinning
  BHs' ring singularities with radii $a=|\bs a|$, and in black the BH horizons.
To obtain the limit of a spinning test BH, we take its mass $m_\mr t$ to be negligible,
  $m_\mr t/m_\mr B\to 0$, while keeping its rescaled spin $\bs a_\mr
  t$ finite.  Taking the spin of the test BH to zero yields a
  monopolar test point mass, following a geodesic in a Kerr
  background.}\label{fig:binaries}
\end{figure*}

Here we begin to analyze the higher-multipole contributions for binary
BHs at 2PM order, with an eye toward including the BHs' complete
multipole series and \textit{resumming} them, as in
Refs.~\cite{Vines:2016qwa,Vines:2017hyw,Siemonsen:2017yux}.  We
continue, as in Refs.~\cite{Damour:2016gwp,Vines:2017hyw}, to
investigate the extent to which PM results for the conservative local-in-time
dynamics of \textit{real} (arbitrary-mass-ratio) binary BHs can be deduced
via simple mappings from results in the \textit{test-body}
limit --- specifically, the \textit{spinning-test-BH} limit, in which the
mass ratio tends to zero while keeping finite the smaller BH's (the test
BH's) mass-rescaled spin or ring radius $a=S/mc$, and thus also all of
its mass-rescaled multipoles. This approach is valuable since
exact solutions for the gravitational field in classical gravity
are known, in particular the Schwarzschild and Kerr metrics, which
makes the test-body case particularly tractable, even nonperturbatively.
Figure \ref{fig:binaries} sketches the limiting cases encountered in the present paper.

Our investigations are greatly simplified by restricting attention
to the \textit{aligned-spin} case, in which the BHs' spin vectors are
parallel (anti-parallel) to the systems' orbital (and thus to its
total) angular momentum vector.  This is one case in which the
directions of all of these vectors are unambiguously defined even in
full general relativity.  The orbital motion is confined to a plane,
the one orthogonal to the constant direction of the angular momentum.

Regarding the conservative contributions to the orbital dynamics (to
the extent that these can be well defined), an aligned-spin binary BH
has effectively the same degrees of freedom as a two-point-mass
system, with only a 2D relative position (and velocity or momentum) in
the orbital plane.  We thus expect the aligned-spin binary BH system
to share the following important properties with the binary point-mass
system, as emphasized in Refs.~\cite{Damour:2016gwp,Damour:2017zjx}. 
For the point-mass system, at 1PM order (apparently to all PN orders)
and at 2PM order (at least through the third subleading PN order),
the complete conservative local-in-time dynamical information is encoded in the
system's \emph{scattering-angle function}: for an unbound system, the
angle by which both masses are scattered in the center-of-mass frame,
as a function of the (rest) masses, the total energy (or the relative
velocity at infinity), and the orbital angular momentum (or the impact
parameter).  The complete conservative information, for both unbound
\emph{and bound} orbits, can be defined, for example, as the part of
the information content of a (perturbative) canonical Hamiltonian
governing the conservative dynamics which is invariant under
(perturbative) canonical transformations.

In this paper, considering an aligned-spin binary BH system instead of
a binary point-mass system, we verify that its scattering angle
function also encodes its complete conservative  local-in-time dynamics, to a similar
level of approximation.  This holds according to all available PN
results (truncated at 2PM order), including in particular the 1PM and
2PM contributions through quadratic order in the BHs' spins, through
subsubleading PN orders. Up to those same levels of approximation, we find a simple mapping
between the scattering-angle functions for a real binary BH and for a
test BH moving in a background Kerr spacetime.  A potentially more general form of this result (still at 2PM order but extending beyond the reach of current PN results) is suggested by considerations of amplitudes-based derivations of classical scattering angles.  Given that the scattering angle fully encodes the conservative local-in-time dynamics, this has
significant implications for constructions of EOB
models for binary BHs.

The paper is organized as follows.  We begin in
Sec.~\ref{sec:NSEOB} with a discussion of two-point-mass scattering at
2PM order.  We point out a simple mapping between the real system and
its test-body limit (geodesic motion in a stationary Schwarzschild
spacetime) which is implicit in the 2PM result for the scattering
angle first derived in Ref.~\cite{Westpfahl:1985}; this generalizes
similar observations at 1PM order made in Ref.~\cite{Damour:2016gwp}. 
After defining the scattering-angle function for an aligned-spin
binary BH in Sec.~\ref{sec:angles}, we briefly review in
Sec.~\ref{sec:1PMspin} the mappings between the real two-body angle
and its test-body limits found at 1PM order in
Ref.~\cite{Vines:2017hyw}.  In Sec.~\ref{sec:2PMspin}, we present and
discuss our generalization of one of those mappings to 2PM order,
valid at least in the restricted 2PM context described above (within
the reach of available PN results). This is the central result of the
present paper. In Sec.~\ref{sec:PN} we derive a dual PN-PM expansion of the
scattering angle from known PN results for canonical Hamiltonians
encoding the conservative  local-in-time dynamics of aligned-spin binary BHs.
Focusing on the 1PM and 2PM (spin-dependent) parts of the PN results, we discuss how
the gauge-invariant information content of a canonical Hamiltonian
(defined modulo canonical transformations) is uniquely determined by
the scattering-angle function.
In Sec.~\ref{sec:test}, we compare the PN-PM expansion of the real
binary-BH scattering angle to PM results which can be obtained in the
limit of test-BH motion in a stationary Kerr spacetime.  This
comparison leads us to the 2PM mapping discussed in
Sec.~\ref{sec:2PMspin}, i.e., to the central result.  We focus on contributions
up to quadratic order in the spin of the test BH, as PN results with
2PM parts are available only up to spin-squared order. 
Section~\ref{sec:2PMspin} also serves as an illustration of the utility of exact BH metrics
in connection with our central result. Finally, we conclude in Sec.~\ref{sec:conclude}.

\section{Nonspinning-black-hole scattering at second post-Minkowskian order}\label{sec:NSEOB}

Pioneering studies of the PM approximation
\cite{Bertotti:1956,Bertotti:1960,Rosenblum:1978zr,Bel:1981be,Damour:1981bh,Portilla:1979xx,Portilla:1980uz,Westpfahl:1979gu,Westpfahl:1985,Schafer:1986,Westpfahl:1987}
applied to the gravitational dynamics of massive bodies culminated in
Westpfahl's computation, to 2PM order, of the scattering-angle
function for an unbound system of two monopolar point masses (which could represent nonspinning BHs), via a
direct assault on the nonlinear field equations in position space
coupled with effective point-particle equations of motion
\cite{Westpfahl:1985}.  In the intervening decades, this result stood
alone and quite separated from primarily PN studies of bound
coalescing binary systems and their GW emissions,
until it was revisited in the latter context in
Ref.~\cite{Damour:2017zjx}.

It was shown in Ref.~\cite{Damour:2017zjx} that the full
gauge-invariant information determining the two-point-mass
conservative local-in-time dynamics (for unbound and bound orbits) through 2PM order
(at least up to 3PN order) is contained in Westpfahl's result for the
center-of-mass-frame scattering-angle function; the information can be
quantified, e.g., by counting coefficients in a dual PN-PM expansion
of the scattering angle and of a Hamiltonian or Lagrangian along with
relevant phase-space diffeomorphisms.  This property had been
discussed at 1PM order in Ref.~\cite{Damour:2016gwp}, where it was
shown that the 1PM scattering angle for a real two-body system can be
deduced, via a simple kinematical mapping, from the scattering angle
for geodesics in a Schwarzschild spacetime (truncated at 1PM order).
Reference \cite{Damour:2017zjx} demonstrated that Westpfahl's 2PM
result agrees with available PN results (up to 3PN order) and that it
correctly reduces in the test-body limit (the zero-mass-ratio limit)
to the 2PM result for Schwarzschild geodesics.  (See Ref.~\cite{Bini:2017wfr} for a calculation of the scattering angle to 4PN order, the order at which one first encounters nonlocal-in-time contributions \cite{Damour:2014jta}.)  Reference
\cite{Damour:2017zjx} did not explicitly discuss any mapping by which
one could recover the real two-body angle from the (much more easily
obtained) 2PM expansion of the Schwarzschild-geodesic angle.

It is nonetheless hard to miss the striking similarity between the two 2PM results.  Westpfahl's \cite{Westpfahl:1985} scattering angle $\chi$ for an arbitrary-mass-ratio two-body system with rest masses $m_1$ and $m_2$ and total center-of-mass-frame energy $E$ and angular momentum $J$, and the scattering angle $\chi_\mr t$ for a test particle of mass $m_\mr t$ with (background-frame) energy $E_\mr t$ and angular momentum $J_\mr t$ following a geodesic in a Schwarzschild background of mass $m_\mr B$ (B for background) are given, as in Eqs.~(2.18)--(2.24) of Ref.~\cite{Damour:2017zjx}, by 
\begin{widetext}
\bse\label{chiEJ}
\begin{alignat}{3}\label{chiEJtwo}
\chi(m_1,m_2,E,J)
&=2\frac{2\gamma^2-1}{\sqrt{\gamma^2-1}}\frac{GM\mu}{J}
+\frac{3\pi}{4}\frac{M}{E}(5\gamma^2-1)\left(\frac{GM\mu}{J}\right)^2
+{\cal O}(G^3),
&\qquad 
\gamma&=1+\frac{E^2-M^2}{2M\mu},
\\\label{chiEJtest}
\chi_\mr t(m_\mr B,m_\mr t,E_\mr t,J_\mr t)
&=2\frac{2\gamma_\mr t^2-1}{\sqrt{\gamma_\mr t^2-1}}\frac{Gm_\mr Bm_\mr t}{J_\mr t}
+\frac{3\pi}{4}(5\gamma_\mr t^2-1)\left(\frac{Gm_\mr Bm_\mr t}{J_\mr t}\right)^2
+{\cal O}(G^3),
&\qquad \gamma_\mr t&=\frac{E_\mr t}{m_\mr t},
\end{alignat}
\ese
\end{widetext}
where $M$ is the two-body system's total rest mass, and $\mu$ is its reduced mass,
\be\label{masses}
M=m_1+m_2,
\qquad 
\mu=\frac{m_1m_2}{M}=\nu M,
\ee
defining also the symmetric mass ratio $\nu=\mu/M$.  The quantities $\gamma$ and $\gamma_\mr t$ here are both denoted $\hat{\mc E}_\mr{eff}$ in Ref.~\cite{Damour:2017zjx}; they correspond to the relative Lorentz factors between the respective pairs of the bodies' rest frames at infinity, as further detailed below.

It was pointed out in Ref.~\cite{Damour:2016gwp} that one mapping by which one can obtain the 1PM part [the ${\cal O}(G^1)$ part] of the real two-body result (\ref{chiEJtwo}) from the 1PM part of the test-body result (\ref{chiEJtest}), which we will refer to as an ``EOB scattering-angle mapping,'' is as follows:
\be\label{chiEJmap}
\begin{array}{c}
\phantom{\Big|}
\mu=m_\mr t,
\quad
M=m_\mr B,
\phantom{\Big|}
\\
\phantom{\Big|}
J=J_\mr,
\quad
\gamma=\gamma_\mr t 
\phantom{\Big|}
\end{array}
\;\Rightarrow\quad 
\chi=\chi_\mr t+ {\cal O}(G^2);
\ee
i.e., the scattering angles will be equal at 1PM order if we use the usual ``Newtonian EOB mapping'' of the rest masses (the test-body mass is the reduced mass, and the background mass is the total mass), if we identify the two angular momenta with one another, and if the two energies are related by
\be\label{energymap}
\gamma=\gamma_\mr t
\quad\Leftrightarrow\quad
E_\mr t=\mu+\frac{E^2-M^2}{2M},
\ee
which is the ``EOB energy map,'' proposed in Ref.~\cite{Buonanno:1998gg} to relate the Hamiltonian of an effective test body in an effective background to the Hamiltonian of a real two-body system.

It is clear from (\ref{chiEJ}) that the mapping (\ref{chiEJmap}) breaks down at 2PM order, specifically because of the factor of $M/E$ in the ${\cal O}(G^2)$ term of (\ref{chiEJtwo}).  We will see presently that an alternative EOB scattering-angle mapping which continues to hold at 2PM order suggests itself when we look at the same results (\ref{chiEJ}) expressed in terms of different (equivalent) variables, in particular, trading the angular momenta for the corresponding impact parameters [see Eq.~(\ref{chi_pm}) below].

\begin{figure*}
$\vcenter{\hbox{\includegraphics[scale=.43]{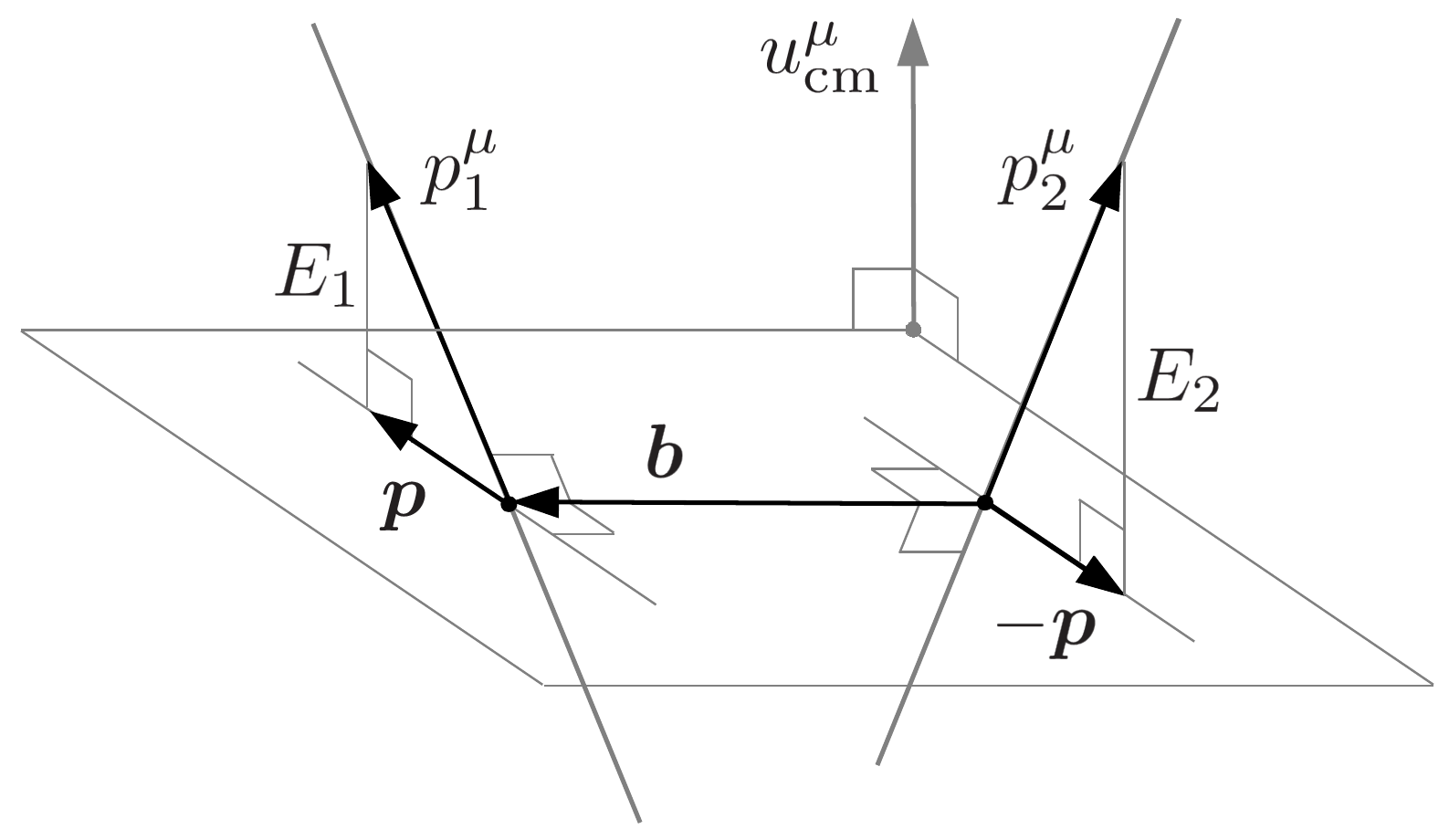}}}\qquad\vcenter{\hbox{\includegraphics[scale=.43]{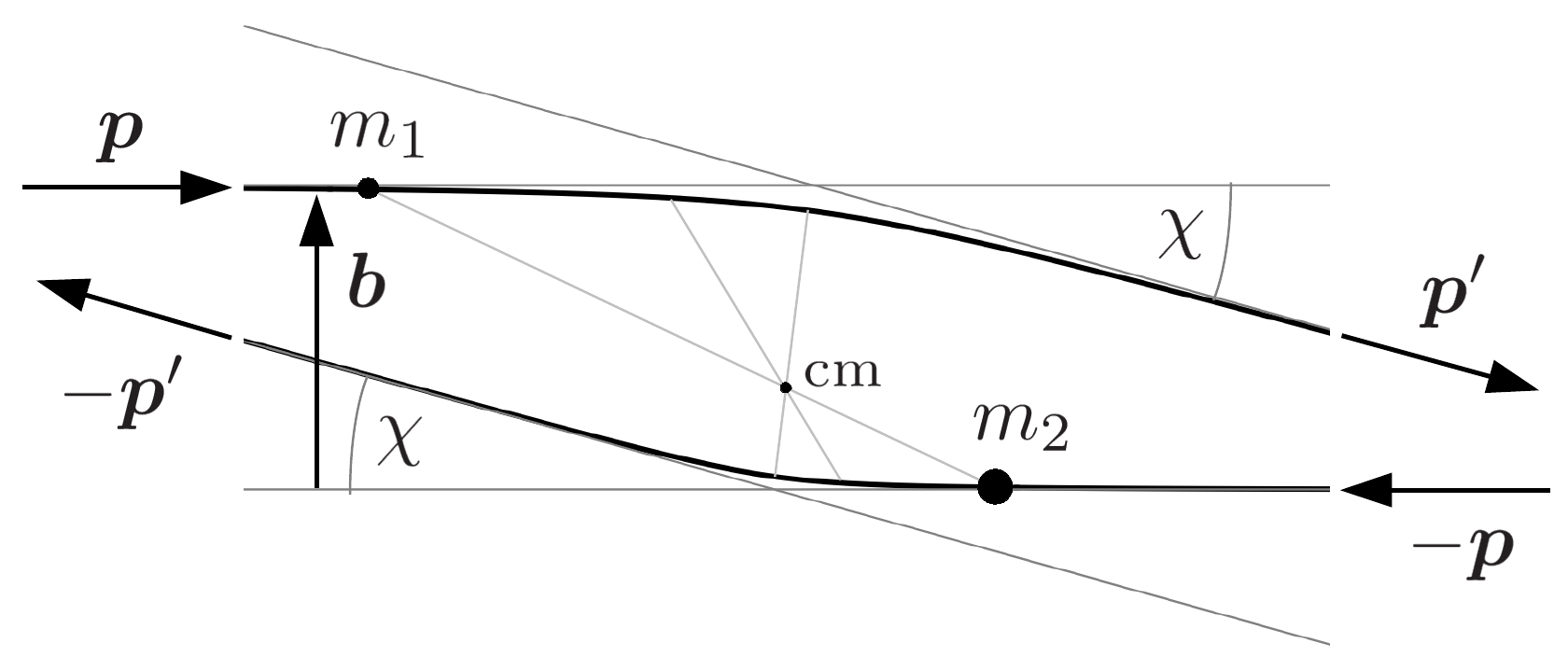}}}$
\caption{
\emph{Above}: the (arbitrary-mass-ratio) two-body case.  \emph{Below}: the test-body case.  \emph{Left}: the Minkowskian geometry of the (incoming) zeroth-order state.  \emph{Right}: the spatial geometry of the scattering plane---in the center-of-mass frame above, and in the background frame below.  Above left, the 4-momenta are decomposed as $p_1^\mu=E_1 u_\mr{cm}^\mu+\bs p^\mu$ and $p_2^\mu=E_2 u_\mr{cm}^\mu-\bs p^\mu$; $E=E_1+E_2$ is the total center-of-mass frame energy of Eq.~(\ref{Egamma}).  The test-body's momentum is decomposed according to $p_\mr t^\mu=E_\mr t u_\mr B^\mu+\bs p_\mr t^\mu$.  The magnitudes of the ``spatial'' momenta are $|\bs p|=m_1m_2\gamma v/E$ and $|\bs p_\mr t|=m_\mr t\gamma v$, so that Eqs.~(\ref{J}) and (\ref{Jt}) can be rewritten as $J=|\bs p|b$ and $J_\mr t=|\bs p_\mr t|b$, respectively.
}\label{fig:twovstest}
$\vcenter{\hbox{\includegraphics[scale=.43]{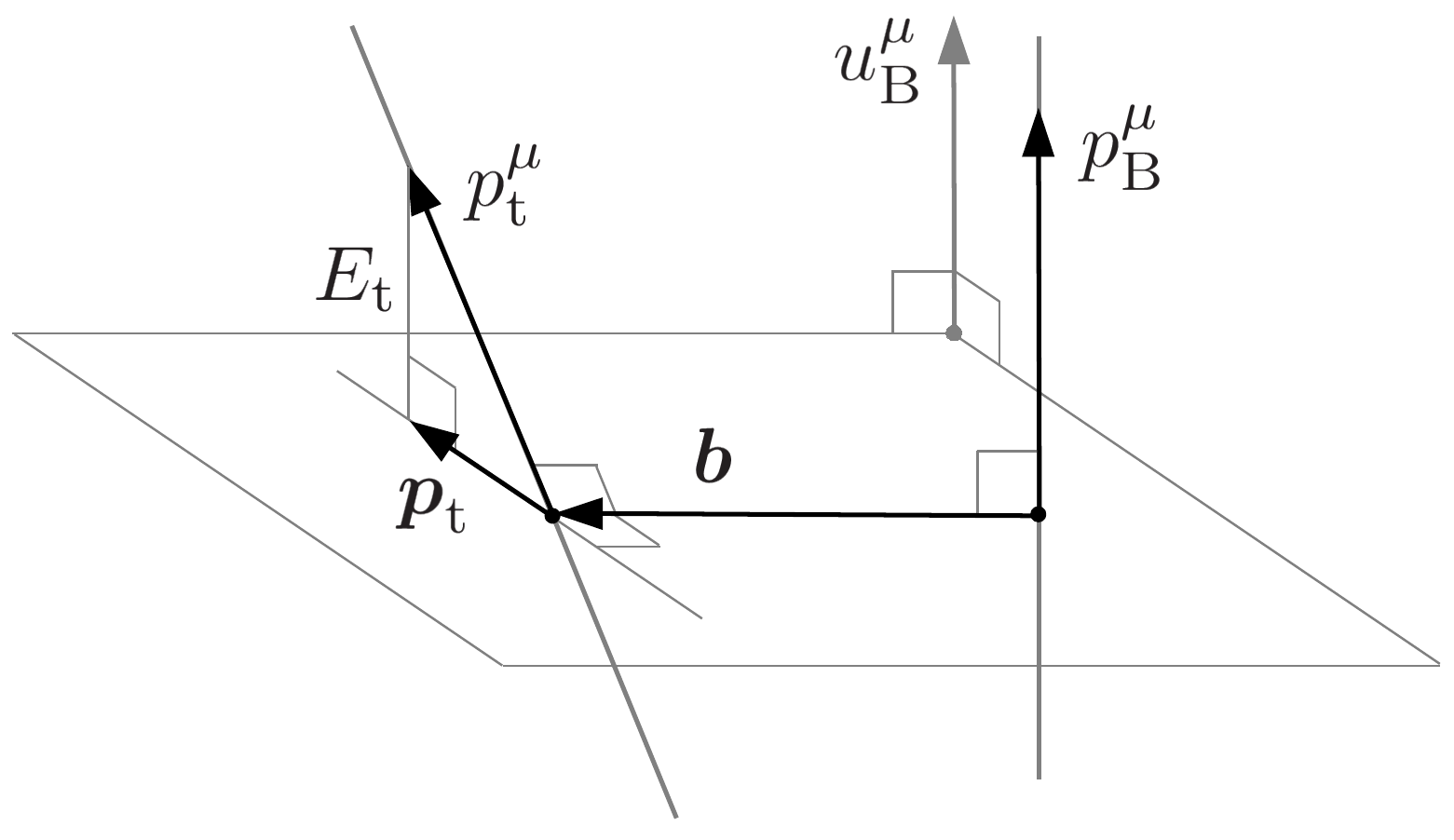}}}\qquad\vcenter{\hbox{\includegraphics[scale=.43]{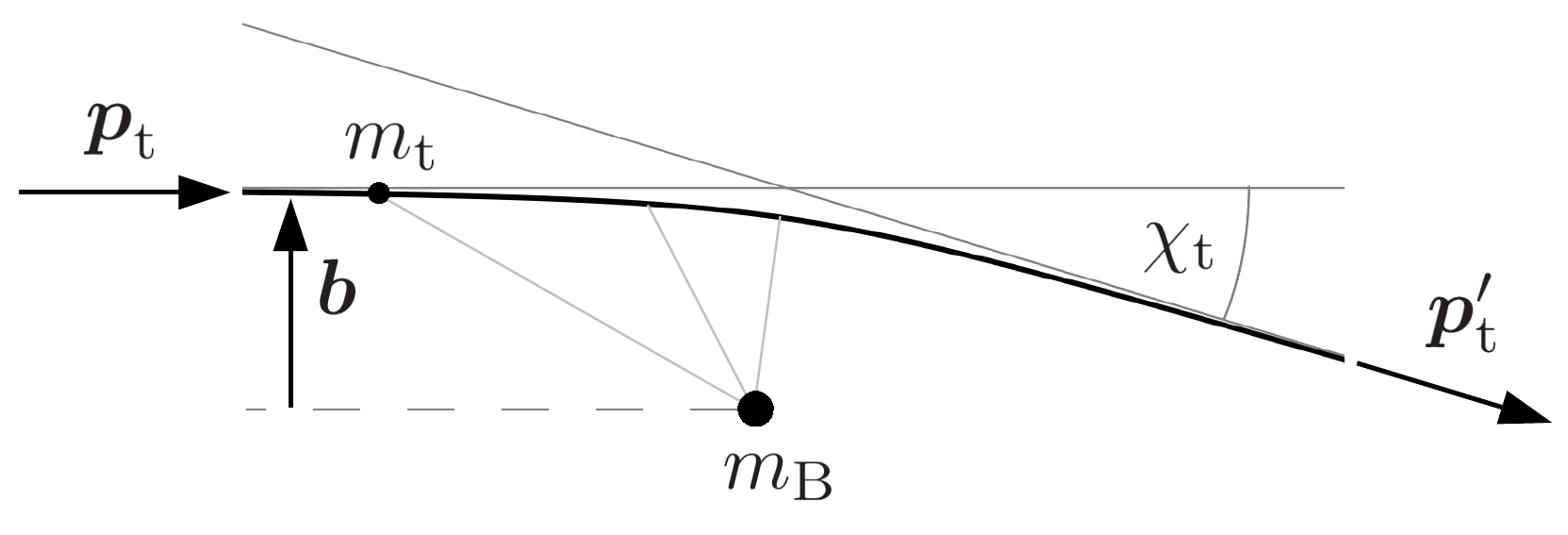}}}$
\end{figure*}

Let us first recall how the energies $E$ and $E_\mr t$ are related to the respective pairs of the bodies' asymptotic 4-momenta, using flat-spacetime kinematics at infinity (see Fig.~\ref{fig:twovstest}).  For the two-body system, with momenta $p_1^\mu$ and $p_2^\mu$ and relative Lorentz factor
\be\label{gamma}
\gamma=\frac{p_1\cdot p_2}{m_1m_2},
\ee
the total center-of-mass-frame energy $E$ is the magnitude of the system's total momentum,
\be\label{Egamma}
E^2=(p_1+p_2)^2=m_1^2+m_2^2+2m_1m_2\gamma,
\ee
which, with Eq.~(\ref{masses}), leads to the expression for $\gamma$ in Eq.~(\ref{chiEJtwo}).  The energy $E_\mr t$ of the test body in the background frame is defined by
\be\label{gammaEt}
(\gamma_\mr t\,\to)\;\;\gamma=\frac{E_\mr t}{m_\mr t}=\frac{p_\mr t\cdot p_\mr B}{m_\mr tm_\mr B},
\ee
where we indicate that we will henceforth drop the subscript t on the relative Lorentz factor for the test-background system.  In both cases, the relative velocity $v$ between the bodies' asymptotic rest frames is related to the Lorentz factor by
\be\label{gammav}
\gamma=\frac{1}{\sqrt{1-v^2}}.
\ee
The 4-momenta, $p_1^\mu$ and $p_2^\mu$, or $p_\mr t^\mu$ and
$p_\mr B^\mu$, here could be either (both) the initial or (both) the final
momenta at infinity, since the rest masses and the energies (and thus
$v$ and $\gamma$) are conserved at the level of approximation we consider.  Note
that we use the $(+,-,-,-)$ signature for the Minkowski metric, with
$p_1^2=m_1^2$, etc., and in this and the following section 
we work in units in which the speed of light $c$ is $1$.

Next, let us recall how the angular momenta $J$ and $J_\mr t$ are related to the (point-mass) bodies' asymptotic worldlines (trajectories), and thus to the respective impact parameters defined at infinity.  For the two-body system, the total relativistic angular momentum tensor about any point $x$ is given by
\be\label{Jmunu}
J^{\mu\nu}(x)=2p_1^{[\mu}(x-z_1)^{\nu]}+2p_2^{[\mu}(x-z_2)^{\nu]},
\ee
where $z_1$ and $z_2$ can be any points on each of the bodies' asymptotic (zeroth-order; say, incoming) worldlines (which are flat-spacetime geodesics), and where square brackets denote antisymmetrization of enclosed indices.  In the center-of-mass frame, with unit 4-velocity $u_\mr{cm}^\mu$, the total angular momentum vector $J^\mu$ is defined by the first line here (and turns out to be independent of $x$), and the second follows from inserting Eq.~(\ref{Jmunu}):
\begin{alignat}{3}
\begin{aligned}\label{Jmu}
J^\mu&=\frac{1}{2}\epsilon^\mu{}_{\nu\rho\sigma} u_\mr{cm}^\nu J^{\rho\sigma} \qquad \mbox{with}
\qquad u_\mr{cm}^\mu=\frac{p_1^\mu+p_2^\mu}{E},
\\
&=-\frac{1}{E}\epsilon^\mu{}_{\nu\rho\sigma}p_1^\nu p_2^\rho(z_1-z_2)^\sigma
\\
&=-\frac{1}{E}\epsilon^\mu{}_{\nu\rho\sigma}p_1^\nu p_2^\rho b^\sigma.
\end{aligned}
\end{alignat}
Here, $b^\mu$ is the vectorial impact parameter, connecting the two worldlines' points of mutual closest approach, equal to the projection of $(z_1-z_2)^\mu$ orthogonal to both $p_1^\mu$ and $p_2^\mu$ (again for any points $z_1$ and $z_2$ on the asymptotic worldlines).  It follows from Eqs.~(\ref{masses}), (\ref{gamma}), (\ref{gammav}), and (\ref{Jmu}) that the magnitude $J$ of the (center-of-mass-frame) angular momentum and the magnitude $b$ of the impact parameter are related by
\be\label{J}
J=\frac{M}{E}\mu\gamma v b.
\ee
On the other hand, consider the angular momentum tensor of (only) the test body, about the background body, assuming the latter to be at rest at the origin for simplicity $(x=0=z_\mr B)$,
\be
J_\mr t^{\mu\nu}=-2p_\mr t^{[\mu}z_\mr t^{\nu]},
\ee
for any point $z_\mr t$ on the test body's asymptotic worldline.  The test body's background-frame angular momentum vector is
\begin{alignat}{3}
\begin{aligned}
J_\mr t^\mu&=\epsilon^\mu{}_{\nu\rho\sigma}u_\mr B^\nu J_\mr t^{\rho\sigma} \qquad \mbox{with}
\qquad u_\mr B^\mu=\frac{p_\mr B^\mu}{m_\mr B},
\\
&=-\frac{1}{m_\mr B}\epsilon^\mu{}_{\nu\rho\sigma}p_\mr B^\nu p_\mr t^\rho z_\mr t^\sigma
\\
&=-\frac{1}{m_\mr B}\epsilon^\mu{}_{\nu\rho\sigma}p_\mr B^\nu p_\mr t^\rho b^\sigma,
\end{aligned}
\end{alignat}
where $b^\mu$ is the impact parameter (which we will not distinguish from that for the two-body system above), orthogonal to both momenta.  The magnitudes are related by
\be\label{Jt}
J_\mr t=m_\mr t\gamma v b,
\ee
having used (\ref{gammaEt}) and (\ref{gammav}). 
Equations (\ref{Egamma}), (\ref{gammaEt}), (\ref{J}) and (\ref{Jt}) allow us to express the energies and angular momenta, $E$ and $J$ for the two-body case, and $E_\mr t$ and $J_\mr t$ for the test-body case, in terms of the rest masses and $v$ and $b$, where, in both cases, $v$ is the relative velocity at infinity and $b$ is the impact parameter.

In terms of these new variables, the scattering angle $\chi$ (\ref{chiEJtwo}) for the arbitrary-mass-ratio two-body system is given by
\bse\label{chi_pm}
\begin{alignat}{3}
\label{chi_two_pm}
&\chi(m_1,m_2,v,b)
\\\nnm
&\quad=2\frac{GE}{v^2 b}\left[1+v^2
+3\pi\frac{4+v^2}{8}\frac{GM}{b}\right]
+{\cal O}(G^3),\phantom{\Bigg|}
\end{alignat}
and its test-body limit $\chi_\mr{t}$ (\ref{chiEJtest}), for Schwarzschild geodesics, is
\begin{alignat}{3}
\label{chi_test_pm}
&\chi_\mr{t}(m_\mr B,v,b)
\\\nnm
&\quad=
2\frac{Gm_\mr{B}}{v^2 b}\left[1+v^2
+3\pi\frac{4+v^2}{8}\frac{Gm_\mr B}{b}\right]
+{\cal O}(G^3).
\end{alignat}
\ese

We now see that one simple way to obtain the two-body result (\ref{chi_two_pm}) from the test-body result (\ref{chi_test_pm}), up to 2PM order, is as follows, directly generalizing Eq.~(92) of Ref.~\cite{Vines:2017hyw} to 2PM order for point masses,
\bse\label{pm_map}
\be
\chi(m_1,m_2,v,b)=\frac{E}{M}\chi_\mr t(M,v,b)+{\cal O}(G^3),
\ee
or
\begin{alignat}{3}
&\chi(m_1,m_2,v,b)\phantom{\frac{1}{2}}
\\\nnm
&\quad=\frac{\sqrt{m_1^2+m_2^2+2m_1m_2\gamma}}{m_1+m_2}\chi_\mr t(m_1+m_2,v,b)+ {\cal O}(G^3).
\end{alignat}
\ese
Comparing this alternative EOB scattering-angle mapping to the original mapping (\ref{chiEJmap}) from Ref.~\cite{Damour:2016gwp}, they both involve the identifying of the relative velocity $v$ (or $\gamma$ factor) at infinity for the test-body system with that for the two-body system, which (along with the Newtonian rest-mass mappings) implies the EOB energy map (\ref{energymap}); the differences are that (\ref{chiEJmap}) similarly identifies the two angular momenta, while (\ref{pm_map}) instead identifies the two impact parameters, and that the two scattering angles are equal in (\ref{chiEJmap}), while they differ by a factor of $E/M$ in (\ref{pm_map}).

 We should be shocked that the two-body result is so simple (and that it is so simply related to the test-body result).  In Westpfahl's calculation \cite{Westpfahl:1985}, one sees many complicated pieces---related to finite retardation effects, the nonlinearity of the (gauge-dependent) field equations, iterating the effective orbital equations of motion to second order, etc.---but in the end, it all boils down to (\ref{chi_two_pm}), which can be obtained from the test-body result (\ref{chi_test_pm}), via the simple mapping (\ref{pm_map}).
This is quite reminiscent of (and not unrelated to) difficult Feynman-diagram calculations boiling down to shockingly simple results for quantum scattering amplitudes.

We can in fact give one explanation for the validity of the 2PM EOB scattering-angle mapping (\ref{pm_map}) based on simple properties of the recent derivation in Ref.~\cite{Bjerrum-Bohr:2018xdl} of Westpfahl's 2PM scattering angle from the leading classical parts of tree (1PM) and one-loop (2PM) amplitudes for massive scalars (becoming monopolar point masses in the eikonal limit) exchanging gravitons, obtained via the on-shell unitarity method \cite{Neill:2013wsa,Bjerrum-Bohr:2013bxa,Vaidya:2014kza,Cachazo:2017jef,Guevara:2017csg}.  The relevant classical part of the total amplitude (at the leading orders in the momentum transfer $q$, those which contribute to the classical scattering angle) is given by Eqs.~(16) and (19) of Ref.~\cite{Bjerrum-Bohr:2018xdl} as
\bse\label{amplitude}
\begin{widetext}
\begin{alignat}{3}\label{mcM}
\begin{array}{c}
p_1-q
\\
\\
\\
\\
\\
\\
p_1
\end{array}\!\!\!\!\!\!\!\!\!\!
\vcenter{\hbox{\includegraphics[scale=.25]{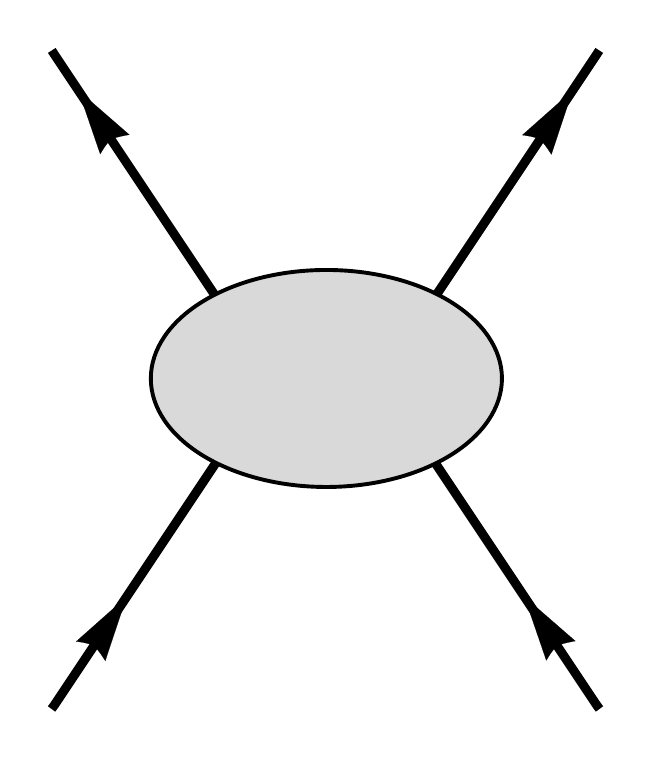}}}
\!\!\!\!\!\!\!\!\!
\begin{array}{c}
p_2+q
\\
\\
\\
\\
\\
\\
p_2
\end{array}
=\mc M=\vcenter{\hbox{\includegraphics[scale=.25]{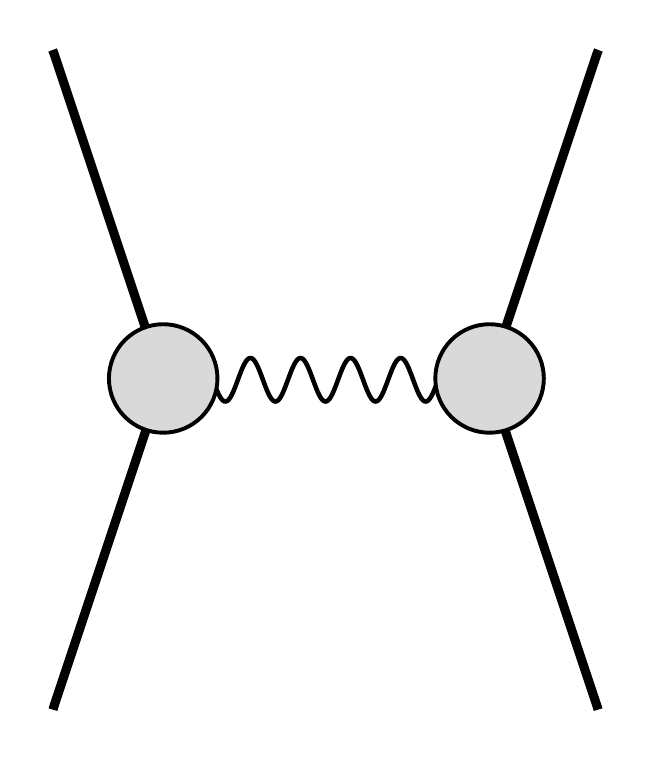}}}
+\vcenter{\hbox{\includegraphics[scale=.25]{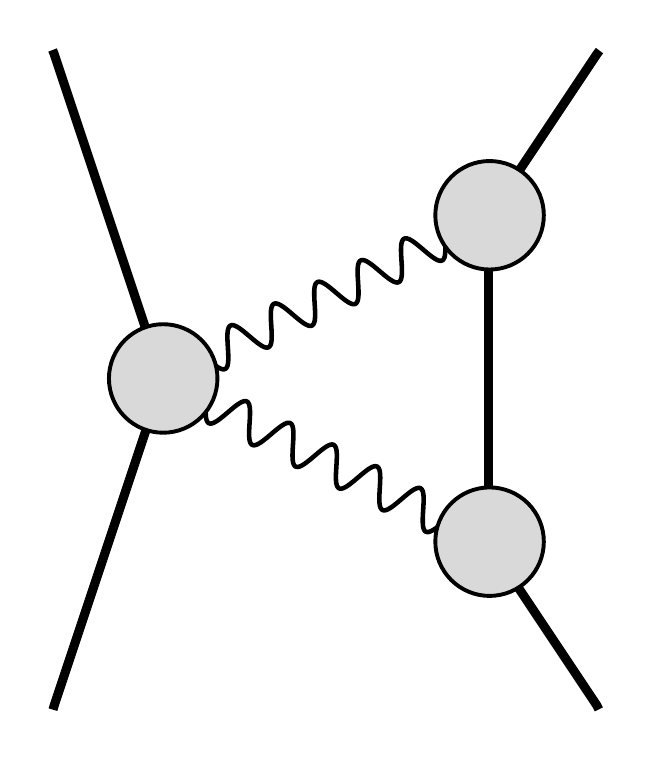}}}+\vcenter{\hbox{\includegraphics[scale=.25]{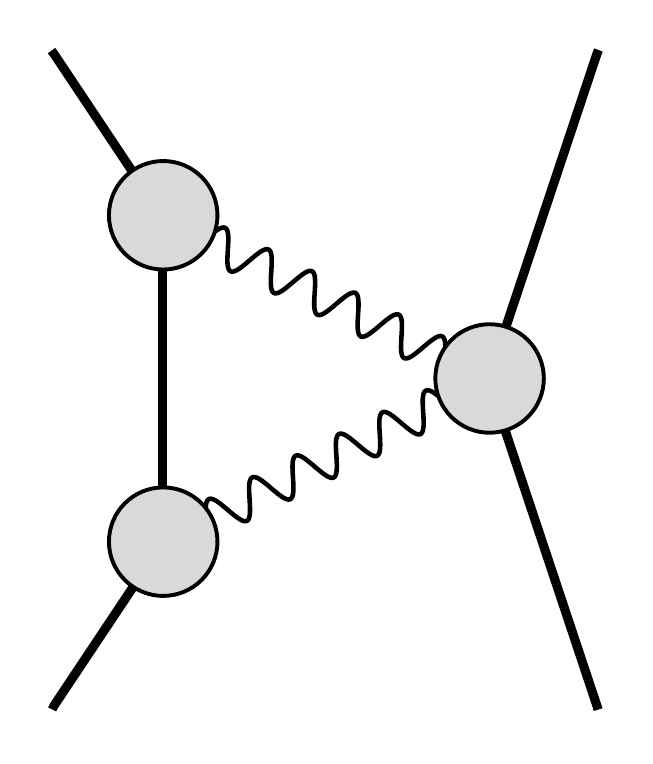}}}+ {\cal O}(G^3),
\end{alignat}
\end{widetext}
with
\begin{alignat}{3}
\vcenter{\hbox{\includegraphics[scale=.18]{fig_amp_tree}}}&=\mc M_\mr{tree}=\frac{16\pi G(m_1m_2)^2}{\hbar|\bs q|^2}(2\gamma^2-1)+{\cal O}(q^0),
\nnm\\\nnm
\vcenter{\hbox{\includegraphics[scale=.18]{fig_amp_ab}}}&=\mc M_\triangleleft=\frac{6\pi^2 G^2(m_1m_2)^2}{\hbar^2|\bs q|}m_2(5\gamma^2-1)+{\cal O}(q^1),
\\
\vcenter{\hbox{\includegraphics[scale=.18]{fig_amp_ba}}}&=\mc M_\triangleright=\frac{6\pi^2 G^2(m_1m_2)^2}{\hbar^2|\bs q|}m_1(5\gamma^2-1)+{\cal O}(q^1),
\end{alignat}
\ese
where $\gamma=p_1\cdot p_2/m_1m_2=(1-v^2)^{-1/2}$ just as in Eqs.~(\ref{gamma}) and (\ref{gammav}) above, and where we have
restored factors of $\hbar$, noting that the amplitude $\mc M$ is dimensionless.  As in Eqs.~(23)--(24) of Ref.~\cite{Bjerrum-Bohr:2018xdl}, the classical scattering angle $\chi$ (called $\theta$ in Ref.~\cite{Bjerrum-Bohr:2018xdl}), through one-loop (2PM) order, is a linear functional of the amplitude $\mc M$ given in our notation (and with some suggestive rearrangement) by
\begin{alignat}{3}\label{chi_amp}
\chi&=2\sin\frac{\chi}{2}+{\cal O}(\chi^3)
\\\nnm
&=\frac{-\hbar E}{(2\gamma v)^2}\frac{\doe}{\doe b}
\int \frac{d^2\bs q}{(2\pi)^2}e^{-i\bs q\cdot\bs b/\hbar}\frac{\mc M(\bs q)}{(m_1m_2)^2}+{\cal O}(G^3),
\end{alignat}
where the integral over the (spacelike) momentum transfer $\bs q$ spans the 2D plane orthogonal to $p_1^\mu$ and $p_2^\mu$, the plane containing the impact parameter vector $\bs b$, and the result of the integral depends only on $b=|\bs b|$.
As in Eq.~(28) of Ref.~\cite{Bjerrum-Bohr:2018xdl}, inserting the amplitude (\ref{amplitude}) into Eq.~(\ref{chi_amp}), dropping the $\hbar$ corrections resulting from the higher-order-in-$q$ terms, yields the two-body scattering angle $\chi$ just as in Eq.~(\ref{chi_two_pm}) or (\ref{chiEJtwo}) above.

The important point we would like to note about this calculation concerns the dependence of the various contributions on the masses $m_1$ and $m_2$, at fixed $v$ (or $\gamma$) and $b$.  Apart from the explicit appearances of $m_1$ and $m_2$ in Eqs.~(\ref{amplitude}) and (\ref{chi_amp}), they otherwise enter only through the total center-of-mass-frame energy $E$ (\ref{Egamma}) in the prefactor of Eq.~(\ref{chi_amp}).  Thus, using the linearity of Eq.~(\ref{chi_amp}), we can fully separate out the mass dependence as follows:
\bse\label{chiMtwo}
\begin{alignat}{3}\label{chiMtwoa}
\chi[\mc M]&=\chi[\mc M_\mr{tree}]+\chi[\mc M_\triangleleft]+\chi[\mc M_\triangleright]+{\cal O}(G^3),
\nnm\\
\chi[\mc M_\mr{tree}]&=E\, f_\mr{tree}(v,b),
\phantom{\Big|}
\nnm\\\nnm
\chi[\mc M_\triangleleft]&=E\,m_2 \, f_\triangleleft(v,b),
\\
\chi[\mc M_\triangleright]&=E \, m_1 \, f_\triangleright(v,b),
\phantom{\Big|}
\end{alignat}
noting, importantly, that $f_\triangleleft=f_\triangleright$, and thus
\be
\chi=E\Big[f_\mr{tree}(v,b)+M f_\triangleleft(v,b)\Big]+ {\cal O}(G^3),
\ee
\ese
where again $M=m_1+m_2$.  If we take the test-body limit, say, $m_1\to 0$ and $m_2\to m_\mr B$, we have from (\ref{Egamma}) that $E\to m_2\to m_\mr B$, and thus
\be\label{chiMtest}
\chi\to\chi_\mr t=m_\mr B f_\mr{tree}(v,b)+m_\mr B^2f_\triangleleft(v,b)+{\cal O}(G^3).
\ee
Here, in the test-body limit, we have lost the contribution from $\mc M_\triangleright$.  But we have not lost the most nontrivial part of the information in $\mc M_\triangleright$, the function $f_\triangleright(v,b)$, due to the symmetry property $f_\triangleright=f_\triangleleft$ and to the fact that $f_\triangleleft(v,b)$ still appears.  Comparing (\ref{chiMtwo}) and (\ref{chiMtest}), we see that, regardless of the precise forms of the $f$ functions, given only that $f_\triangleright=f_\triangleleft$, the arbitrary-mass-ratio two-body result (\ref{chiMtwo}) can be obtained from the test-body result (\ref{chiMtest}) via the mapping (\ref{pm_map}).\footnote{It is further suggested by this discussion that the EOB scattering-angle mapping (\ref{pm_map}) can only be expected to hold up to quadratic order in $G$, i.e., up to 2PM (or one-loop) order.  If, hypothetically, a linear formula like (\ref{chi_amp}) continued to hold at ${\cal O}(G^3)$, we could naively extrapolate the pattern of mass dependence, $\chi[\mc M_\mr{tree}]\propto E$ and $\chi[\{\mc M_\tr{one-loop}\}]\propto E\{m_1,m_2\}$, to continue as $\chi[\{\mc M_\tr{two-loop}\}]\propto E\{m_1^2,m_1m_2,m_2^2\}$.  Continuing to naively extrapolate, we would then lose both the $m_1^2$ and $m_1m_2$ terms in the test-body limit $m_1\to0$; while one might expect to recover the $m_1^2$ terms from symmetry with the $m_2^2$ terms, there would be no such hope for the $m_1m_2$ terms.  At least the conclusion of this hand-waving argument, that the mapping (\ref{pm_map}) will break down at 3PM order (where complete PM results are currently unavailable), is confirmed by results from the PN approximation, as we will see in Sec.~\ref{sec:PN} below. However, by the same reasoning, one might expect that results at the next order in the small-mass-ratio approximation (beyond the test-BH case) can be mapped to the generic-mass case at 3PM order.}

This kind of reasoning, about the interplay between scattering angles, scattering amplitudes, and the test-body limit, will guide us in our analysis below, where we consider scattering not of two (monopolar) point masses but of two spinning BHs (including higher-multipole contributions).

\section{Effective-one-body scattering-angle mappings for an aligned-spin two-black-hole system}\label{sec:SEOB}

As shown in Ref.~\cite{Vines:2017hyw}, a direct analog of the EOB (test-body to two-body) scattering-angle mapping (\ref{pm_map}) holds for a two-spinning-BH system, with aligned spins, at 1PM order.  This is expressed in Eq.~(\ref{1PMtwospin}) below.  
Furthermore, the two-spinning-BH scattering angles, the two-body version $\chi$ and its test-body limit $\chi_\mr t$, can both be obtained at 1PM order from the scattering angle $\chi_\mr g$ for geodesics in the equatorial plane of a stationary Kerr spacetime.  The mappings and the three scattering-angle functions are given in Eqs.~(\ref{chig1})--(\ref{1PMtwospingeod}) below.

Direct analogs of those 1PM EOB aligned-spin scattering-angle mappings do not hold for binary BHs at 2PM order.  We find that there is a straightforward 2PM generalization of the former mapping (from a spinning test BH in Kerr to the real binary BH), but not the latter (from equatorial Kerr geodesics to either of the two-spinning-BH cases).  This, our central result, is expressed in Eq.~(\ref{2PMtwospin}) below.  As discussed in the introduction, we do not prove Eq.~(\ref{2PMtwospin}) to its full potential extent at 2PM order, but instead verify it against available PN results, in Sec.~\ref{sec:PN} below. While we originally arrived at the mapping (\ref{2PMtwospin}) via the manipulations of PN results described in Sec.~\ref{sec:PN}, we motivate it here, in Sec.~\ref{sec:2PMspin}, with (conjectural) arguments about derivations of aligned-spin binary-BH scattering angles from (classical limits of) quantum scattering amplitudes for massive spinning particles exchanging gravitons.

Let us emphasize again that the utility of such mappings is rooted in the existence of analytic expressions for BH metrics (in particular the Schwarzschild and Kerr metrics), with two important implications: (i) a calculation of the test-BH scattering angle is a much more tractable problem than of the generic binary BH (see Sec.~\ref{sec:test}), and (ii) the test-BH scattering angle can be obtained without restriction on the impact parameter or velocity and hence is nonperturbative from the PN and PM perspectives.

\subsection{Aligned-spin scattering angles}\label{sec:angles}

As in the two-monopole case, the scattering angle for an aligned-spin two-BH system can be expressed as a function of the relative velocity $v$ at infinity, an impact parameter $b$, and the masses $m_1$ and $m_2$, but now with an extra dependence on the BH's spins $S_1$ and $S_2$.  These (aligned) spin components $S_1$ and $S_2$ are positive if they are aligned with the orbital angular momentum, negative if antialigned.

 Now we must also more precisely define the impact parameter (at infinity), in relation to each BH's total angular momentum tensor field.  For the initial and final asymptotic states (or zeroth-order states, effectively in flat spacetime at infinity), we define, for each BH, its ``proper'' center-of-mass(-energy) worldline to be the set of points $x$ about which its proper mass-dipole vector $\propto J^{\mu\nu}(x)p_\nu$ vanishes, where $J^{\mu\nu}(x)$ is the \emph{single} BH's total relativistic angular momentum about $x$, and $p^\mu$ is its momentum.  The impact parameter we use here is the one orthogonally separating the two BH's proper worldlines (asymptotically).  In other words, we employ here the ``covariant'' or Tulczyjew-Dixon ``spin supplementary condition'' to define the representative trajectory of each BH \cite{Tulczyjew:1959,Dixon:1979,Schattner:1979vp,Vines:2017hyw}.  The BHs' (Pauli-Lubanski) spin vectors $S^\mu$ are each defined by $S^\mu=\epsilon^\mu{}_{\nu\rho\sigma}J^{\rho\sigma}p^\nu/2m$, and their magnitudes are the scalars $(\pm)S$ (see, e.g., Secs.~II.H and III of Ref.~\cite{Vines:2017hyw} for further details).

We can then express the scattering angle for an aligned-spin binary BH as
\begin{alignat}{3}
\begin{aligned}
\chi\big((m_1,a_1),(m_2,a_2),v,b\big)&
\\
=\chi\big((m_2,a_2),(m_1,a_1),v,b\big)&
\end{aligned}
\end{alignat}
where
\be
a_1=\dfrac{S_1}{m_1},\qquad a_2=\dfrac{S_2}{m_2}
\ee
are the (oriented/signed) radii of the BHs' ring singularities, or their mass-rescaled spins (sometimes also referred to simply as the spins below).

Let 
\be
\chi_\mr{t}(m_\mr{B},a_\mr{B},a_\mr{t},v,b)
\ee
be the scattering angle for a test BH---in a way, a naked ring singularity of finite radius $a_\mr t$ and negligible mass (see Fig.~\ref{fig:binaries})---in a background  Kerr spacetime with mass $m_\mr B$ and spin $m_\mr Ba_\mr B$, in the aligned-spin configuration.  It is obtained as the test-body limit of the two-body angle $\chi$, with $m_\mr t\to 0$ at fixed $a_\mr t$,
\begin{alignat}{3}
\begin{aligned}
&\chi_\mr{t}(m_\mr B,a_\mr B,a_t,v,b)
\\
&\quad=
\chi\big((m_\mr B,a_\mr B),(m_\mr t\to0,a_\mr t),v,b\big).
\end{aligned}
\end{alignat}
An important assumption implicit throughout our analysis is that this limit is well-defined and finite; we find that this is true at least through the orders considered here.
Then, with $a_\mr t\to0$,
\be
\chi_\mr g(m,a,v,b)=\chi_\mr t(m,a,0,v,b)
\ee
is the scattering angle for geodesics in the equatorial plane of a Kerr background with mass $m$ and spin $ma$.

\subsection{At first post-Minkowskian order}\label{sec:1PMspin}
 
From Eq.~(94) of Ref.~\cite{Vines:2017hyw}, the scattering angle for equatorial-Kerr geodesics is given to 1PM order by
\begin{alignat}{3}
\label{chig1}
\begin{aligned}
\chi_\mr g
&=\frac{2Gm}{b}\frac{2\gamma^2-1-2\gamma\sqrt{\gamma^2-1}\,a/b}{(\gamma^2-1)(1-{a^2}/{b^2})}+{\cal O}(G^2)
\\
&=\frac{2Gm}{v^2}\frac{(1+v^2)b-2va}{b^2-a^2}+{\cal O}(G^2)
\\
&=\frac{Gm}{v^2}\left(\frac{(1+v)^2}{b+a}+\frac{(1-v)^2}{b-a}\right)+{\cal O}(G^2).
\end{aligned}
\end{alignat}
The spinning-test-BH-in-Kerr angle $\chi_\mr t$,
\be
\chi_\mr t=\frac{Gm_\mr B}{v^2}\left(\frac{(1+v)^2}{b+a_\mr B+a_\mr t}+\frac{(1-v)^2}{b-a_\mr B-a_\mr t}\right)+{\cal O}(G^2),
\ee
is obtained from the geodesic result via the mapping
\begin{alignat}{3}\label{1PMtestspin}
\begin{aligned}
&\chi_\mr t(m_\mr B,a_\mr B,a_\mr t,v,b)\phantom{\frac{1}{2}}
\\
&\quad=\chi_\mr{g}(m_\mr B,\,a_\mr B+a_\mr t,\,v,\,b)+{\cal O}(G^2),
\end{aligned}
\end{alignat}
as in Eq.~(90) of Ref.~\cite{Vines:2017hyw}.
Furthermore, the angle $\chi$ for the arbitrary-mass-ratio two-body case,
\be
\chi=\frac{GE}{v^2}\left(\frac{(1+v)^2}{b+a_1+a_2}+\frac{(1-v)^2}{b-a_1-a_2}\right)+{\cal O}(G^2),
\ee
is obtained from $\chi_\mr t$, much like in Eq.~(\ref{pm_map}) for the nonspinning case, or from $\chi_\mr g$, via the mappings
\begin{alignat}{3}\label{1PMtwospin}
&\chi\big((m_1,a_1),(m_2,a_2),v,b\big)
\nonumber\\
&\quad=\frac{E}{M}\chi_\mr{t}(M,a_1,a_2,v,b)+{\cal O}(G^2)
\\
&\quad=\frac{E}{M}\chi_\mr{t}(M,a_2,a_1,v,b)+{\cal O}(G^2)
\nonumber\\\label{1PMtwospingeod}
&\quad=\frac{E}{M}\chi_\mr{g}(M,a_1+a_2,v,b)+{\cal O}(G^2),
\end{alignat}
as in Eq.~(93) of Ref.~\cite{Vines:2017hyw}.

\subsection{At second post-Minkowskian order}\label{sec:2PMspin}

\nin At 2PM order, neither Eq.~(\ref{1PMtwospin}) nor Eq.~(\ref{1PMtestspin}) [nor (\ref{1PMtwospingeod})] holds. There seems to be no directly straightforward generalization to 2PM order of the relation (\ref{1PMtestspin}) which determines the spinning-test-BH angle $\chi_\mr t$ from the geodesic angle $\chi_\mr g$.  Unlike at 1PM order, from 2PM order onward, $\chi_\mr t$ does not depend only on the combination $a_\mr B+a_\mr t$, (dropping the $(v,b)$ dependence)
\begin{alignat}{3}
\chi_\mr t(m_\mr B,\,a_\mr B,\,a_\mr t)
\ne
\chi_\mr t(m_\mr B,\,a_\mr B+a_\mr t,\,0).
\end{alignat}
Furthermore, also unlike its 1PM version, $\chi_\mr t$ is not symmetric under $a_\mr B\leftrightarrow a_\mr t$,
\begin{alignat}{3}
\chi_\mr t(m_\mr B,\,a_\mr B,\,a_\mr t)
\ne
\chi_\mr t(m_\mr B,\,a_\mr t,\,a_\mr B).
\end{alignat}

But there is a slight modification of the mapping (\ref{1PMtwospin}), determining the real binary-BH angle $\chi$ from the test-BH-in-Kerr angle $\chi_\mr t$, which does hold up to 2PM order---according to the 1PM and 2PM parts of the aligned-spin scattering angles derived from all known PN results for binary-BH dynamics, as we will show in Sec.~\ref{sec:PN}.

We can (hand-wavingly) motivate the form of this mapping as follows, extrapolating from the discussion of classical limits of quantum scattering amplitudes for monopolar (scalar) masses at the end of Sec.~\ref{sec:NSEOB}.  It is suggested, most directly by the results of Ref.~\cite{Guevara:2017csg}, that when considering particles/bodies with spin and higher (spin-induced) multipoles, it will continue to be true that the relevant classical-limit scattering amplitude $\mc M$, through 2PM/one-loop order, will be sufficiently described by a sum of contributions precisely as appearing in Eq.~(\ref{mcM}), with one tree-level $\mc M_\mr{tree}$ contribution and two one-loop ``triangles'' $\mc M_\triangleleft$ and $\mc M_\triangleright$ related by interchange of the two massive (now spinning) particles.  Due to the correspondence between the effective degrees of freedom for a two-monopole system and an \emph{aligned-spin} two-BH system, we are led to conjecture that there exists a functional analogous to (\ref{chi_amp}) which \emph{linearly} produces the aligned-spin scattering angle $\chi$ from (an appropriate form of) the amplitude $\mc M$.  We can further conjecture, extrapolating from (\ref{chiMtwoa}), that the contributions to the scattering angle will take the following form, in particular regarding their dependences on the masses $m_1$ and $m_2$ (and the energy $E$),
\begin{alignat}{3}\label{chiMspin}
\chi[\mc M]&=\chi[\mc M_\mr{tree}]+\chi[\mc M_\triangleleft]+\chi[\mc M_\triangleright]+{\cal O}(G^3),
\nnm\\
\chi[\mc M_\mr{tree}]&=E\, f_\mr{tree}(v,b,a_1,a_2),
\phantom{\Big|}
\nnm\\\nnm
\chi[\mc M_\triangleleft]&=E\,m_2 \, f_\triangleleft(v,b,a_1,a_2),
\\
\chi[\mc M_\triangleright]&=E \, m_1 \, f_\triangleright(v,b,a_1,a_2).
\phantom{\Big|}
\end{alignat}
This generalizes (\ref{chiMtwoa}) only by adding a dependence of the $f$ functions on the \emph{mass-rescaled} spins $a_1$ and $a_2$ (which crucially differs from having an analogous dependence on, e.g., the full spins $S=ma$, or the dimensionless spins $\hat a=a/Gm$); this is motivated by Eqs.~(4.12), (4.20) and (A.10) of Ref.~\cite{Guevara:2017csg}.    If we assume (\ref{chiMspin}), then the inherent $(m_1,a_1)\leftrightarrow(m_2,a_2)$ symmetry of the two triangle-loop contributions implies
\be\label{fspin}
f_\triangleleft(v,b,a_1,a_2)
= f_\triangleright(v,b,a_2,a_1).
\ee
This would mean that the information in $f_\triangleright$, which would be lost in the test-body limit $m_1\to0$, could be recovered from the information in $f_\triangleleft$ which remains.
Thus, \emph{if} Eq.~(\ref{chiMspin}) and thus Eq.~(\ref{fspin}) were to validly apply to the aligned-spin scattering angle for a two-spinning-BH system, to some or all orders in spin, \emph{then} one would be able to conclude that the EOB scattering-angle mapping stated in the following paragraph holds to all orders in $1/c$ and to some or all orders in the BHs' spins, at 2PM order.

As we will show in the following section, given the scattering angle $\chi_t(m_\mr B,a_\mr B,a_\mr t,v,b)$ for a spinning test BH with ring radius $a_\mr t$ in a background Kerr spacetime with mass $m_\mr B$ and spin $m_\mr Ba_\mr B$, the scattering angle $\chi$ for an arbitrary-mass-ratio aligned-spin binary BH is given, \emph{at least} to the accuracy indicated here, by
\begin{alignat}{3}\label{2PMtwospin}
&\chi\big((m_1,a_1),(m_2,a_2),v,b\big)\phantom{\frac{1}{2}}
\\\nnm
&\quad=\frac{E}{M}\bigg[\frac{m_1}{M}\chi_\mr{t}(M,a_1,a_2,v,b)
\\\nnm
&\quad\qquad+\frac{m_2}{M}\chi_\mr{t}(M,a_2,a_1,v,b)\bigg]
\\\nnm
&\quad\quad+{\cal O}(\tr{4PN }a^0)+{\cal O}(\tr{4.5PN }a^1)+{\cal O}(\tr{5PN }a^2)\phantom{\frac{1}{2}}
\\\nnm
&\quad\quad+{\cal O}(G^2a^3)+{\cal O}(G^3).
\end{alignat}
This reduces to Eq.~(\ref{1PMtwospin}) at 1PM order because $\chi_\mr t$ is symmetric under $a_1\leftrightarrow a_2$ at 1PM order.  There is no such symmetry at 2PM order, and we see in Eq.~(\ref{2PMtwospin}) a rest-mass-weighted average of the angles for each of $a_1$ and $a_2$ playing the roles of the background Kerr BH's spin, with the other as the test BH's spin.  We also see the same overall factor of $E/M$ appearing in all of the above scattering-angle mappings, with that factor carrying the only (other) dependence on the mass ratio, at fixed $(M,a_1,a_2,b,v)$.

We have indicated in Eq.~(\ref{2PMtwospin}) the levels of 
approximation up to which we have verified this 2PM EOB aligned-spin
scattering-angle mapping against available PN results for binary BH
conservative local-in-time dynamics (see Sec.~\ref{sec:HPN}).  This includes\footnote{Here we use the
  standard PN order counting for rapidly rotating BHs.  Restoring
  factors of $c$, we assume spin magnitudes $S=mca$ with the BHs'
  ring radii $a$ being on the order of their gravitational radii,
  $a=(Gm/c^2)\hat a$, with the dimensionless spins $\hat a$ being of
  order unity; i.e., we assume the BHs are nearly extremally spinning.
  Then, using $1/c^2$ as the formal PN expansion parameter, an $n$PN
  contribution (to the scattering angle, or to a Hamiltonian, as
  below) comes with a factor of $1/c^{2n}$ relative to Newtonian (0PN)
  order, when the results are expressed in terms of the dimensionless
  spins $\hat a$.  We use the shorthands LO for leading order (the
  leading PN order at a given order in spin), NLO for next-to-leading
  order (or subleading order), etc., where ``next-to-'' or ``N''
  always increments by ``one PN order,'' i.e., by a factor of
  $1/c^2$.  See Footnote 2 of Ref.~\cite{Vines:2017hyw} for comments on the PM order counting for spin effects being used here.}
\begin{itemize}
\item the point-mass results through NNNLO, 3PN \cite{Jaranowski:1997ky,Jaranowski:1999ye,Damour:2000kk,Jaranowski:1999qd,Blanchet:2000nv,Blanchet:2000ub,Blanchet:2000nu,Blanchet:2000cw,Damour:2000ni,deAndrade:2000gf,Damour:2001bu,Blanchet:2003gy,Blanchet:2002mb,Itoh:2003fy,Itoh:2003fz,Foffa:2011ub}, which are complete at (are fully determined by the results at) 4PM order,
\item the spin-orbit (linear-in-spin) results, at
\begin{itemize}
\item LO, 1.5PN \cite{Tulczyjew:1959,Barker:1970zr,Barker:1975ae,D'Eath:1975vw,Barker:1979,Damour:1982,Thorne:1984mz,Damour:1988mr}, complete at 1PM,
\item NLO, 2.5PN \cite{Tagoshi:2000zg,Faye:2006gx,Damour:2007nc,Steinhoff:2008zr,Perrodin:2010dy,Porto:2010tr,Levi:2010zu}, complete at 2PM,
\item NNLO, 3.5PN \cite{Hartung:2011te,Hartung:2013dza,Marsat:2012fn,Bohe:2012mr,Levi:2015uxa,Levi:2015msa}, complete at 3PM,
\end{itemize}
\item and the quadratic-in-spin results, at
\begin{itemize}
\item LO, 2PN \cite{Barker:1975ae,D'Eath:1975vw,Barker:1979,Thorne:1984mz,Poisson:1997ha,Damour:2001tu,Racine:2008qv}, complete at 1PM,
\item NLO, 3PN \cite{Steinhoff:2007mb,Porto:2008tb,Porto:2006bt,Levi:2008nh,Porto:2008jj,Steinhoff:2008ji,Hergt:2010pa,Hergt:2011ik,Bohe:2015ana}, complete at 2PM,
\item NNLO, 4PN \cite{Hartung:2011ea,Levi:2011eq,Hartung:2013dza,Levi:2014sba,Levi:2015ixa,Levi:2015msa}, complete at 3PM.
 \end{itemize}
\end{itemize}
This list includes all currently known PN results for spin-dependent contributions with 2PM parts, with 2PM parts first appearing at the NLO PN levels.  The 1PM and 2PM parts constitute the complete LO and NLO PN results at each order in spin.  No NLO PN results for real binary BHs are currently available at cubic and higher orders in spin.  The mapping (\ref{2PMtwospin}) holds to all orders in spin at 1PM order [where it simplifies to (\ref{1PMtwospin})], and thus at the LO PN levels at all orders in spin, according to the results of Refs.~\cite{Vines:2017hyw,Vines:2016qwa}.
According to Westpfahl's 2PM point-mass scattering angle \cite{Westpfahl:1985}, and according to the scattering angle derived from Bini and Damour's 2PM spin-orbit results \cite{Bini:2018ywr}, the mapping (\ref{2PMtwospin}) also holds for the 1PM and 2PM parts at all PN orders up to linear order in the BHs' spins.

A canonical Hamiltonian (in a certain gauge) encoding the known
aligned-spin binary BH dynamics at all the PN orders listed above
is shown in Sec.~\ref{sec:HPN} below.  We derive from this a dual PN-PM
expansion of the real binary BH scattering-angle function, through the
same orders, in Sec.~\ref{sec:chiPN}.  We show that the complete 1PM
and 2PM parts of those PN-expanded scattering angles are indeed
obtained from the mapping (\ref{2PMtwospin}) applied to results for a
test BH in a background Kerr spacetime presented in
Sec.~\ref{sec:test}.  The 2PM test-BH results are contained in Eq.~(\ref{chitestPM}) below.

We also argue in Sec.~\ref{sec:HPN} that Eqs.~(\ref{2PMtwospin}) and (\ref{chitestPM}) allow one
to reconstruct the original Hamiltonians at the considered order (up to a gauge or canonical transformation).
Hence Eq.~(\ref{2PMtwospin}) applied to Eq.~(\ref{chitestPM}) encodes a number of rather lengthy PN results---the complete 1PM and 2PM parts of all the PN results, thus including the complete PN results through NLOs, for aligned spins---in a strikingly compact manner.

\section{The post-Newtonian--\\post-Minkowskian  expansion of the scattering angle}\label{sec:PN}

Here we take PN results for canonical Hamiltonians governing the conservative local-in-time dynamics of arbitrary-mass-ratio binary BHs, specialized to the aligned-spin case, and derive from them a PN-PM expansion of the scattering-angle function.  This generalizes to spin-squared order similar calculations in Ref.~\cite{Bini:2017wfr} through linear order in spin.  We also discuss how this process can be run in reverse, to deduce from the scattering angle a complete aligned-spin Hamiltonian (valid for both unbound and bound orbits), modulo phase-space gauge freedom, at least at the considered PN orders.

We begin in Sec.~\ref{sec:Htochi} with a general discussion of canonical Hamiltonians for binary-BH conservative local-in-time dynamics, the specialization to the aligned-spin case, and the procedure for deriving the scattering angle from an aligned-spin Hamiltonian.  An important ingredient, discussed in Sec.~\ref{sec:translate}, is the translation from the ``canonical variables'' associated with the canonical Hamiltonian---specifically, the orbital angular momentum and the corresponding impact parameter---to the ``covariant variables'' (those used in Sec.~\ref{sec:angles}) in terms of which the spin-dependent parts of the scattering angle take their simplest forms.  In Sec.~\ref{sec:HPN}, we display an aligned-spin Hamiltonian in a ``quasi-isotropic'' gauge, at all of the PN orders listed in Sec.~\ref{sec:2PMspin}, derived via canonical transformations from the Hamiltonians given in Refs.~\cite{Damour:2000we,Levi:2015uxa,Levi:2016ofk}.  We present the resultant PN-PM-expanded scattering angle in Sec.~\ref{sec:chiPN}.  We restore in this section factors of the speed of light $c$ which were set to 1 in the previous two sections.

\subsection{Canonical Hamiltonians, aligned spins, and the scattering angle}\label{sec:Htochi}

A canonical Hamiltonian encoding the conservative local-in-time dynamics of a generic binary BH in the center-of-mass frame, with arbitrary spin orientations,
\be\label{Hgeneric}
H\Big((m_1,\bs a_1),(m_2,\bs a_2),\bs R,\bs P\Big),
\ee
depends on the (constant) rest masses $m_1$ and $m_2$, a canonical relative position $\bs R(t)$ and its conjugate momentum $\bs P(t)$, and the canonical spin vectors $\bs S_1(t)$ and $\bs S_2(t)$ with rescaled versions
\be
\bs a_{1}=\frac{\bs S_{1}}{m_{1}c},\qquad
\bs a_{2}=\frac{\bs S_{2}}{m_{2}c},
\ee
having dimensions of length.
The Hamiltonian determines the canonical equations of motion,
\begin{alignat}{3}\label{Hameqs}
\begin{aligned}
&\dot {\bs R}=\frac{\doe H}{\doe\bs P},
\qquad
\dot{\bs P}=-\frac{\doe H}{\doe\bs R},
\\
&\quad\dot{\bs S}_A=-\bs S_A\times \frac{\doe H}{\doe\bs S_A},
\end{aligned}
\end{alignat}
with $A=1,2$ (no sum implied), via the canonical Poisson brackets
\be\label{brackets}
\{R^i,P_j\}=\delta^i_j,\qquad \{S_A^i,S_A^j\}=\epsilon^{ij}{}_kS_A^k,
\ee
with all others vanishing.

In the aligned-spin configuration, both spin vectors are constant and parallel to the constant (canonical) orbital angular momentum vector
\be\label{introL}
\bs L=\bs R\times \bs P=L\,\hat{\bs L},
\ee
with $L=|\bs L|$,
\be\label{alignedspins}
\bs a_1=a_1\, \hat{\bs L},\qquad \bs a_2=a_2\,\hat{\bs L}.
\ee
The orbit is confined to the plane orthogonal to $\hat{\bs L}$, in which we can use polar coordinates $(R,\phi)$, where $R=|\bs R|$, with canonically conjugate momenta $(P_R,P_\phi)$, where $P_\phi=L$ is the (canonical) orbital angular momentum, with
\be
\bs P^2=P_R^2+\frac{L^2}{R^2}.
\ee
Note that we are implicitly employing a flat background Euclidean 3-metric here.  An aligned-spin binary-BH canonical Hamiltonian takes the form
\be\label{alignedH}
H\Big((m_1,a_1),(m_2,a_2),R,P_R,L\Big),
\ee
with the equations of motion
\begin{alignat}{3}\label{Hampolar}
\begin{aligned}
\dot R&=\dfrac{\doe H}{\doe P_R},
\qquad&
\dot{P}_R&=-\dfrac{\doe H}{\doe R},
\\
\dot\phi&=\dfrac{\doe H}{\doe L},
&
\dot{L}&=-\dfrac{\doe H}{\doe\phi}=0,
\end{aligned}
\end{alignat}
where $L=P_\phi$ is a constant of motion due to the system's axial symmetry.  
For the generic Hamiltonians of the form (\ref{Hgeneric}) which we employ, from Refs.~\cite{Damour:2000we,Levi:2015uxa,Levi:2016ofk}, the corresponding aligned-spin Hamiltonian of the form (\ref{alignedH}) can be obtained simply by inserting the aligned-spin relations (\ref{alignedspins}) and simplifying.\footnote{This involves, e.g., taking $\bs a\cdot\bs L\to a L$ (as is the spin dependence of all linear-in-spin terms) and setting to zero the quantities $\bs a\cdot\bs R$ and $\bs a\cdot \bs P$.  In general, with spin-squared terms, one should be concerned that this process may not commute with the process of obtaining the equations of motion from the Hamiltonian (involving derivatives).   However, one can verify that these processes do commute for the generic Hamiltonians we employ, most notably because the quantities $\bs a\cdot\bs R$ and $\bs a\cdot \bs P$ which vanish for aligned spins never appear as lone factors in a given term in the Hamiltonian; rather, they always appear multiplied by a second such factor.}

Apart from the spins $a_1$ and $a_2$ appearing as constant parameters, the aligned-spin binary-BH Hamiltonian and equations of motion (\ref{alignedH})--(\ref{Hampolar}) are identical in form to those for a two-point-mass system.
It follows from the equations of motion (\ref{Hampolar}), as shown e.g.\ in Ref.~\cite{Damour:2017zjx}, that the gauge-invariant scattering angle $\chi$---the total change in the angle coordinate, $
\Delta\phi$, minus $\pi$ (minus $\Delta\phi$ for $G\to 0$)---can be found by solving the relation $H=H(R,P_R,L)$ giving the Hamiltonian for the relation $P_R=P_R(H,L,R)$ giving the radial momentum (up to a sign), and then evaluating
\be\label{chi}
\chi(H,L)=-\int \mr dR\,\frac{\doe}{\doe L}\,P_R(H,L,R)-\pi
\ee
along the appropriate path through the phase space---namely: first, with $P_R<0$, from $R=\infty$ down to $R_\mr{min}$ ($>0$) where $P_R=0$, and then back to infinity with $P_R>0$.  Here we are assuming that $L$ and $H$ are such that $P_R$ is real as $R\to\infty$ (such that the orbit is unbound).  Note that we have suppressed in Eq.~(\ref{chi}) the dependence on the constant masses and spins.  The total change in the angle coordinate $\phi$ will correspond to the physical center-of-mass-frame scattering angle as long as the Hamiltonian $H$ reduces to a standard form for a free system in Minkowski space as $R\to\infty$---specifically, with that (0PM) Hamiltonian depending only on $\bs P^2=P_R^2+L^2/R^2$ (and on the constant rest masses), with $H$ and $L$ being the physical center-of-mass-frame total energy and (canonical) orbital angular momentum, respectively.

We note that for a scattering orbit, as opposed to a bound (or, more precisely, nearly circular) orbit, the velocity and the gravitational field strength become independent.  Hence we expand the scattering angle (\ref{chi}) independently in $G$ (PM expansion) and $c^{-1}$ (PN expansion).  When this expansion is applied to the integrand of Eq.~(\ref{chi}), the individual parts of the integral become simple to evaluate (but one has to deal with singularities).  We refer the reader to Appendix B of Ref.~\cite{Damour:1988mr} for an explanation of this perturbative integration method.

\subsection{Orbital angular momenta and impact parameters: ``canonical'' versus ``covariant'' variables}\label{sec:translate}

The $L$ above is the magnitude of the \emph{canonical} orbital angular momentum
\be
\bs L\equiv\bs L_\mr{can}=\bs R\times\bs P,
\ee
given in terms of the canonical $\bs R$ and $\bs P$ from the Hamiltonian.  This corresponds \cite{Barausse:2009aa,Vines:2016unv,Vines:2017hyw} to the physical orbital angular momentum (at least at infinity) defined in terms of the BHs' worldlines which are specified by canonical (or Pryce-Newton-Wigner \cite{Pryce:1935ibt, Pryce:1948pf, Newton:1949cq}) spin supplementary conditions for each BH with respect to the system's center-of-mass frame.

Referring the reader to Ref.~\cite{Vines:2017hyw} for further details, in the aligned-spin case, a simple way to relate the canonical orbital angular momentum $L=L_\mr{can}$ to the ``covariant'' orbital angular momentum $L_\mr{cov}$ (the one defined in terms of the BHs' Tulczyjew-Dixon worldlines as discussed in Sec.~\ref{sec:angles}), is to note their respective relationships (at infinity) to the invariant magnitude $J_\mr{tot}$ of the two-BH system's center-of-mass-frame total angular momentum and to the (rescaled) spins $a_1$ and $a_2$.  From Eqs.~(96a) and (106d) of Ref.~\cite{Vines:2017hyw}, we have
\begin{equation}
\begin{split}\label{Jtot}
J_\mr{tot}&=L_\mr{can}+m_1c\,a_1+m_2c\,a_2
\\
&=L_\mr{cov}+\frac{E_1}{c}a_1+\frac{E_2}{c}a_2,
\end{split}
\end{equation}
where $E_1$ and $E_2$ are the BHs' individual energies in the center-of-mass frame.  These energies are defined, at infinity, by $E_1=p_1\cdot u_\mr{cm}$ and $E_2=p_2\cdot u_\mr{cm}$, where $u_\mr{cm}^\mu=(p_1^\mu+p_2^\mu)/H$ is the 4-velocity of the center-of-mass frame.  Like the relative speed $v$ at infinity and the Lorentz factor $\gamma$, the energies $E_1$ and $E_2$ can be expressed solely in terms of the rest masses $m_1$ and $m_2$ and the total center-of-mass-frame energy $H$, equal to the quantity $E$ from (\ref{Egamma}),
\bse\label{HE}
\be\label{HEM}
H=E=Mc^2\sqrt{1+2\nu(\gamma-1)}=E_1+E_2,
\ee
as follows.  Let us define the total energy per total rest-mass energy,
\be\label{Gamma}
\Gamma=\frac{H}{Mc^2}=\sqrt{1+2\nu(\gamma-1)},
\ee
so that
\be
\gamma=\frac{1}{\sqrt{1-v^2/c^2}}=1+\frac{\Gamma^2-1}{2\nu},
\ee
recalling the definitions of the total rest mass $M$, the reduced mass $\mu$, and the symmetric mass ratio $\nu$, and introducing the antisymmetric mass ratio $\delta$,
\begin{alignat}{3}
\begin{aligned}
M&=m_1+m_2,\qquad \mu=\frac{m_1m_2}{M}=\nu M,
\\
\delta&=\frac{m_1-m_2}{M}=\sqrt{1-4\nu}\frac{m_1-m_2}{|m_1-m_2|}.
\end{aligned}
\end{alignat}
Then the individual energies can be expressed as
\begin{alignat}{5}
\begin{aligned}
E_1&=\sqrt{m_1^2c^4+|\bs p|^2c^2}=\frac{Mc^2}{2}\left(\Gamma+\frac{\delta}{\Gamma}\right),
\\
E_2&=\sqrt{m_2^2c^4+|\bs p|^2c^2}=\frac{Mc^2}{2}\left(\Gamma-\frac{\delta}{\Gamma}\right),
\end{aligned}
\end{alignat}
where
\be
|\bs p|=\frac{\mu\gamma v}{\Gamma}
\ee 
\ese
is the magnitude of the (physical) relative momentum $\bs p$, orthogonal to $u_\mr{cm}^\mu$, as defined in Sec.~II.H.1 of Ref.~\cite{Vines:2017hyw} where it is called $p_{\perp}^\mu$, such that $p_1^\mu=E_1u_\mr{cm}^\mu+\bs p^\mu$ and $p_2^\mu=E_2 u_\mr{cm}^\mu-\bs p^\mu$ asymptotically (see Fig.~\ref{fig:twovstest}).  Note that $|\bs p|$ generally differs from the magnitude $|\bs P|$ of the canonical momentum $\bs P$ (at infinity) in the Hamiltonian.  Finally, one finds from (\ref{Jtot})--(\ref{HE}) that the relationship between the (aligned-spin) canonical and covariant orbital angular momenta can be expressed as
\begin{alignat}{3}
\begin{aligned}
L_\mr{can}&=L_\mr{cov}+\frac{E_1-m_1c^2}{c}a_1+\frac{E_2-m_2c^2}{c}a_2
\\
&=L_\mr{cov}+M c\frac{\Gamma-1}{2}\left(a_+-\frac{\delta}{\Gamma}a_-\right),
\end{aligned}
\end{alignat}
where we define
\be
a_+=a_1+a_2,\qquad a_-=a_1-a_2.
\ee

The impact parameters---the distances (in either BH's rest frame or in the center-of-mass frame) orthogonally separating the BHs' asymptotic worldlines, $b_\mr{can}$ for the center-of-mass-frame Pryce-Newton-Wigner worldlines, and $b_\mr{cov}\equiv b$ for the Tulczyjew-Dixon worldlines---are related to the orbital angular momenta by
\begin{alignat}{3}\label{bs}
\begin{aligned}
b_\mr{can}&=\frac{L_\mr{can}}{|\bs p|},
\\
b\equiv b_\mr{cov}&=\frac{L_\mr{cov}}{|\bs p|},
\end{aligned}
\end{alignat}
as in Eq.~(67) of Ref.~\cite{Vines:2017hyw}.  Thus, from (\ref{HE})--(\ref{bs}), the canonical orbital angular momentum $L=L_\mr{can}$, appearing in the aligned-spin Hamiltonian (\ref{alignedH}), is related to the covariant impact parameter $b=b_\mr{cov}$ by
\begin{alignat}{3}\label{key}
\begin{aligned}
  L=L_\mr{can}
&=\frac{\mu\gamma vb}{\Gamma}+M c\frac{\Gamma-1}{2}\left(a_+-\frac{\delta}{\Gamma}a_-\right).
\end{aligned}
\end{alignat}
Using this key relation to express $L$ in terms of $b$ leads to significant simplifications of the spin-dependent parts of the PN-PM expansion of the scattering angle.

Note that, given fixed rest masses, the quantities $v$, $\gamma$, and $\Gamma$ are each determined by the total energy $H$, and vice versa, via (\ref{HEM})--(\ref{Gamma}).  We can thus trade $H$ for $v$, and $L$ for $b$, as the independent variables in the scattering-angle function.  Then (\ref{chi}) is replaced by
\be\label{chib}
\chi(v,b)=-\frac{\Gamma}{\mu\gamma v}\int \mr dR\,\frac{\doe}{\doe b}\,P_R(v,b,R)-\pi,
\ee
where $P_R(v,b,R)$ is found by solving the Hamiltonian relation $H=H(R,P_R,L)$ while using (\ref{HE}) and (\ref{key}) to eliminate $H$ and $L$ in favor of $v$ and $b$.

\subsection{The post-Newtonian Hamiltonian in a quasi-isotropic gauge}\label{sec:HPN}

In the following we collect the PN Hamiltonians that enter the calculation of the scattering angle.
A canonical Hamiltonian for a binary BH including the NNLO-PN contributions up to quadratic order in the BHs' spins is given in Refs.~\cite{Levi:2015uxa,Levi:2016ofk}.  The NNNLO (3PN) point-mass contributions can be found in Ref.~\cite{Damour:2000we}.
Since the Hamiltonians are given in different gauges (or canonical coordinates) in the literature, we need to take special care to transform them to the same canonical variables.

Using PN perturbative canonical transformations as discussed in, e.g., Refs.~\cite{Buonanno:1998gg,Damour:2000we,Levi:2015uxa}, one finds that the aligned-spin Hamiltonian can be brought into a form such that it depends on the momentum $\bs P$ only through $\bs P^2=P_R^2+L^2/R^2$ (not separately on $P_R$ and $L$), except in the odd-in-spin terms where one has single factors of $\bs L\cdot\bs a\to La$.  This defines a ``quasi-isotropic'' gauge.  
One finds, in fact, that these requirements fix the gauge of the Hamiltonian up to a family of canonical transformations determined by one function of one (dynamical) variable (and rest masses), and that this freedom is fixed once a 0PM Hamiltonian of the form $H_\mr{0PM}(\bs P^2;m_1,m_2)$ is fixed---at the least, at the PN orders considered here.  It can be easily verified that the scattering angle derived from (\ref{chib}) is invariant under this family of canonical transformations, as is a consequence of the less easily verified fact that the scattering angle is invariant under arbitrary canonical transformations of the Hamiltonian preserving the property that $H_\mr{0PM}$ depends only on $\bs P^2$ (and rest masses).  These facts, along with counting coefficients in the angle and in the Hamiltonian, demonstrate that the complete (phase-space-)gauge-invariant information of the Hamiltonian is encoded in the scattering angle.  At least perturbatively, at the orders considered here, one can deduce a valid Hamiltonian, modulo gauge, by posing an ansatz with undetermined coefficients, computing the scattering angle, and matching coefficients.  Specifically, for example, one can start from the Hamiltonians given precisely as in Eqs.~(\ref{Ha0}), (\ref{Ha1}) and (\ref{Ha2}) below, with \emph{unknown} coefficients $\alpha_{\ldots}$, compute the scattering angle from that Hamiltonian, and set it equal to the final result (\ref{chiPNfinal}); this yields a redundant system of equations with a family of solutions for the Hamiltonian coefficients having one constant (mass-dependent) free coefficient at each (point-mass) PN order starting from 1PN, and this remaining freedom can be fixed by demanding a particular form for the 0PM Hamiltonian as a function of $\bs P^2$.  We present in the following the results for the Hamiltonian in a certain quasi-isotropic gauge, with the remaining freedom fixed as described in the following subsection.

\subsubsection{Point-mass contributions}
At 4PN order the point-mass or spin-independent Hamiltonian becomes nonlocal in time \cite{Damour:2014jta}; see Ref.~\cite{Bini:2017wfr} for a calculation of the scattering angle to this order.  Since the present paper is concerned with spin contributions, we restrict our attention here to the simpler local-in-time point-mass Hamiltonian up to 3PN.  The point-mass contributions to the Hamiltonian, in an isotropic gauge, can be expressed as follows,
\begin{widetext}
\begin{alignat}{15}\label{Ha0}
H_{a^0}\;=&&Mc^2 \quad&
\nnm\\
&&+\mu\bigg(
\dfrac{{\bs P}^2}{2\mu^2}
&
-\dfrac{GM}{R}\bigg)
&&&&&&&&
\;\,:\;\mr{0PN}
\nnm\\
&&\quad
+\dfrac{\mu}{c^2}\bigg(
\alpha_{10}\dfrac{\bs P^4}{\mu^4}
&
+\alpha_{11}\dfrac{\bs P^2}{\mu^2}\dfrac{GM}{R}
&&
+\alpha_{12}\dfrac{(GM)^2}{R^2}
\bigg)
&&&&&&
\;\,:\;\mr{1PN}
\nnm\\
&&\quad
+\dfrac{\mu}{c^4}\bigg(
\alpha_{20}\dfrac{\bs P^6}{\mu^6}
&
+\alpha_{21}\dfrac{\bs P^4}{\mu^4}\dfrac{GM}{R}
&&
+\alpha_{22}\dfrac{\bs P^2}{\mu^2}\dfrac{(GM)^2}{R^2}
&&
+\alpha_{23}\dfrac{(GM)^3}{R^3}
\bigg)
&&&&
\;\,:\;\mr{2PN}
\\\nnm
&&\quad\underbrace{
\phantom{\Bigg|}
+\dfrac{\mu}{c^6}\bigg(
\alpha_{30}\dfrac{\bs P^8}{\mu^8}
}_\mr{0PM}
&\underbrace{
+\,\alpha_{31}\dfrac{\bs P^6}{\mu^6}\dfrac{GM}{R}
\phantom{\Bigg|}
}_\mr{1PM}
&&\underbrace{
+\,\alpha_{32}\dfrac{\bs P^4}{\mu^4}\dfrac{(GM)^2}{R^2}
\phantom{\Bigg|}
}_\mr{2PM}
&&\underbrace{
+\,\alpha_{33}\dfrac{\bs P^2}{\mu^2}\dfrac{(GM)^3}{R^3}
\phantom{\Bigg|}}_\mr{3PM}
&&\underbrace{
+\,\alpha_{34}\dfrac{(GM)^4}{R^4}
\bigg)\phantom{\Bigg|}}_\mr{4PM}
&&
\;\,:\;\mr{3PN},
\end{alignat}
\end{widetext}
with the 1PN coefficients,
\bse\label{pointmasscoefficients}
\be
\bpm
\alpha_{10} \\ \alpha_{11} \\ \alpha_{12}
\epm
=
\bpm
-1/8 & -1/8
\\
-3/2 & 1/2
\\
1/2 & -1/2
\epm
\bpm
1 \\ \nu
\epm,
\ee
the 2PN coefficients,
\be
\bpm
\alpha_{20} \\ \alpha_{21} \\ \alpha_{22} \\ \alpha_{23}
\epm
=
\bpm
1/16 & 1/16 & 1/16
\\
5/8 & 5/8 & -3/8
\\
5/2 & -1/4 & 3/4
\\
-1/4 & 0 & -1/2
\epm
\bpm
1 \\ \nu \\ \nu^2
\epm,
\ee
and the 3PN coefficients,
\begin{alignat}{3}
\bpm
\alpha_{30} \\ \alpha_{31} \\ \alpha_{32} \\ \alpha_{33}
\epm
&=
\bpm
-5/2^7 & -5/2^7 & -3/2^6 & -5/2^7
\\
-7/2^4 & -7/2^4 & -3/8 & 5/2^4
\\
-27/2^4 & -15/2^4 & 3/8 & -15/2^4
\\
-25/8 & 5/8 & 3/8 & 5/4
\epm
\bpm
1 \\ \nu \\ \nu^2 \\ \nu^3
\epm,
\nonumber\\
\alpha_{34}&=\frac{1}{8}+\left(\frac{235}{24}-\frac{41\pi^2}{64}\right)\nu-\frac{1}{4}\nu^2 -\frac{5}{8}\nu^3.
\end{alignat}
\ese

The gauge of the Hamiltonian here has been fixed by requiring that it is isotropic, depending on the
canonical momentum $\bs P$ only through ${\bs P}^2$, and (to fix the remaining gauge freedom discussed above) by requiring that the 0PM column of Eq.~(\ref{Ha0}) matches the expansion in $1/c^2$ of
\be\label{0PMcolumn}
H_\mr{0PM}=\sqrt{M^2c^4+2Mc^2\left(\sqrt{\mu^2c^4+\bs P^2c^2}-\mu c^2\right)},
\ee
which is the result of the EOB energy map (\ref{energymap}) being applied, with $E\to H$, to the Hamiltonian $E_\mr t\to H_\mr t=\sqrt{\mu^2c^4+\bs P^2c^2}$ for a free particle of mass $\mu$ in flat spacetime.\footnote{The 1PN part of the Hamiltonian in the standard Arnowitt-Deser-Misner (ADM) gauge \cite{Blanchet:2013haa} reads
\begin{alignat}{3}
H_\mr{1PN}^\mr{ADM}&=\frac{\mu}{c^2}\bigg[-\frac{1-3\nu}{8}\frac{\bs P^4}{\mu^4}-\frac{3+\nu}{2}\frac{L^2}{\mu^2R^2}\frac{GM}{R}
\nnm\\
&\qquad\quad-\frac{3+2\nu}{2}\frac{P_R^2}{\mu^2}\frac{GM}{R}+\frac{G^2M^2}{2R^2}\bigg].
\end{alignat}
Using the generating function
\be
\mc G_\mr{1PN}=\bs P\cdot\bs R\frac{1}{c^2}\bigg(\beta_1\frac{\bs P^2}{\mu^2}+\beta_2\frac{GM}{R}\bigg),
\ee
a perturbative canonical transformation yields the transformed 1PN Hamiltonian $H'_\mr{1PN}=H_\mr{1PN}^\mr{ADM}+\{\mc G_\mr{1PN},H_\mr{0PN}\}$,
\begin{alignat}{3}
H'_\mr{1PN}&=\frac{\mu}{c^2}\bigg[-\frac{1-3\nu+8\beta_1}{8}\frac{\bs P^4}{\mu^4}
\nnm\\
&\qquad\quad-\frac{3+\nu-2\beta_1+2\beta_2}{2}\frac{L^2}{\mu^2R^2}\frac{GM}{R}
\nnm\\
&\qquad\quad-\frac{3+2\nu-6\beta_1}{2}\frac{P_R^2}{\mu^2}\frac{GM}{R}+\frac{1+2\beta_2}{2}\frac{G^2M^2}{R^2}\bigg].
\end{alignat}
We see that the second two terms will combine into one term proportional to $\bs P^2=L^2/R^2+P_R^2$ if
\be
\beta_2=\frac{\nu-4\beta_1}{2}\quad\Rightarrow\quad\tr{an isotropic gauge},
\ee
but there is still the freedom of choosing $\beta_1$, so that we have a one-parameter family of isotropic gauges up to the 1PN point-mass level.  While continuing to enforce the isotropic condition, one similarly encounters one new free parameter at each $n$PN point-mass level with $n\ge1$.  In (\ref{pointmasscoefficients}), we have used $\beta_1=\nu/2$ at 1PN and corresponding choices at higher PN orders which make the 0PM column of Eq.~(\ref{Ha0}) match the $1/c^2$ expansion of Eq.~(\ref{0PMcolumn}), yielding an isotropic EOB gauge for the point-mass Hamiltonian.  One can instead make the 0PM column match 
\be
H_\mr{0PM}'=\sqrt{m_1^2c^4+\bs P^2c^2}+\sqrt{m_2^2c^4+\bs P^2c^2},
\ee
as in the 0PM part of the ADM-gauge Hamiltonian \cite{Blanchet:2013haa}, or as in the isotropic-gauge Hamiltonian of Ref.~\cite{Cheung:2018wkq}, by using $\beta_1=0$ and corresponding choices at higher PN orders.
}
This defines a unique isotropic EOB gauge for the point-mass Hamiltonian $H_{a^0}(R,\bs P^2;m_1,m_2)$.  It also fixes a unique quasi-isotropic gauge for the aligned-spin Hamiltonian $H(R,\bs P^2,L;m_1,m_2,a_1,a_2)$.  Once the remaining freedom (among isotropic gauges) has been fixed in this way for the point-mass part of the Hamiltonian, no further gauge freedom is present in the following aligned-spin contributions, if we impose the quasi-isotropic conditions discussed above.

\subsubsection{Spin-orbit contributions}
The linear-in-spin, or spin-orbit, Hamiltonians up to NNLO from Ref.~\cite{Levi:2015uxa} can be brought into a quasi-isotropic gauge through a canonical transformation, yielding the form
\begin{widetext}
\begin{alignat}{15}\label{Ha1}
H_{a^1}=&\;\frac{L}{cR^2}\bigg(\frac{7}{4}a_++\frac{\delta}{4}a_-\bigg)\frac{GM}{R}
&&&&&&
\;\,:\;\tr{LO (1.5PN)}
\nnm\\\nnm
&+\frac{L}{c^3R^2}\bpm a_+ & \delta a_- \epm\Bigg[
\bpm
\alpha_{11+} \\ \alpha_{11-}
\epm
\frac{\bs P^2}{\mu^2}\frac{GM}{R}
&&
+
\bpm
\alpha_{12+} \\ \alpha_{12-}
\epm
\frac{(GM)^2}{R^2}
\Bigg]
&&&&
\;\,:\;\tr{NLO (2.5PN)}
\\\nnm
&
\underbrace{
+\frac{L}{c^5R^2}\bpm a_+ & \delta a_- \epm\Bigg[
\bpm
\alpha_{21+} \\ \alpha_{21-}
\epm
\frac{\bs P^4}{\mu^4}\frac{GM}{R}
}_\mr{1PM}
&&
\underbrace{
+
\bpm
\alpha_{22+} \\ \alpha_{22-}
\epm
\frac{\bs P^2}{\mu^2}\frac{(GM)^2}{R^2}
\phantom{\Bigg|}
}_\mr{2PM}
&&
\underbrace{
+
\bpm
\alpha_{23+} \\ \alpha_{23-}
\epm
\frac{(GM)^3}{R^3}
\Bigg]
}_\mr{3PM}
&&
\;\,:\;\tr{NNLO (3.5PN)},
\\
\end{alignat}
\end{widetext}
recalling $a_\pm=a_1\pm a_2$ and $\delta=\dfrac{m_1-m_2}{M}$.

With the point-mass part in the isotropic EOB gauge, the spin-orbit coefficients in the unique quasi-isotropic EOB gauge are as follows: at NLO,
\be
\bpm
\alpha_{11+}
\\
\alpha_{11-}
\\
\alpha_{12+}
\\
\alpha_{12-}
\epm
=
\bpm
-5/2^4 & -29/2^4
\\
5/2^4 & -5/2^4
\\
-11/2 & 2
\\
-1/2 & 1/2
\epm
\bpm 1 \\ \nu \epm,
\ee
and at NNLO,
\be
\bpm
\alpha_{21+}
\\
\alpha_{21-}
\\
\alpha_{22+}
\\
\alpha_{22-}
\\
\alpha_{23+}
\\
\alpha_{23-}
\epm
=
\bpm
7/2^5 & 11/2^4 & 59/2^5
\\
-7/2^5 & -5/2^4 & 11/2^5
\\
27/2^4 & 123/2^4 & -9/2
\\
-27/2^4 & 21/2^4 & -9/8
\\
159/2^4 & -99/2^4 & 45/2^4
\\
9/2^4 & -11/2^4 & 15/2^4
\epm
\bpm 1 \\ \nu \\ \nu^2 \epm.
\ee

\subsubsection{Spin-squared and quadrupole contributions}

The spin-squared parts of the Hamiltonians up to NNLO from Refs.~\cite{Levi:2015uxa,Levi:2016ofk} include contributions from the bodies' quadrupole moments, which depend on the internal structure.  The quadrupole moments here are specialized to those of BHs.  After performing a perturbative canonical transformation, the spin-squared Hamiltonian in the quasi-isotropic EOB gauge reads
\begin{widetext}
\begin{alignat}{15}\label{Ha2}
H_{a^2}=&\;-\mu\frac{a_+^2}{2R^2}\frac{GM}{R}
&&&&&&
\;\,:\;\tr{LO (2PN)}
\nnm\\
&+\frac{\mu}{c^2R^2}
\bpm 
a_+^2 & \delta a_+a_- & a_-^2
\epm
\Bigg[
\bpm
\alpha_{11++} \\ \alpha_{11+-} \\ \alpha_{11--}
\epm
\frac{\bs P^2}{\mu^2}\frac{GM}{R}
&&
+
\bpm
\alpha_{12++} \\ \alpha_{12+-} \\ \alpha_{12--}
\epm
\frac{(GM)^2}{R^2}
\Bigg]
&&&&
\;\,:\;\tr{NLO (3PN)}
\nnm\\\nnm
&
\underbrace{
+\frac{\mu}{c^4R^2}
\bpm 
a_+^2 & \delta a_+a_- & a_-^2
\epm
\Bigg[
\bpm
\alpha_{21++} \\ \alpha_{21+-} \\ \alpha_{21--}
\epm
\frac{\bs P^4}{\mu^4}\frac{GM}{R}
}_\mr{1PM}
&&
\underbrace{
+
\bpm
\alpha_{22++} \\ \alpha_{22+-} \\ \alpha_{22--}
\epm
\frac{\bs P^2}{\mu^2}\frac{(GM)^2}{R^2}
}_\mr{2PM}
&&
\underbrace{
+
\bpm
\alpha_{23++} \\ \alpha_{23+-} \\ \alpha_{23--}
\epm
\frac{(GM)^3}{R^3}
\Bigg]
}_\mr{3PM}
&&
\;\,:\;\tr{NNLO (4PN)}
\\
\end{alignat}
\end{widetext}
with the NLO coefficients
\bse
\be
\bpm
\alpha_{11++} \\ \alpha_{11+-} \\ \alpha_{11--}
\\
\alpha_{12++} \\ \alpha_{12+-} \\ \alpha_{12--}
\epm
=
\bpm
-9/2^5 & 3/4
\\
-7/2^4 & 0
\\
-1/2^5 & 1/8
\\
83/2^5 & -1
\\
-3/2^4 & 0
\\
3/2^5 & 1/8
\epm
\bpm 1 \\ \nu \epm,
\ee

\newpage

\noindent and the NNLO coefficients
$$
\bpm
\alpha_{21++} \\ \alpha_{21+-} \\ \alpha_{21--}
\\
\alpha_{22++} \\ \alpha_{22+-} \\ \alpha_{22--}
\\
\alpha_{23++} \\ \alpha_{23+-} \\ \alpha_{23--}
\epm
=
\bpm
5/2^5 & 11/2^5 & -15/2^4
\\
3/2^4 & 25/2^5 & 0
\\
-1/2^5 & 3/2^4 & -1/4
\\
93/2^6 & -493/2^6 & 21/8
\\
105/2^5 & -77/2^6 & 0
\\
9/2^6 & -17/2^5 & 3/8
\\
-425/2^6 & 241/2^6 & -15/8
\\
31/2^5 & -45/2^6 & 0
\\
-21/2^6 & -7/8 & 1/2
\epm
\bpm 1 \\ \nu \\ \nu^2 \epm.
$$
\ese

\subsection{The scattering angle}\label{sec:chiPN}

Solving the expression of the Hamiltonian for the radial momentum $P_R$ as discussed below (\ref{chib}), inserting this into (\ref{chib}), and integrating (using, e.g., the method described in \cite{Damour:1988mr}) yields the scattering angle as follows.  Factoring out the quantity $\Gamma=H/M$ seen in the numerator of the prefactor in (\ref{chib}), we find the spin$^0$ part,
\begin{widetext}
\bse\label{chiPNfinal}
\begin{alignat}{5}
\frac{\chi_{a^0}}{\Gamma}&=\frac{GM}{v^2b}\bigg[2+2\frac{v^2}{c^2}+{\cal O}(\frac{v^8}{c^8})\bigg]
\\\nnm
&\quad
+\pi \left(\frac{GM}{v^2b}\right)^2\left[
3\frac{v^2}{c^2}
+\frac{3}{4}\frac{v^4}{c^4}
+{\cal O}(\frac{v^8}{c^8})
\right]
\\\nnm
&\quad
+\left(\frac{GM}{v^2b}\right)^3\left[
-\frac{2}{3}
+2\frac{15-\nu}{3}\frac{v^2}{c^2}
+\frac{60-13\nu}{2}\frac{v^4}{c^4}
+\frac{40-227\nu}{12}\frac{v^6}{c^6}
+{\cal O}(\frac{v^8}{c^8})
\right]
\\\nnm
&\quad
+\pi \left(\frac{GM}{v^2b}\right)^4\left[
15\frac{7-2\nu}{4}\frac{v^4}{c^4}
+\left(\frac{105}{4}-\frac{437}{8}\nu+\frac{123}{128}\pi^2\nu\right)\frac{v^6}{c^6}
+{\cal O}(\frac{v^8}{c^8})
\right]
\\\nnm
&\quad+\mc O(G^5),\phantom{\bigg|}
\end{alignat}
the spin$^1$ part,
\begin{alignat}{5}
\frac{\chi_{a^1}}{\Gamma}
&=
\frac{v}{c}\frac{\bpm a_+ & \delta a_-\epm}{b}\Bigg\{\frac{GM}{v^2b}
\bigg[
\bpm -4 \\ 0 \epm +{\cal O}(\frac{v^6}{c^6})
\bigg]
\\\nnm
&\quad
+\pi \left(\frac{GM}{v^2b}\right)^2
\left[
{-}\frac{1}{2}\bpm 7 \\ 1 \epm
-\frac{3}{4}\bpm 7 \\ 1 \epm\frac{v^2}{c^2}
+{\cal O}(\frac{v^6}{c^6})
\right]
\\\nnm
&\quad
+\left(\frac{GM}{v^2b}\right)^3
\left[
{-}2\bpm 5 \\ 1 \epm
-20\bpm 5-\nu/2 \\ 1 \epm\frac{v^2}{c^2}
-10\bpm 5-77\nu/20 \\ 1 \epm\frac{v^4}{c^4}
+{\cal O}(\frac{v^6}{c^6})
\right]\Bigg\}
\\\nnm
&\quad+\mc O(G^4),\phantom{\bigg|}
\end{alignat}
and the spin$^2$ part,
\begin{alignat}{5}
\frac{\chi_{a^2}}{\Gamma}
&=
\frac{\bpm a_+^2 & \delta a_+a_- & a_-^2\epm}{b^2}
\Bigg\{
\frac{GM}{v^2b}
\bigg[
\bpm 2 \\ 0 \\ 0 \epm 
+\bpm 2 \\ 0 \\ 0 \epm \frac{v^2}{c^2}
+{\cal O}(\frac{v^6}{c^6})
\bigg]
\\\nnm
&\quad
+\pi \left(\frac{GM}{v^2b}\right)^2
\bigg[
\frac{3}{2}\bpm 1 \\ 0 \\ 0 \epm
+\frac{3}{16}\bpm 59 \\ 14 \\ -1 \epm\frac{v^2}{c^2}
+\frac{3}{64}\bpm 47 \\ 14 \\ -1 \epm\frac{v^4}{c^4}
+{\cal O}(\frac{v^6}{c^6})
\bigg]
\\\nnm
&\quad
+\left(\frac{GM}{v^2b}\right)^3
\bigg[
\bpm 4 \\ 0 \\ 0 \epm
+4\bpm 35-\nu \\ 10 \\ -2\nu \epm\frac{v^2}{c^2}
+\bpm 220-45\nu  \\ 80 \\ -12\nu \epm\frac{v^4}{c^4}
+{\cal O}(\frac{v^6}{c^6})
\bigg]\Bigg\}
\\\nnm
&\quad+\mc O(G^4).\phantom{\bigg|}
\end{alignat}
\ese
\end{widetext}
We can already see here that, in these forms, in terms of these variables, remarkably, the 1PM and 2PM parts (of the right-hand sides) are independent of the symmetric mass ratio $\nu$ and linear in the antisymmetric mass ratio $\delta$.  In the test-body limit, say, $m_2\to 0$, we have $\nu\to0$, $\delta\to 1$, and $\Gamma\to 1$.  One can then verify directly from Eq.~(\ref{chiPNfinal}) and its test-body limit 
that our main result, the EOB scattering-angle mapping (\ref{2PMtwospin}), holds up to these PN orders.  At 1PM order, also the dependence on $\delta$ drops out, and the simpler map in Eq.~(\ref{1PMtwospin}) is valid.

As discussed above, the scattering angle $\chi$ determines the Hamiltonian $H$ (for arbitrary mass ratios) up to gauge, at these PN orders (not just at 1PM and 2PM orders, but also including the 3PM and 4PM parts seen here).  The process of deriving $\chi$ from $H$ projects out precisely the gauge information, and $H$ can be fully recovered from $\chi$, modulo phase-space gauge freedom, at these orders.  While we have started here from a PN Hamiltonian, one could also start from independent results for the scattering angle and deduce a valid Hamiltonian.

Such independent results for the scattering angle can be obtained up to 2PM order by applying the 2PM EOB scattering-angle mapping (\ref{2PMtwospin}) to results for spinning test BH in a background Kerr spacetime presented in the following section.  The test-BH results below are in fact valid for arbitrary $v/c$, i.e.\ to all PN orders at a given PM order.  While we have shown conclusively only that the mapping (\ref{2PMtwospin}) produces correct arbitrary-mass-ratio results up to certain PN orders and certain orders in spin, we conjecture that its validity extends beyond these orders.

\section{Test-black-hole scattering in a background Kerr spacetime}\label{sec:test}

Above we have worked with the dynamics of arbitrary-mass-ratio binary BHs as calculated in the PN (weak-field and slow-motion) approximation, in particular having computed the binary BH scattering angle as a dual expansion in $G$ and in $1/c^2$, to orders accessible by use of previous derivations of PN Hamiltonians.  Here we present analogous PM (weak-field, arbitrary-speed) results which can be obtained in a certain test-body limit (a limit where the mass ratio tends to zero) of the binary BH problem.  We verify that (i) when we take the test-body limit of the PN results from Sec.~\ref{sec:chiPN}, we obtain the PN expansions (expansions in $1/c^2$) of the test-body results presented here, and (ii) the 1PM and  2PM parts of the arbitrary-mass-ratio PN results are fully recovered from the PN expansion of the mapping (\ref{2PMtwospin}), our central result, applied to the test-body results.  We emphasize again that the (PM, or even strong-field) test-body computations are significantly more easily accomplished than the arbitrary-mass-ratio (PN) computations, while our mapping (\ref{2PMtwospin}) allows one to obtain the latter from the former, up to 2PM order, to the extent that PN results are available.  The PM test-body results below are in fact obtained by PM-expanding exact (nonperturbative, strong-field) equations governing the motion of a test body (a test BH) in a background Kerr spacetime, at low orders in the multipole expansion of the test body.  
We again set $c=1$ in this section.

We consider in particular an \emph{extended}-test-body limit, in which the mass of one body (and thus its influence on the gravitational field) becomes negligible, but in which it retains a finite spatial extent; even as its mass tends to zero, the extended test body's mass-rescaled multipole moments remain finite, and influence its motion.  Such a test body, moving in an arbitrary (possibly strong-field) fixed background spacetime with metric $g_{\mu\nu}$, can be described by a (physical or effective) stress-energy tensor $T^{\mu\nu}$ which is conserved according to $\nabla_\mu T^{\mu\nu}=0$, where $\nabla_\mu$ is the covariant derivative for the background $g_{\mu\nu}$. Following the early analyses of Mathisson \cite{Mathisson:1937,Mathisson:2010} and Papapetrou \cite{Papapetrou:1951pa} at pole-dipole order (see also \cite{Tulczyjew:1959}), it was most rigorously demonstrated by Dixon \cite{Dixon:1979,Dixon:2015vxa} (see also \cite{Ehlers:1977,Schattner:1979vp,Harte:2011ku,Harte:2014wya}) that such a test body must obey translational and rotational equations of motion of the following form, obtained via a multipole expansion of the body's stress-energy distribution, the so-called Mathisson-Papapetrou-Dixon (MPD) equations,
\begin{alignat}{5}\label{MPD}
\frac{Dp_\mu}{d\sigma}+\frac{1}{2}R_{\mu\nu\rho\sigma}\dot z^\nu S^{\rho\sigma}&=-\frac{1}{6}\nabla_\mu R_{\nu\rho\sigma\tau}J^{\nu\rho\sigma\tau}+\ldots\phantom{\Bigg|}
\nnm\\
\frac{DS^{\mu\nu}}{d\sigma}-2p^{[\mu}\dot z^{\nu]}&=\frac{4}{3}R^{[\mu}{}_{\rho\sigma\tau}J^{\nu]\rho\sigma\tau}+\ldots\phantom{\Bigg|}
\\\nnm
S^{\mu\nu}p_\nu&=0.\phantom{\bigg|}
\end{alignat}
The MPD equations (the first two lines) govern the evolution of the test body's linear momentum vector $p^\mu(\sigma)$ and angular momentum (or spin) tensor $S^{\mu\nu}(\sigma)$ along a worldline $x=z(\sigma)$ with tangent $\dot z^\mu=dz^\mu/d\sigma$, where $\sigma$ is an arbitrary parameter.  The last line is the Tulczyjew-Dixon supplementary condition \cite{Tulczyjew:1959,Dixon:1979,Schattner:1979vp}, which fixes a choice for the body's centroid worldline by setting to zero its mass-dipole vector $(\propto S^{\mu\nu}p_\nu)$ about that worldline as defined in the body's own local rest frame.  The equations further depend only on the background spacetime (through the metric $g_{\mu\nu}$ and its covariant curvature tensors, the Riemann tensor $R_{\mu\nu\rho\sigma}$ and its covariant derivatives) and on the body's higher (relativistic) multipole moments, beginning with the quadrupole $J^{\mu\nu\rho\sigma}$.  

We refer the reader to \cite{Dixon:1979,Dixon:2015vxa,Harte:2011ku,Harte:2014wya} for detailed discussions of Dixon's definitions of the multipoles, the monopole $p_\mu$, the dipole $S_{\mu\nu}$, etc., in terms of its stress-energy $T^{\mu\nu}$, noting here only the following two properties.  Firstly, in the absence of spacetime curvature, the definitions of $p_\mu$ and $S_{\mu\nu}$ reduce to the standard definitions for an isolated body in flat spacetime.  Secondly, given any Killing vector $\xi^\mu$ of the background, the quantity
\be\label{Qconserved}
\mc Q=p_\mu \xi^\mu+\frac{1}{2}S^\mu{}_\nu\nabla_\mu\xi^\nu,
\ee
is exactly conserved, to all orders in the multipole expansion \cite{Ehlers:1977,Dixon:1979,Harte:2014wya}.

We are interested here in taking an extended-test-body limit for a spinning BH, to obtain a ``spinning test BH.''  By this we understand that we neglect the influence of the test BH on the curvature of spacetime (its mass is small compared to the scale of the ``background'' curvature), while we keep its spatial extent finite.  This can be achieved by taking the limit $m_{\mr t} / m_{\mr B} \rightarrow 0$ while keeping fixed the ring radius $a_t$ of the test BH.  Strictly speaking, this does not describe a BH (with a ring singularity hidden behind an event horizon), but a ``naked'' ring singularity of negligible mass.  But the mass-rescaled multipoles, and hence the equations of motion, of both the naked ring singularity and the BH ring singularity are identically determined as a function of the ring radius $a_{\mr t}$ \cite{Hansen:1974zz} at the level of approximation that we are interested in; these are the spin-induced multipole moments.  (We neglect tidally induced multipole moments, including absorption or ``tidal heating'' effects from the horizon, which would contribute at orders beyond those considered here.)  At the quadrupolar level in the multipole expansion, a worldline action including a generic spin-induced quadrupole was derived in Ref.~\cite{Porto:2008jj}, and this can be related to the MPD equations \cite{Steinhoff:2010zz,Steinhoff:2012rw,Buonanno:2012rv,Bini:2013nw,Bini:2013uwa,Marsat:2014xea,Vines:2016unv} and specialized to a BH (with appropriate matching to the Kerr solution), leading to
\be
J^{\mu\nu\rho\sigma}=3\frac{p\cdot\dot z}{(p^2)^2}p^{[\mu}S^{\nu]\kappa}p^{[\rho}S^{\sigma]}{}_\kappa
\ee
as the relativistic quadrupole moment appropriate for a spinning BH in the MPD equations with the Tulczyjew-Dixon condition (\ref{MPD}).

Let us now discuss the motion of a test BH in the background spacetime of a large BH described by the Kerr metric.  The Kerr metric possesses two Killing vectors, a timelike Killing vector $t^\mu$ (time translation symmetry) and an axial Killing vector $\phi^\mu$ (rotation symmetry about the spin axis), leading respectively to the test body's conserved energy $E$ and total angular momentum $J$ via Eq.~(\ref{Qconserved}), for an aligned-spin equatorial orbit,
\begin{alignat}{3}\label{EJconserved}
\begin{aligned}
m_\mr t\gamma=E&=p_\mu t^\mu+\frac{1}{2}S^\mu{}_\nu\nabla_\mu t^\nu,
\\
m_\mr t\gamma( v b+ a_\mr t)=J&=-p_\mu \phi^\mu-\frac{1}{2}S^\mu{}_\nu\nabla_\mu \phi^\nu .
\end{aligned}
\end{alignat}
[See Eq.~(\ref{gammaEt}) and the test-body ($\nu\to0$) limits of Eqs.~(\ref{Jtot}) and (\ref{key}), and discussion, e.g., in Refs.~\cite{Bini:2017pee,Vines:2017hyw}, for the identifications on far-left-hand sides.]
These conservation laws allow one to integrate the equations of motion for the aligned-spin case \cite{Steinhoff:2012rw,Bini:2017pee}:  we have three independent equations from the supplementary condition $S^{\mu\nu} p_{\nu} = 0$, another three from its time derivative, the normalization $\dot z^\mu \dot z_\mu = 1$, the (approximately) conserved mass of the test BH related to $p^\mu p_\mu$ [see Eq.~(64) in Ref.~\cite{Steinhoff:2012rw} or Ref.~\cite{Bini:2017pee}], one equation restricting the motion to the equatorial plane, three equations for the (constant) magnitude and aligned direction of the test spin, and the two conservation laws (\ref{EJconserved}).  These 14 algebraic equations can be solved for the 14 independent components of $p_\mu$, $\dot z^\mu$, and $S^{\mu\nu} = - S^{\nu\mu}$.

Having an algebraic solution for $\dot z^\mu$, one can integrate it to yield $z^\mu(\sigma)$.  Since we are only interested in the scattering angle, we can directly integrate $d \phi / d r = \dot z^\phi / \dot z^r$ given by Eq.~(66) from Ref.~\cite{Bini:2017pee} between the initial and final state.
Expanding the integrand in spins and in $G$ (PM expansion) to the same levels as in the last section (but without PN expansion in $v$) leads to the following result for the aligned-spin scattering angle for a test BH with ring radius $a_{\mr t}$ in a Kerr background with mass $m_{\mr B}\equiv M$ and spin ${M} a_{\mr B}$,
\begin{widetext}
\begin{align}\label{chitestPM}
  \frac{\chi_\mr t}{2} &= \bigg[ \frac{GM}{b}\frac{1+v^2}{v^2}+\frac{3\pi}{8}\frac{(GM)^2}{b^2}\frac{4+v^2}{v^2}
    +\frac{1}{3}\frac{(GM)^3}{b^3}\frac{-1+15v^2+45 v^4+5v^6}{v^6}
    +\frac{105\pi}{128}\frac{(GM)^4}{b^4}\frac{16+16v^2+v^4}{v^4}+{\cal O}(G^5) \bigg] \nonumber \\
&\quad +\bigg[ -2\frac{GM}{b^2}\frac{1}{v}a_+-\frac{\pi}{8}\frac{(GM)^2}{b^3}\frac{2+3v^2}{v^3}(7a_++a_-)
    -\frac{(GM)^3}{b^4}\frac{1+10v^2+5v^4}{v^5}(5a_++a_-)+{\cal O}(G^4) \bigg] \nonumber \\
&\quad + \bigg[ \frac{GM}{b^3}\frac{1+v^2}{v^2}a_+^2
    +\frac{3\pi}{128}\frac{(GM)^2}{b^4}\bigg(\frac{32+236v^2+47v^4}{v^4}a_+^2 +\frac{4+v^2}{v^2}a_-(14a_+-a_-)\bigg) \nonumber \\
&\quad\qquad +2\frac{(GM)^3}{b^5}\bigg(\frac{1+35v^2+55v^4+5v^6}{v^6}a_+^2 +2\frac{5+10v^2+v^4}{v^4}a_+a_-\bigg)+{\cal O}(G^4) \bigg] + {\cal O}(a_\pm^3) ,
\end{align}
\end{widetext}
where, here,
\be
a_\pm =   a_{\mr B} \pm a_{\mr t}.
\ee
Firstly, one can verify that the test-body limit of the PN scattering angle given by Eqs.~(\ref{chiPNfinal}), to the given PN orders, matches this result derived from the test-BH MPD equations.
Finally and most importantly, applying the mapping (\ref{2PMtwospin}) to the test-BH-in-Kerr scattering angle (\ref{chitestPM}), and PN-expanding the result, one obtains precisely the 1PM and 2PM parts of the PN-PM results (\ref{chiPNfinal}).

\section{Conclusions}\label{sec:conclude}

The encounter of two BHs is a fundamental process in our Universe, from the inspiral of astrophysical BHs, observed through GWs, to the (hypothetical) scattering of two BHs in analogy to particle physics experiments.
In the present paper, we proposed that the scattering-angle function for two spinning BHs at 2PM order is related in a particularly simple way to the scattering angle for a test BH in a stationary BH spacetime, for the case of aligned spins [see Eq.~(\ref{2PMtwospin})].
While we were unable to verify this mapping to its full potential extent at 2PM order and at all orders in spins, we checked its validity against all available results for the conservative local-in-time dynamics of binary BHs in the PM, PN, and test-BH approximations:
the PN Hamiltonian including subsubleading-order results up to quadratic order in spin \cite{Damour:2000we,Levi:2015uxa,Levi:2016ofk}, the 1PM scattering angle at all orders in spin \cite{Vines:2017hyw} (implying agreement with the LO PN Hamiltonian to all orders in spin \cite{Vines:2016qwa}), the 2PM spin-orbit (linear-in-spin) scattering angle \cite{Bini:2018ywr}, and the scattering angle for a test BH in a Kerr background to quadratic order in spin \cite{Bini:2017pee}.

This result is interesting not only for the scattering of BHs, but also for BHs in bound orbits.
The reason is that, for aligned spins and at least to 2PM order (at least up to the subsubleading PN orders), the scattering angle uniquely encodes the Hamiltonian dynamics (more precisely, an equivalence class of Hamiltonians subject to canonical transformations). 
An important possible application of our result is the 2PM resummation of conservative spin effects in the EOB gravitational waveform model for inspiraling BHs.
The challenge here is to find a suitable gauge (canonical representation) for the EOB Hamiltonian informed by the 2PM scattering angle.

\begin{comment}
{\color{blue}A caveat is that the tail contribution to the Hamiltonian constructed from the scattering angle alone will not be correct for bound orbits: while for bound orbits the tails effects were shown to contribute at 2PM in a small eccentricity expansion, they vanish at 2PM for (``highly eccentric'') scattering orbits.
The reconstruction of tail effects from a scattering process (eventually involving outgoing GWs) is an important task for future work.
Nevertheless, the tail effects at 2PM do not involve the spins, so that the spin-dependent part of the Hamiltonians obtained from a scattering angle at 2PM are valid for bound orbits, too.}
\end{comment}

It is not uncommon that elegant resummations at lower orders have extrapolated to new results.
For instance, it was shown in Ref.~\cite{Vines:2016qwa} that the simple ``EOB spin map'' employed in Ref.~\cite{Damour:2008qf} and some subsequent EOB models (identifying the ring radius of an effective [$\nu$-deformed] Kerr spacetime with the sum of the ring radii of the individual BHs, as first suggested in Ref.~\cite{Damour:2001tu}), while intending to resum LO PN results at quadratic order in spin only, in fact led to the correct dynamics at the leading PN orders for all even orders in spin.
If the map (\ref{2PMtwospin}) were to hold to some further extent at 2PM order, say, through subleading PN orders at some higher orders in spin, and if one had results for the test-BH-in-Kerr scattering angle at higher orders in spin, then this would provide new and complete sub-LO PN results beyond quadratic order in spin (for aligned spins).  Furthermore, since the scattering angle for a test BH (assuming that this is well-defined at higher orders in spin at 2PM order) could be obtained exactly, based on exact solutions like the Kerr metric, the map (\ref{2PMtwospin}) represents a PM-nonperturbative resummation in the spirit of EOB models.

The simplicity of the BH scattering-angle map at 2PM order is reminiscent of the elegance of on-shell methods used to calculate scattering amplitudes for elementary particles, using consistency with relativistic kinematics, gauge symmetries, locality, and unitarity \cite{Parke:1986gb,Bern:1994zx,Bern:1994cg,Britto:2004nc,Britto:2004ap,Britto:2005fq,Bern:2008qj,Bern:2010yg,Bern:2010ue,Guevara:2017csg,Arkani-Hamed:2017jhn,Ochirov:2018uyq}. This is in sharp contrast to the lengthy calculations of scattering amplitudes through Feynman diagrams, rules, and integrals.
Likewise, we suspect that a proof of the BH scattering-angle map at 2PM (at least up to the PN and/or spin orders where it has been shown to hold here, if not higher) is possible using simple generic principles, instead of going through lengthy iterative solutions of the equations of motion (which should in complete generality involve asymptotic matching to perturbed BH spacetimes).
In fact, it is conceivable that the simplest demonstration of some (approximate or exact) version of the scattering-angle map might come from a classical limit of a seemingly more complicated problem, namely, calculation of the one-loop quantum scattering amplitude involving two quantum BHs.

\acknowledgements

We are grateful to Mohammed Khalil for his help in verifying the changes in this new version of this paper, where we have corrected misstatements concerning the amount of freedom in canonical transformations between different quasi-isotropic gauges for aligned-spin canonical Hamiltonians.

% to generate .bbl file, uncomment this and run make script
\bibliography{testspin}

%merlin.mbs apsrev4-1.bst 2010-07-25 4.21a (PWD, AO, DPC) hacked
%Control: key (0)
%Control: author (0) dotless jnrlst
%Control: editor formatted (1) identically to author
%Control: production of article title (0) allowed
%Control: page (1) range
%Control: year (0) verbatim
%Control: production of eprint (0) enabled
\begin{thebibliography}{150}%
\makeatletter
\providecommand \@ifxundefined [1]{%
 \@ifx{#1\undefined}
}%
\providecommand \@ifnum [1]{%
 \ifnum #1\expandafter \@firstoftwo
 \else \expandafter \@secondoftwo
 \fi
}%
\providecommand \@ifx [1]{%
 \ifx #1\expandafter \@firstoftwo
 \else \expandafter \@secondoftwo
 \fi
}%
\providecommand \natexlab [1]{#1}%
\providecommand \enquote  [1]{``#1''}%
\providecommand \bibnamefont  [1]{#1}%
\providecommand \bibfnamefont [1]{#1}%
\providecommand \citenamefont [1]{#1}%
\providecommand \href@noop [0]{\@secondoftwo}%
\providecommand \href [0]{\begingroup \@sanitize@url \@href}%
\providecommand \@href[1]{\@@startlink{#1}\@@href}%
\providecommand \@@href[1]{\endgroup#1\@@endlink}%
\providecommand \@sanitize@url [0]{\catcode `\\12\catcode `\$12\catcode
  `\&12\catcode `\#12\catcode `\^12\catcode `\_12\catcode `\%12\relax}%
\providecommand \@@startlink[1]{}%
\providecommand \@@endlink[0]{}%
\providecommand \url  [0]{\begingroup\@sanitize@url \@url }%
\providecommand \@url [1]{\endgroup\@href {#1}{\urlprefix }}%
\providecommand \urlprefix  [0]{URL }%
\providecommand \Eprint [0]{\href }%
\providecommand \doibase [0]{http://dx.doi.org/}%
\providecommand \selectlanguage [0]{\@gobble}%
\providecommand \bibinfo  [0]{\@secondoftwo}%
\providecommand \bibfield  [0]{\@secondoftwo}%
\providecommand \translation [1]{[#1]}%
\providecommand \BibitemOpen [0]{}%
\providecommand \bibitemStop [0]{}%
\providecommand \bibitemNoStop [0]{.\EOS\space}%
\providecommand \EOS [0]{\spacefactor3000\relax}%
\providecommand \BibitemShut  [1]{\csname bibitem#1\endcsname}%
\let\auto@bib@innerbib\@empty
%</preamble>
\bibitem [{\citenamefont {Misner}\ \emph {et~al.}(1973)\citenamefont {Misner},
  \citenamefont {Thorne},\ and\ \citenamefont {Wheeler}}]{Misner:1974qy}%
  \BibitemOpen
  \bibfield  {author} {\bibinfo {author} {\bibfnamefont {Charles~W.}\
  \bibnamefont {Misner}}, \bibinfo {author} {\bibfnamefont {K.~S.}\
  \bibnamefont {Thorne}}, \ and\ \bibinfo {author} {\bibfnamefont {J.~A.}\
  \bibnamefont {Wheeler}},\ }\href@noop {} {\emph {\bibinfo {title}
  {{Gravitation}}}}\ (\bibinfo  {publisher} {W. H. Freeman},\ \bibinfo
  {address} {San Francisco},\ \bibinfo {year} {1973})\BibitemShut {NoStop}%
%%CITATION = INSPIRE-95654;%%
\bibitem [{\citenamefont {Parke}\ and\ \citenamefont
  {Taylor}(1986)}]{Parke:1986gb}%
  \BibitemOpen
  \bibfield  {author} {\bibinfo {author} {\bibfnamefont {Stephen~J.}\
  \bibnamefont {Parke}}\ and\ \bibinfo {author} {\bibfnamefont {T.~R.}\
  \bibnamefont {Taylor}},\ }\bibfield  {title} {\enquote {\bibinfo {title} {{An
  Amplitude for $n$ Gluon Scattering}},}\ }\href {\doibase
  10.1103/PhysRevLett.56.2459} {\bibfield  {journal} {\bibinfo  {journal}
  {Phys. Rev. Lett.}\ }\textbf {\bibinfo {volume} {56}},\ \bibinfo {pages}
  {2459} (\bibinfo {year} {1986})}\BibitemShut {NoStop}%
%%CITATION = PRLTA,56,2459;%%
\bibitem [{\citenamefont {Bern}\ \emph {et~al.}(1994)\citenamefont {Bern},
  \citenamefont {Dixon}, \citenamefont {Dunbar},\ and\ \citenamefont
  {Kosower}}]{Bern:1994zx}%
  \BibitemOpen
  \bibfield  {author} {\bibinfo {author} {\bibfnamefont {Zvi}\ \bibnamefont
  {Bern}}, \bibinfo {author} {\bibfnamefont {Lance~J.}\ \bibnamefont {Dixon}},
  \bibinfo {author} {\bibfnamefont {David~C.}\ \bibnamefont {Dunbar}}, \ and\
  \bibinfo {author} {\bibfnamefont {David~A.}\ \bibnamefont {Kosower}},\
  }\bibfield  {title} {\enquote {\bibinfo {title} {{One loop n point gauge
  theory amplitudes, unitarity and collinear limits}},}\ }\href {\doibase
  10.1016/0550-3213(94)90179-1} {\bibfield  {journal} {\bibinfo  {journal}
  {Nucl. Phys.}\ }\textbf {\bibinfo {volume} {B425}},\ \bibinfo {pages}
  {217--260} (\bibinfo {year} {1994})},\ \Eprint
  {http://arxiv.org/abs/hep-ph/9403226} {arXiv:hep-ph/9403226 [hep-ph]}
  \BibitemShut {NoStop}%
%%CITATION = HEP-PH/9403226;%%
\bibitem [{\citenamefont {Bern}\ \emph {et~al.}(1995)\citenamefont {Bern},
  \citenamefont {Dixon}, \citenamefont {Dunbar},\ and\ \citenamefont
  {Kosower}}]{Bern:1994cg}%
  \BibitemOpen
  \bibfield  {author} {\bibinfo {author} {\bibfnamefont {Zvi}\ \bibnamefont
  {Bern}}, \bibinfo {author} {\bibfnamefont {Lance~J.}\ \bibnamefont {Dixon}},
  \bibinfo {author} {\bibfnamefont {David~C.}\ \bibnamefont {Dunbar}}, \ and\
  \bibinfo {author} {\bibfnamefont {David~A.}\ \bibnamefont {Kosower}},\
  }\bibfield  {title} {\enquote {\bibinfo {title} {{Fusing gauge theory tree
  amplitudes into loop amplitudes}},}\ }\href {\doibase
  10.1016/0550-3213(94)00488-Z} {\bibfield  {journal} {\bibinfo  {journal}
  {Nucl. Phys.}\ }\textbf {\bibinfo {volume} {B435}},\ \bibinfo {pages}
  {59--101} (\bibinfo {year} {1995})},\ \Eprint
  {http://arxiv.org/abs/hep-ph/9409265} {arXiv:hep-ph/9409265 [hep-ph]}
  \BibitemShut {NoStop}%
%%CITATION = HEP-PH/9409265;%%
\bibitem [{\citenamefont {Britto}\ \emph
  {et~al.}(2005{\natexlab{a}})\citenamefont {Britto}, \citenamefont {Cachazo},\
  and\ \citenamefont {Feng}}]{Britto:2004nc}%
  \BibitemOpen
  \bibfield  {author} {\bibinfo {author} {\bibfnamefont {Ruth}\ \bibnamefont
  {Britto}}, \bibinfo {author} {\bibfnamefont {Freddy}\ \bibnamefont
  {Cachazo}}, \ and\ \bibinfo {author} {\bibfnamefont {Bo}~\bibnamefont
  {Feng}},\ }\bibfield  {title} {\enquote {\bibinfo {title} {{Generalized
  unitarity and one-loop amplitudes in N=4 super-Yang-Mills}},}\ }\href
  {\doibase 10.1016/j.nuclphysb.2005.07.014} {\bibfield  {journal} {\bibinfo
  {journal} {Nucl. Phys.}\ }\textbf {\bibinfo {volume} {B725}},\ \bibinfo
  {pages} {275--305} (\bibinfo {year} {2005}{\natexlab{a}})},\ \Eprint
  {http://arxiv.org/abs/hep-th/0412103} {arXiv:hep-th/0412103 [hep-th]}
  \BibitemShut {NoStop}%
%%CITATION = HEP-TH/0412103;%%
\bibitem [{\citenamefont {Britto}\ \emph
  {et~al.}(2005{\natexlab{b}})\citenamefont {Britto}, \citenamefont {Cachazo},\
  and\ \citenamefont {Feng}}]{Britto:2004ap}%
  \BibitemOpen
  \bibfield  {author} {\bibinfo {author} {\bibfnamefont {Ruth}\ \bibnamefont
  {Britto}}, \bibinfo {author} {\bibfnamefont {Freddy}\ \bibnamefont
  {Cachazo}}, \ and\ \bibinfo {author} {\bibfnamefont {Bo}~\bibnamefont
  {Feng}},\ }\bibfield  {title} {\enquote {\bibinfo {title} {{New recursion
  relations for tree amplitudes of gluons}},}\ }\href {\doibase
  10.1016/j.nuclphysb.2005.02.030} {\bibfield  {journal} {\bibinfo  {journal}
  {Nucl. Phys.}\ }\textbf {\bibinfo {volume} {B715}},\ \bibinfo {pages}
  {499--522} (\bibinfo {year} {2005}{\natexlab{b}})},\ \Eprint
  {http://arxiv.org/abs/hep-th/0412308} {arXiv:hep-th/0412308 [hep-th]}
  \BibitemShut {NoStop}%
%%CITATION = HEP-TH/0412308;%%
\bibitem [{\citenamefont {Britto}\ \emph
  {et~al.}(2005{\natexlab{c}})\citenamefont {Britto}, \citenamefont {Cachazo},
  \citenamefont {Feng},\ and\ \citenamefont {Witten}}]{Britto:2005fq}%
  \BibitemOpen
  \bibfield  {author} {\bibinfo {author} {\bibfnamefont {Ruth}\ \bibnamefont
  {Britto}}, \bibinfo {author} {\bibfnamefont {Freddy}\ \bibnamefont
  {Cachazo}}, \bibinfo {author} {\bibfnamefont {Bo}~\bibnamefont {Feng}}, \
  and\ \bibinfo {author} {\bibfnamefont {Edward}\ \bibnamefont {Witten}},\
  }\bibfield  {title} {\enquote {\bibinfo {title} {{Direct proof of tree-level
  recursion relation in Yang-Mills theory}},}\ }\href {\doibase
  10.1103/PhysRevLett.94.181602} {\bibfield  {journal} {\bibinfo  {journal}
  {Phys. Rev. Lett.}\ }\textbf {\bibinfo {volume} {94}},\ \bibinfo {pages}
  {181602} (\bibinfo {year} {2005}{\natexlab{c}})},\ \Eprint
  {http://arxiv.org/abs/hep-th/0501052} {arXiv:hep-th/0501052 [hep-th]}
  \BibitemShut {NoStop}%
%%CITATION = HEP-TH/0501052;%%
\bibitem [{\citenamefont {Bern}\ \emph {et~al.}(2008)\citenamefont {Bern},
  \citenamefont {Carrasco},\ and\ \citenamefont {Johansson}}]{Bern:2008qj}%
  \BibitemOpen
  \bibfield  {author} {\bibinfo {author} {\bibfnamefont {Z.}~\bibnamefont
  {Bern}}, \bibinfo {author} {\bibfnamefont {J.~J.~M.}\ \bibnamefont
  {Carrasco}}, \ and\ \bibinfo {author} {\bibfnamefont {Henrik}\ \bibnamefont
  {Johansson}},\ }\bibfield  {title} {\enquote {\bibinfo {title} {{New
  Relations for Gauge-Theory Amplitudes}},}\ }\href {\doibase
  10.1103/PhysRevD.78.085011} {\bibfield  {journal} {\bibinfo  {journal} {Phys.
  Rev.}\ }\textbf {\bibinfo {volume} {D78}},\ \bibinfo {pages} {085011}
  (\bibinfo {year} {2008})},\ \Eprint {http://arxiv.org/abs/0805.3993}
  {arXiv:0805.3993 [hep-ph]} \BibitemShut {NoStop}%
%%CITATION = ARXIV:0805.3993;%%
\bibitem [{\citenamefont {Bern}\ \emph
  {et~al.}(2010{\natexlab{a}})\citenamefont {Bern}, \citenamefont {Dennen},
  \citenamefont {Huang},\ and\ \citenamefont {Kiermaier}}]{Bern:2010yg}%
  \BibitemOpen
  \bibfield  {author} {\bibinfo {author} {\bibfnamefont {Zvi}\ \bibnamefont
  {Bern}}, \bibinfo {author} {\bibfnamefont {Tristan}\ \bibnamefont {Dennen}},
  \bibinfo {author} {\bibfnamefont {Yu-tin}\ \bibnamefont {Huang}}, \ and\
  \bibinfo {author} {\bibfnamefont {Michael}\ \bibnamefont {Kiermaier}},\
  }\bibfield  {title} {\enquote {\bibinfo {title} {{Gravity as the Square of
  Gauge Theory}},}\ }\href {\doibase 10.1103/PhysRevD.82.065003} {\bibfield
  {journal} {\bibinfo  {journal} {Phys. Rev.}\ }\textbf {\bibinfo {volume}
  {D82}},\ \bibinfo {pages} {065003} (\bibinfo {year} {2010}{\natexlab{a}})},\
  \Eprint {http://arxiv.org/abs/1004.0693} {arXiv:1004.0693 [hep-th]}
  \BibitemShut {NoStop}%
%%CITATION = ARXIV:1004.0693;%%
\bibitem [{\citenamefont {Bern}\ \emph
  {et~al.}(2010{\natexlab{b}})\citenamefont {Bern}, \citenamefont {Carrasco},\
  and\ \citenamefont {Johansson}}]{Bern:2010ue}%
  \BibitemOpen
  \bibfield  {author} {\bibinfo {author} {\bibfnamefont {Zvi}\ \bibnamefont
  {Bern}}, \bibinfo {author} {\bibfnamefont {John Joseph~M.}\ \bibnamefont
  {Carrasco}}, \ and\ \bibinfo {author} {\bibfnamefont {Henrik}\ \bibnamefont
  {Johansson}},\ }\bibfield  {title} {\enquote {\bibinfo {title} {{Perturbative
  Quantum Gravity as a Double Copy of Gauge Theory}},}\ }\href {\doibase
  10.1103/PhysRevLett.105.061602} {\bibfield  {journal} {\bibinfo  {journal}
  {Phys. Rev. Lett.}\ }\textbf {\bibinfo {volume} {105}},\ \bibinfo {pages}
  {061602} (\bibinfo {year} {2010}{\natexlab{b}})},\ \Eprint
  {http://arxiv.org/abs/1004.0476} {arXiv:1004.0476 [hep-th]} \BibitemShut
  {NoStop}%
%%CITATION = ARXIV:1004.0476;%%
\bibitem [{\citenamefont {Guevara}(2017)}]{Guevara:2017csg}%
  \BibitemOpen
  \bibfield  {author} {\bibinfo {author} {\bibfnamefont {Alfredo}\ \bibnamefont
  {Guevara}},\ }\bibfield  {title} {\enquote {\bibinfo {title} {{Holomorphic
  Classical Limit for Spin Effects in Gravitational and Electromagnetic
  Scattering}},}\ }\href@noop {} {\  (\bibinfo {year} {2017})},\ \Eprint
  {http://arxiv.org/abs/1706.02314} {arXiv:1706.02314 [hep-th]} \BibitemShut
  {NoStop}%
%%CITATION = ARXIV:1706.02314;%%
\bibitem [{\citenamefont {Arkani-Hamed}\ \emph {et~al.}(2017)\citenamefont
  {Arkani-Hamed}, \citenamefont {Huang},\ and\ \citenamefont
  {Huang}}]{Arkani-Hamed:2017jhn}%
  \BibitemOpen
  \bibfield  {author} {\bibinfo {author} {\bibfnamefont {Nima}\ \bibnamefont
  {Arkani-Hamed}}, \bibinfo {author} {\bibfnamefont {Tzu-Chen}\ \bibnamefont
  {Huang}}, \ and\ \bibinfo {author} {\bibfnamefont {Yu-tin}\ \bibnamefont
  {Huang}},\ }\bibfield  {title} {\enquote {\bibinfo {title} {{Scattering
  Amplitudes For All Masses and Spins}},}\ }\href@noop {} {\  (\bibinfo {year}
  {2017})},\ \Eprint {http://arxiv.org/abs/1709.04891} {arXiv:1709.04891
  [hep-th]} \BibitemShut {NoStop}%
%%CITATION = ARXIV:1709.04891;%%
\bibitem [{\citenamefont {Ochirov}(2018)}]{Ochirov:2018uyq}%
  \BibitemOpen
  \bibfield  {author} {\bibinfo {author} {\bibfnamefont {Alexander}\
  \bibnamefont {Ochirov}},\ }\bibfield  {title} {\enquote {\bibinfo {title}
  {{Helicity amplitudes for QCD with massive quarks}},}\ }\href {\doibase
  10.1007/JHEP04(2018)089} {\bibfield  {journal} {\bibinfo  {journal} {JHEP}\
  }\textbf {\bibinfo {volume} {04}},\ \bibinfo {pages} {089} (\bibinfo {year}
  {2018})},\ \Eprint {http://arxiv.org/abs/1802.06730} {arXiv:1802.06730
  [hep-ph]} \BibitemShut {NoStop}%
%%CITATION = ARXIV:1802.06730;%%
\bibitem [{\citenamefont {Abbott}\ \emph {et~al.}(2016)\citenamefont {Abbott}
  \emph {et~al.}}]{Abbott:2016blz}%
  \BibitemOpen
  \bibfield  {author} {\bibinfo {author} {\bibfnamefont {B.~P.}\ \bibnamefont
  {Abbott}} \emph {et~al.} (\bibinfo {collaboration} {Virgo, LIGO
  Scientific}),\ }\bibfield  {title} {\enquote {\bibinfo {title} {{Observation
  of Gravitational Waves from a Binary Black Hole Merger}},}\ }\href {\doibase
  10.1103/PhysRevLett.116.061102} {\bibfield  {journal} {\bibinfo  {journal}
  {Phys. Rev. Lett.}\ }\textbf {\bibinfo {volume} {116}},\ \bibinfo {pages}
  {061102} (\bibinfo {year} {2016})},\ \Eprint
  {http://arxiv.org/abs/1602.03837} {arXiv:1602.03837 [gr-qc]} \BibitemShut
  {NoStop}%
%%CITATION = ARXIV:1602.03837;%%
\bibitem [{\citenamefont {Vaidya}(2015)}]{Vaidya:2014kza}%
  \BibitemOpen
  \bibfield  {author} {\bibinfo {author} {\bibfnamefont {Varun}\ \bibnamefont
  {Vaidya}},\ }\bibfield  {title} {\enquote {\bibinfo {title} {{Gravitational
  spin Hamiltonians from the S matrix}},}\ }\href {\doibase
  10.1103/PhysRevD.91.024017} {\bibfield  {journal} {\bibinfo  {journal} {Phys.
  Rev.}\ }\textbf {\bibinfo {volume} {D91}},\ \bibinfo {pages} {024017}
  (\bibinfo {year} {2015})},\ \Eprint {http://arxiv.org/abs/1410.5348}
  {arXiv:1410.5348 [hep-th]} \BibitemShut {NoStop}%
%%CITATION = ARXIV:1410.5348;%%
\bibitem [{\citenamefont {Blanchet}(2014)}]{Blanchet:2013haa}%
  \BibitemOpen
  \bibfield  {author} {\bibinfo {author} {\bibfnamefont {Luc}\ \bibnamefont
  {Blanchet}},\ }\bibfield  {title} {\enquote {\bibinfo {title} {{Gravitational
  Radiation from Post-Newtonian Sources and Inspiralling Compact Binaries}},}\
  }\href {\doibase 10.12942/lrr-2014-2} {\bibfield  {journal} {\bibinfo
  {journal} {Living Rev. Rel.}\ }\textbf {\bibinfo {volume} {17}},\ \bibinfo
  {pages} {2} (\bibinfo {year} {2014})},\ \Eprint
  {http://arxiv.org/abs/1310.1528} {arXiv:1310.1528 [gr-qc]} \BibitemShut
  {NoStop}%
%%CITATION = ARXIV:1310.1528;%%
\bibitem [{\citenamefont {Schäfer}\ and\ \citenamefont
  {Jaranowski}(2018)}]{Schafer:2018kuf}%
  \BibitemOpen
  \bibfield  {author} {\bibinfo {author} {\bibfnamefont {Gerhard}\ \bibnamefont
  {Schäfer}}\ and\ \bibinfo {author} {\bibfnamefont {Piotr}\ \bibnamefont
  {Jaranowski}},\ }\bibfield  {title} {\enquote {\bibinfo {title} {{Hamiltonian
  formulation of general relativity and post-Newtonian dynamics of compact
  binaries}},}\ }\href {\doibase 10.1007/s41114-018-0016-5} {\bibfield
  {journal} {\bibinfo  {journal} {Living Rev. Rel.}\ }\textbf {\bibinfo
  {volume} {21}},\ \bibinfo {pages} {7} (\bibinfo {year} {2018})},\ \Eprint
  {http://arxiv.org/abs/1805.07240} {arXiv:1805.07240 [gr-qc]} \BibitemShut
  {NoStop}%
%%CITATION = ARXIV:1805.07240;%%
\bibitem [{\citenamefont {Futamase}\ and\ \citenamefont
  {Itoh}(2007)}]{Futamase:2007zz}%
  \BibitemOpen
  \bibfield  {author} {\bibinfo {author} {\bibfnamefont {Toshifumi}\
  \bibnamefont {Futamase}}\ and\ \bibinfo {author} {\bibfnamefont {Yousuke}\
  \bibnamefont {Itoh}},\ }\bibfield  {title} {\enquote {\bibinfo {title} {{The
  post-Newtonian approximation for relativistic compact binaries}},}\ }\href
  {\doibase 10.12942/lrr-2007-2} {\bibfield  {journal} {\bibinfo  {journal}
  {Living Rev. Rel.}\ }\textbf {\bibinfo {volume} {10}},\ \bibinfo {pages} {2}
  (\bibinfo {year} {2007})}\BibitemShut {NoStop}%
%%CITATION = 00222,10,2;%%
\bibitem [{\citenamefont {Poisson}\ and\ \citenamefont
  {Will}(2014)}]{Poisson:2014}%
  \BibitemOpen
  \bibfield  {author} {\bibinfo {author} {\bibfnamefont {E.}~\bibnamefont
  {Poisson}}\ and\ \bibinfo {author} {\bibfnamefont {C.M.}\ \bibnamefont
  {Will}},\ }\href {https://books.google.de/books?id=PZ5cAwAAQBAJ} {\emph
  {\bibinfo {title} {Gravity: Newtonian, Post-Newtonian, Relativistic}}}\
  (\bibinfo  {publisher} {Cambridge University Press},\ \bibinfo {year}
  {2014})\BibitemShut {NoStop}%
\bibitem [{\citenamefont {Goldberger}(2007)}]{Goldberger:2007hy}%
  \BibitemOpen
  \bibfield  {author} {\bibinfo {author} {\bibfnamefont {Walter~D.}\
  \bibnamefont {Goldberger}},\ }\bibfield  {title} {\enquote {\bibinfo {title}
  {{Les Houches lectures on effective field theories and gravitational
  radiation}},}\ }in\ \href@noop {} {\emph {\bibinfo {booktitle} {{Les Houches
  Summer School - Session 86: Particle Physics and Cosmology: The Fabric of
  Spacetime Les Houches, France, July 31-August 25, 2006}}}}\ (\bibinfo {year}
  {2007})\ \Eprint {http://arxiv.org/abs/hep-ph/0701129} {arXiv:hep-ph/0701129
  [hep-ph]} \BibitemShut {NoStop}%
%%CITATION = HEP-PH/0701129;%%
\bibitem [{\citenamefont {Foffa}\ and\ \citenamefont
  {Sturani}(2014)}]{Foffa:2013qca}%
  \BibitemOpen
  \bibfield  {author} {\bibinfo {author} {\bibfnamefont {Stefano}\ \bibnamefont
  {Foffa}}\ and\ \bibinfo {author} {\bibfnamefont {Riccardo}\ \bibnamefont
  {Sturani}},\ }\bibfield  {title} {\enquote {\bibinfo {title} {{Effective
  field theory methods to model compact binaries}},}\ }\href {\doibase
  10.1088/0264-9381/31/4/043001} {\bibfield  {journal} {\bibinfo  {journal}
  {Class. Quant. Grav.}\ }\textbf {\bibinfo {volume} {31}},\ \bibinfo {pages}
  {043001} (\bibinfo {year} {2014})},\ \Eprint {http://arxiv.org/abs/1309.3474}
  {arXiv:1309.3474 [gr-qc]} \BibitemShut {NoStop}%
%%CITATION = ARXIV:1309.3474;%%
\bibitem [{\citenamefont {Porto}(2016)}]{Porto:2016pyg}%
  \BibitemOpen
  \bibfield  {author} {\bibinfo {author} {\bibfnamefont {Rafael~A.}\
  \bibnamefont {Porto}},\ }\bibfield  {title} {\enquote {\bibinfo {title} {{The
  effective field theoristÕs approach to gravitational dynamics}},}\ }\href
  {\doibase 10.1016/j.physrep.2016.04.003} {\bibfield  {journal} {\bibinfo
  {journal} {Phys. Rept.}\ }\textbf {\bibinfo {volume} {633}},\ \bibinfo
  {pages} {1--104} (\bibinfo {year} {2016})},\ \Eprint
  {http://arxiv.org/abs/1601.04914} {arXiv:1601.04914 [hep-th]} \BibitemShut
  {NoStop}%
%%CITATION = ARXIV:1601.04914;%%
\bibitem [{\citenamefont {Rothstein}(2014)}]{Rothstein:2014sra}%
  \BibitemOpen
  \bibfield  {author} {\bibinfo {author} {\bibfnamefont {Ira~Z.}\ \bibnamefont
  {Rothstein}},\ }\bibfield  {title} {\enquote {\bibinfo {title} {{Progress in
  effective field theory approach to the binary inspiral problem}},}\ }\href
  {\doibase 10.1007/s10714-014-1726-y} {\bibfield  {journal} {\bibinfo
  {journal} {Gen. Rel. Grav.}\ }\textbf {\bibinfo {volume} {46}},\ \bibinfo
  {pages} {1726} (\bibinfo {year} {2014})}\BibitemShut {NoStop}%
%%CITATION = GRGVA,46,1726;%%
\bibitem [{\citenamefont {Levi}(2018)}]{Levi:2018nxp}%
  \BibitemOpen
  \bibfield  {author} {\bibinfo {author} {\bibfnamefont {Michele}\ \bibnamefont
  {Levi}},\ }\bibfield  {title} {\enquote {\bibinfo {title} {{Effective Field
  Theories of Post-Newtonian Gravity: A comprehensive review}},}\ }\href@noop
  {} {\  (\bibinfo {year} {2018})},\ \Eprint {http://arxiv.org/abs/1807.01699}
  {arXiv:1807.01699 [hep-th]} \BibitemShut {NoStop}%
%%CITATION = ARXIV:1807.01699;%%
\bibitem [{\citenamefont {{Bertotti}}(1956)}]{Bertotti:1956}%
  \BibitemOpen
  \bibfield  {author} {\bibinfo {author} {\bibfnamefont {B.}~\bibnamefont
  {{Bertotti}}},\ }\bibfield  {title} {\enquote {\bibinfo {title} {{On
  gravitational motion}},}\ }\href {\doibase 10.1007/BF02746175} {\bibfield
  {journal} {\bibinfo  {journal} {Il Nuovo Cimento}\ }\textbf {\bibinfo
  {volume} {4}},\ \bibinfo {pages} {898--906} (\bibinfo {year}
  {1956})}\BibitemShut {NoStop}%
\bibitem [{\citenamefont {{Bertotti}}\ and\ \citenamefont
  {{Plebanski}}(1960)}]{Bertotti:1960}%
  \BibitemOpen
  \bibfield  {author} {\bibinfo {author} {\bibfnamefont {B.}~\bibnamefont
  {{Bertotti}}}\ and\ \bibinfo {author} {\bibfnamefont {J.}~\bibnamefont
  {{Plebanski}}},\ }\bibfield  {title} {\enquote {\bibinfo {title} {{Theory of
  gravitational perturbations in the fast motion approximation}},}\ }\href
  {\doibase 10.1016/0003-4916(60)90132-9} {\bibfield  {journal} {\bibinfo
  {journal} {Annals of Physics}\ }\textbf {\bibinfo {volume} {11}},\ \bibinfo
  {pages} {169--200} (\bibinfo {year} {1960})}\BibitemShut {NoStop}%
\bibitem [{\citenamefont {Rosenblum}(1978)}]{Rosenblum:1978zr}%
  \BibitemOpen
  \bibfield  {author} {\bibinfo {author} {\bibfnamefont {A.}~\bibnamefont
  {Rosenblum}},\ }\bibfield  {title} {\enquote {\bibinfo {title}
  {{Gravitational Radiation Energy Loss in Scattering Problems and the Einstein
  Quadrupole Formula}},}\ }\href {\doibase 10.1103/PhysRevLett.41.1003,
  10.1103/PhysRevLett.41.1140.2} {\bibfield  {journal} {\bibinfo  {journal}
  {Phys. Rev. Lett.}\ }\textbf {\bibinfo {volume} {41}},\ \bibinfo {pages}
  {1003} (\bibinfo {year} {1978})},\ \bibinfo {note} {[Erratum: Phys. Rev.
  Lett.41,1140(1978)]}\BibitemShut {NoStop}%
%%CITATION = PRLTA,41,1003;%%
\bibitem [{\citenamefont {Bel}\ \emph {et~al.}(1981)\citenamefont {Bel},
  \citenamefont {Damour}, \citenamefont {Deruelle}, \citenamefont {Ibanez},\
  and\ \citenamefont {Martin}}]{Bel:1981be}%
  \BibitemOpen
  \bibfield  {author} {\bibinfo {author} {\bibfnamefont {Luis}\ \bibnamefont
  {Bel}}, \bibinfo {author} {\bibfnamefont {T.}~\bibnamefont {Damour}},
  \bibinfo {author} {\bibfnamefont {N.}~\bibnamefont {Deruelle}}, \bibinfo
  {author} {\bibfnamefont {J.}~\bibnamefont {Ibanez}}, \ and\ \bibinfo {author}
  {\bibfnamefont {J.}~\bibnamefont {Martin}},\ }\bibfield  {title} {\enquote
  {\bibinfo {title} {{Poincaré-invariant gravitational field and equations of
  motion of two pointlike objects: The postlinear approximation of general
  relativity}},}\ }\href {\doibase 10.1007/BF00756073} {\bibfield  {journal}
  {\bibinfo  {journal} {Gen. Rel. Grav.}\ }\textbf {\bibinfo {volume} {13}},\
  \bibinfo {pages} {963--1004} (\bibinfo {year} {1981})}\BibitemShut {NoStop}%
%%CITATION = GRGVA,13,963;%%
\bibitem [{\citenamefont {Damour}\ and\ \citenamefont
  {Deruelle}(1981)}]{Damour:1981bh}%
  \BibitemOpen
  \bibfield  {author} {\bibinfo {author} {\bibfnamefont {Thibault}\
  \bibnamefont {Damour}}\ and\ \bibinfo {author} {\bibfnamefont {Nathalie}\
  \bibnamefont {Deruelle}},\ }\bibfield  {title} {\enquote {\bibinfo {title}
  {{Radiation Reaction and Angular Momentum Loss in Small Angle Gravitational
  Scattering}},}\ }\href {\doibase 10.1016/0375-9601(81)90567-3} {\bibfield
  {journal} {\bibinfo  {journal} {Phys. Lett.}\ }\textbf {\bibinfo {volume}
  {A87}},\ \bibinfo {pages} {81} (\bibinfo {year} {1981})}\BibitemShut
  {NoStop}%
%%CITATION = PHLTA,A87,81;%%
\bibitem [{\citenamefont {Portilla}(1979)}]{Portilla:1979xx}%
  \BibitemOpen
  \bibfield  {author} {\bibinfo {author} {\bibfnamefont {M.}~\bibnamefont
  {Portilla}},\ }\bibfield  {title} {\enquote {\bibinfo {title} {Momentum and
  angular momentum of two gravitating particles},}\ }\href {\doibase
  10.1088/0305-4470/12/7/025} {\bibfield  {journal} {\bibinfo  {journal} {J.
  Phys.}\ }\textbf {\bibinfo {volume} {A12}},\ \bibinfo {pages} {1075--1090}
  (\bibinfo {year} {1979})}\BibitemShut {NoStop}%
%%CITATION = JPAGA,A12,1075;%%
\bibitem [{\citenamefont {Portilla}(1980)}]{Portilla:1980uz}%
  \BibitemOpen
  \bibfield  {author} {\bibinfo {author} {\bibfnamefont {M.}~\bibnamefont
  {Portilla}},\ }\bibfield  {title} {\enquote {\bibinfo {title} {Scattering of
  two gravitating particles: Classical approach},}\ }\href {\doibase
  10.1088/0305-4470/13/12/017} {\bibfield  {journal} {\bibinfo  {journal} {J.
  Phys.}\ }\textbf {\bibinfo {volume} {A13}},\ \bibinfo {pages} {3677--3683}
  (\bibinfo {year} {1980})}\BibitemShut {NoStop}%
%%CITATION = JPAGA,A13,3677;%%
\bibitem [{\citenamefont {Westpfahl}\ and\ \citenamefont
  {Goller}(1979)}]{Westpfahl:1979gu}%
  \BibitemOpen
  \bibfield  {author} {\bibinfo {author} {\bibfnamefont {K.}~\bibnamefont
  {Westpfahl}}\ and\ \bibinfo {author} {\bibfnamefont {M.}~\bibnamefont
  {Goller}},\ }\bibfield  {title} {\enquote {\bibinfo {title} {Gravitational
  scattering of two relativistic particles in postlinear approximation},}\
  }\href {\doibase 10.1007/BF02817047} {\bibfield  {journal} {\bibinfo
  {journal} {Lett. Nuovo Cim.}\ }\textbf {\bibinfo {volume} {26}},\ \bibinfo
  {pages} {573--576} (\bibinfo {year} {1979})}\BibitemShut {NoStop}%
%%CITATION = NCLTA,26,573;%%
\bibitem [{\citenamefont {{Westpfahl}}(1985)}]{Westpfahl:1985}%
  \BibitemOpen
  \bibfield  {author} {\bibinfo {author} {\bibfnamefont {K.}~\bibnamefont
  {{Westpfahl}}},\ }\bibfield  {title} {\enquote {\bibinfo {title} {{High-Speed
  Scattering of Charged and Uncharged Particles in General Relativity}},}\
  }\href {\doibase 10.1002/prop.2190330802} {\bibfield  {journal} {\bibinfo
  {journal} {Fortschritte der Physik}\ }\textbf {\bibinfo {volume} {33}},\
  \bibinfo {pages} {417--493} (\bibinfo {year} {1985})}\BibitemShut {NoStop}%
\bibitem [{\citenamefont {{Sch{\"a}fer}}(1986)}]{Schafer:1986}%
  \BibitemOpen
  \bibfield  {author} {\bibinfo {author} {\bibfnamefont {Gerhard}\ \bibnamefont
  {{Sch{\"a}fer}}},\ }\bibfield  {title} {\enquote {\bibinfo {title} {{The ADM
  Hamiltonian at the postlinear approximation.}}}\ }\href {\doibase
  10.1007/BF00765886} {\bibfield  {journal} {\bibinfo  {journal} {General
  Relativity and Gravitation}\ }\textbf {\bibinfo {volume} {18}},\ \bibinfo
  {pages} {255--270} (\bibinfo {year} {1986})}\BibitemShut {NoStop}%
\bibitem [{\citenamefont {{Westpfahl}}\ \emph {et~al.}(1987)\citenamefont
  {{Westpfahl}}, \citenamefont {{Mohles}},\ and\ \citenamefont
  {{Simonis}}}]{Westpfahl:1987}%
  \BibitemOpen
  \bibfield  {author} {\bibinfo {author} {\bibfnamefont {K.}~\bibnamefont
  {{Westpfahl}}}, \bibinfo {author} {\bibfnamefont {R.}~\bibnamefont
  {{Mohles}}}, \ and\ \bibinfo {author} {\bibfnamefont {H.}~\bibnamefont
  {{Simonis}}},\ }\bibfield  {title} {\enquote {\bibinfo {title}
  {{Energy-momentum conservation for gravitational two-body scattering in the
  post-linear approximation}},}\ }\href {\doibase 10.1088/0264-9381/4/5/006}
  {\bibfield  {journal} {\bibinfo  {journal} {Classical and Quantum Gravity}\
  }\textbf {\bibinfo {volume} {4}},\ \bibinfo {pages} {L185--L188} (\bibinfo
  {year} {1987})}\BibitemShut {NoStop}%
\bibitem [{\citenamefont {Ledvinka}\ \emph {et~al.}(2008)\citenamefont
  {Ledvinka}, \citenamefont {Sch{\"a}fer},\ and\ \citenamefont
  {Bicak}}]{Ledvinka:2008tk}%
  \BibitemOpen
  \bibfield  {author} {\bibinfo {author} {\bibfnamefont {Tomas}\ \bibnamefont
  {Ledvinka}}, \bibinfo {author} {\bibfnamefont {Gerhard}\ \bibnamefont
  {Sch{\"a}fer}}, \ and\ \bibinfo {author} {\bibfnamefont {Jiri}\ \bibnamefont
  {Bicak}},\ }\bibfield  {title} {\enquote {\bibinfo {title} {{Relativistic
  Closed-Form Hamiltonian for Many-Body Gravitating Systems in the
  Post-Minkowskian Approximation}},}\ }\href {\doibase
  10.1103/PhysRevLett.100.251101} {\bibfield  {journal} {\bibinfo  {journal}
  {Phys. Rev. Lett.}\ }\textbf {\bibinfo {volume} {100}},\ \bibinfo {pages}
  {251101} (\bibinfo {year} {2008})},\ \Eprint {http://arxiv.org/abs/0807.0214}
  {arXiv:0807.0214 [gr-qc]} \BibitemShut {NoStop}%
%%CITATION = ARXIV:0807.0214;%%
\bibitem [{\citenamefont {Foffa}(2014)}]{Foffa:2013gja}%
  \BibitemOpen
  \bibfield  {author} {\bibinfo {author} {\bibfnamefont {Stefano}\ \bibnamefont
  {Foffa}},\ }\bibfield  {title} {\enquote {\bibinfo {title} {{Gravitating
  binaries at 5PN in the post-Minkowskian approximation}},}\ }\href {\doibase
  10.1103/PhysRevD.89.024019} {\bibfield  {journal} {\bibinfo  {journal} {Phys.
  Rev.}\ }\textbf {\bibinfo {volume} {D89}},\ \bibinfo {pages} {024019}
  (\bibinfo {year} {2014})},\ \Eprint {http://arxiv.org/abs/1309.3956}
  {arXiv:1309.3956 [gr-qc]} \BibitemShut {NoStop}%
%%CITATION = ARXIV:1309.3956;%%
\bibitem [{\citenamefont {Damour}(2016)}]{Damour:2016gwp}%
  \BibitemOpen
  \bibfield  {author} {\bibinfo {author} {\bibfnamefont {Thibault}\
  \bibnamefont {Damour}},\ }\bibfield  {title} {\enquote {\bibinfo {title}
  {{Gravitational scattering, post-Minkowskian approximation and Effective
  One-Body theory}},}\ }\href {\doibase 10.1103/PhysRevD.94.104015} {\bibfield
  {journal} {\bibinfo  {journal} {Phys. Rev.}\ }\textbf {\bibinfo {volume}
  {D94}},\ \bibinfo {pages} {104015} (\bibinfo {year} {2016})},\ \Eprint
  {http://arxiv.org/abs/1609.00354} {arXiv:1609.00354 [gr-qc]} \BibitemShut
  {NoStop}%
%%CITATION = ARXIV:1609.00354;%%
\bibitem [{\citenamefont {Bini}\ and\ \citenamefont
  {Damour}(2017{\natexlab{a}})}]{Bini:2017xzy}%
  \BibitemOpen
  \bibfield  {author} {\bibinfo {author} {\bibfnamefont {Donato}\ \bibnamefont
  {Bini}}\ and\ \bibinfo {author} {\bibfnamefont {Thibault}\ \bibnamefont
  {Damour}},\ }\bibfield  {title} {\enquote {\bibinfo {title} {{Gravitational
  spin-orbit coupling in binary systems, post-Minkowskian approximation and
  effective one-body theory}},}\ }\href {\doibase 10.1103/PhysRevD.96.104038}
  {\bibfield  {journal} {\bibinfo  {journal} {Phys. Rev.}\ }\textbf {\bibinfo
  {volume} {D96}},\ \bibinfo {pages} {104038} (\bibinfo {year}
  {2017}{\natexlab{a}})},\ \Eprint {http://arxiv.org/abs/1709.00590}
  {arXiv:1709.00590 [gr-qc]} \BibitemShut {NoStop}%
%%CITATION = ARXIV:1709.00590;%%
\bibitem [{\citenamefont {Vines}(2018)}]{Vines:2017hyw}%
  \BibitemOpen
  \bibfield  {author} {\bibinfo {author} {\bibfnamefont {Justin}\ \bibnamefont
  {Vines}},\ }\bibfield  {title} {\enquote {\bibinfo {title} {{Scattering of
  two spinning black holes in post-Minkowskian gravity, to all orders in spin,
  and effective-one-body mappings}},}\ }\href {\doibase
  10.1088/1361-6382/aaa3a8} {\bibfield  {journal} {\bibinfo  {journal} {Class.
  Quant. Grav.}\ }\textbf {\bibinfo {volume} {35}},\ \bibinfo {pages} {084002}
  (\bibinfo {year} {2018})},\ \Eprint {http://arxiv.org/abs/1709.06016}
  {arXiv:1709.06016 [gr-qc]} \BibitemShut {NoStop}%
%%CITATION = ARXIV:1709.06016;%%
\bibitem [{\citenamefont {Damour}(2018)}]{Damour:2017zjx}%
  \BibitemOpen
  \bibfield  {author} {\bibinfo {author} {\bibfnamefont {Thibault}\
  \bibnamefont {Damour}},\ }\bibfield  {title} {\enquote {\bibinfo {title}
  {{High-energy gravitational scattering and the general relativistic two-body
  problem}},}\ }\href {\doibase 10.1103/PhysRevD.97.044038} {\bibfield
  {journal} {\bibinfo  {journal} {Phys. Rev.}\ }\textbf {\bibinfo {volume}
  {D97}},\ \bibinfo {pages} {044038} (\bibinfo {year} {2018})},\ \Eprint
  {http://arxiv.org/abs/1710.10599} {arXiv:1710.10599 [gr-qc]} \BibitemShut
  {NoStop}%
%%CITATION = ARXIV:1710.10599;%%
\bibitem [{\citenamefont {Bini}\ and\ \citenamefont
  {Damour}(2018)}]{Bini:2018ywr}%
  \BibitemOpen
  \bibfield  {author} {\bibinfo {author} {\bibfnamefont {Donato}\ \bibnamefont
  {Bini}}\ and\ \bibinfo {author} {\bibfnamefont {Thibault}\ \bibnamefont
  {Damour}},\ }\bibfield  {title} {\enquote {\bibinfo {title} {{Gravitational
  spin-orbit coupling in binary systems at the second post-Minkowskian
  approximation}},}\ }\href {\doibase 10.1103/PhysRevD.98.044036} {\bibfield
  {journal} {\bibinfo  {journal} {Phys. Rev.}\ }\textbf {\bibinfo {volume}
  {D98}},\ \bibinfo {pages} {044036} (\bibinfo {year} {2018})},\ \Eprint
  {http://arxiv.org/abs/1805.10809} {arXiv:1805.10809 [gr-qc]} \BibitemShut
  {NoStop}%
%%CITATION = ARXIV:1805.10809;%%
\bibitem [{\citenamefont {Blanchet}\ and\ \citenamefont
  {Fokas}(2018)}]{Blanchet:2018yvb}%
  \BibitemOpen
  \bibfield  {author} {\bibinfo {author} {\bibfnamefont {Luc}\ \bibnamefont
  {Blanchet}}\ and\ \bibinfo {author} {\bibfnamefont {Athanassios~S.}\
  \bibnamefont {Fokas}},\ }\bibfield  {title} {\enquote {\bibinfo {title}
  {{Equations of motion of self-gravitating $N$-body systems in the first
  post-Minkowskian approximation}},}\ }\href {\doibase
  10.1103/PhysRevD.98.084005} {\bibfield  {journal} {\bibinfo  {journal} {Phys.
  Rev.}\ }\textbf {\bibinfo {volume} {D98}},\ \bibinfo {pages} {084005}
  (\bibinfo {year} {2018})},\ \Eprint {http://arxiv.org/abs/1806.08347}
  {arXiv:1806.08347 [gr-qc]} \BibitemShut {NoStop}%
%%CITATION = ARXIV:1806.08347;%%
\bibitem [{\citenamefont {Bjerrum-Bohr}\ \emph {et~al.}(2014)\citenamefont
  {Bjerrum-Bohr}, \citenamefont {Donoghue},\ and\ \citenamefont
  {Vanhove}}]{Bjerrum-Bohr:2013bxa}%
  \BibitemOpen
  \bibfield  {author} {\bibinfo {author} {\bibfnamefont {N.~E.~J.}\
  \bibnamefont {Bjerrum-Bohr}}, \bibinfo {author} {\bibfnamefont {John~F.}\
  \bibnamefont {Donoghue}}, \ and\ \bibinfo {author} {\bibfnamefont {Pierre}\
  \bibnamefont {Vanhove}},\ }\bibfield  {title} {\enquote {\bibinfo {title}
  {{On-shell Techniques and Universal Results in Quantum Gravity}},}\ }\href
  {\doibase 10.1007/JHEP02(2014)111} {\bibfield  {journal} {\bibinfo  {journal}
  {JHEP}\ }\textbf {\bibinfo {volume} {02}},\ \bibinfo {pages} {111} (\bibinfo
  {year} {2014})},\ \Eprint {http://arxiv.org/abs/1309.0804} {arXiv:1309.0804
  [hep-th]} \BibitemShut {NoStop}%
%%CITATION = ARXIV:1309.0804;%%
\bibitem [{\citenamefont {Holstein}\ and\ \citenamefont
  {Donoghue}(2004)}]{Holstein:2004dn}%
  \BibitemOpen
  \bibfield  {author} {\bibinfo {author} {\bibfnamefont {Barry~R.}\
  \bibnamefont {Holstein}}\ and\ \bibinfo {author} {\bibfnamefont {John~F.}\
  \bibnamefont {Donoghue}},\ }\bibfield  {title} {\enquote {\bibinfo {title}
  {{Classical physics and quantum loops}},}\ }\href {\doibase
  10.1103/PhysRevLett.93.201602} {\bibfield  {journal} {\bibinfo  {journal}
  {Phys. Rev. Lett.}\ }\textbf {\bibinfo {volume} {93}},\ \bibinfo {pages}
  {201602} (\bibinfo {year} {2004})},\ \Eprint
  {http://arxiv.org/abs/hep-th/0405239} {arXiv:hep-th/0405239 [hep-th]}
  \BibitemShut {NoStop}%
%%CITATION = HEP-TH/0405239;%%
\bibitem [{\citenamefont {Bjerrum-Bohr}\ \emph {et~al.}(2018)\citenamefont
  {Bjerrum-Bohr}, \citenamefont {Damgaard}, \citenamefont {Festuccia},
  \citenamefont {Planté},\ and\ \citenamefont
  {Vanhove}}]{Bjerrum-Bohr:2018xdl}%
  \BibitemOpen
  \bibfield  {author} {\bibinfo {author} {\bibfnamefont {N.~E.~J.}\
  \bibnamefont {Bjerrum-Bohr}}, \bibinfo {author} {\bibfnamefont {Poul~H.}\
  \bibnamefont {Damgaard}}, \bibinfo {author} {\bibfnamefont {Guido}\
  \bibnamefont {Festuccia}}, \bibinfo {author} {\bibfnamefont {Ludovic}\
  \bibnamefont {Planté}}, \ and\ \bibinfo {author} {\bibfnamefont {Pierre}\
  \bibnamefont {Vanhove}},\ }\bibfield  {title} {\enquote {\bibinfo {title}
  {{General Relativity from Scattering Amplitudes}},}\ }\href {\doibase
  10.1103/PhysRevLett.121.171601} {\bibfield  {journal} {\bibinfo  {journal}
  {Phys. Rev. Lett.}\ }\textbf {\bibinfo {volume} {121}},\ \bibinfo {pages}
  {171601} (\bibinfo {year} {2018})},\ \Eprint
  {http://arxiv.org/abs/1806.04920} {arXiv:1806.04920 [hep-th]} \BibitemShut
  {NoStop}%
%%CITATION = ARXIV:1806.04920;%%
\bibitem [{\citenamefont {Neill}\ and\ \citenamefont
  {Rothstein}(2013)}]{Neill:2013wsa}%
  \BibitemOpen
  \bibfield  {author} {\bibinfo {author} {\bibfnamefont {Duff}\ \bibnamefont
  {Neill}}\ and\ \bibinfo {author} {\bibfnamefont {Ira~Z.}\ \bibnamefont
  {Rothstein}},\ }\bibfield  {title} {\enquote {\bibinfo {title} {{Classical
  Space-Times from the S Matrix}},}\ }\href {\doibase
  10.1016/j.nuclphysb.2013.09.007} {\bibfield  {journal} {\bibinfo  {journal}
  {Nucl. Phys.}\ }\textbf {\bibinfo {volume} {B877}},\ \bibinfo {pages}
  {177--189} (\bibinfo {year} {2013})},\ \Eprint
  {http://arxiv.org/abs/1304.7263} {arXiv:1304.7263 [hep-th]} \BibitemShut
  {NoStop}%
%%CITATION = ARXIV:1304.7263;%%
\bibitem [{\citenamefont {Cheung}\ \emph {et~al.}(2018)\citenamefont {Cheung},
  \citenamefont {Rothstein},\ and\ \citenamefont {Solon}}]{Cheung:2018wkq}%
  \BibitemOpen
  \bibfield  {author} {\bibinfo {author} {\bibfnamefont {Clifford}\
  \bibnamefont {Cheung}}, \bibinfo {author} {\bibfnamefont {Ira~Z.}\
  \bibnamefont {Rothstein}}, \ and\ \bibinfo {author} {\bibfnamefont
  {Mikhail~P.}\ \bibnamefont {Solon}},\ }\bibfield  {title} {\enquote {\bibinfo
  {title} {{From Scattering Amplitudes to Classical Potentials in the
  Post-Minkowskian Expansion}},}\ }\href@noop {} {\  (\bibinfo {year}
  {2018})},\ \Eprint {http://arxiv.org/abs/1808.02489} {arXiv:1808.02489
  [hep-th]} \BibitemShut {NoStop}%
%%CITATION = ARXIV:1808.02489;%%
\bibitem [{\citenamefont {Duff}(1973)}]{Duff:1973zz}%
  \BibitemOpen
  \bibfield  {author} {\bibinfo {author} {\bibfnamefont {M.~J.}\ \bibnamefont
  {Duff}},\ }\bibfield  {title} {\enquote {\bibinfo {title} {{Quantum Tree
  Graphs and the Schwarzschild Solution}},}\ }\href {\doibase
  10.1103/PhysRevD.7.2317} {\bibfield  {journal} {\bibinfo  {journal} {Phys.
  Rev.}\ }\textbf {\bibinfo {volume} {D7}},\ \bibinfo {pages} {2317--2326}
  (\bibinfo {year} {1973})}\BibitemShut {NoStop}%
%%CITATION = PHRVA,D7,2317;%%
\bibitem [{\citenamefont {Iwasaki}(1971)}]{Iwasaki:1971vb}%
  \BibitemOpen
  \bibfield  {author} {\bibinfo {author} {\bibfnamefont {Y.}~\bibnamefont
  {Iwasaki}},\ }\bibfield  {title} {\enquote {\bibinfo {title} {{Quantum theory
  of gravitation vs. classical theory. - fourth-order potential}},}\ }\href
  {\doibase 10.1143/PTP.46.1587} {\bibfield  {journal} {\bibinfo  {journal}
  {Prog. Theor. Phys.}\ }\textbf {\bibinfo {volume} {46}},\ \bibinfo {pages}
  {1587--1609} (\bibinfo {year} {1971})}\BibitemShut {NoStop}%
%%CITATION = PTPKA,46,1587;%%
\bibitem [{\citenamefont {Kosower}\ \emph {et~al.}(2018)\citenamefont
  {Kosower}, \citenamefont {Maybee},\ and\ \citenamefont
  {O'Connell}}]{Kosower:2018adc}%
  \BibitemOpen
  \bibfield  {author} {\bibinfo {author} {\bibfnamefont {David~A.}\
  \bibnamefont {Kosower}}, \bibinfo {author} {\bibfnamefont {Ben}\ \bibnamefont
  {Maybee}}, \ and\ \bibinfo {author} {\bibfnamefont {Donal}\ \bibnamefont
  {O'Connell}},\ }\bibfield  {title} {\enquote {\bibinfo {title} {{Amplitudes,
  Observables, and Classical Scattering}},}\ }\href@noop {} {\  (\bibinfo
  {year} {2018})},\ \Eprint {http://arxiv.org/abs/1811.10950} {arXiv:1811.10950
  [hep-th]} \BibitemShut {NoStop}%
%%CITATION = ARXIV:1811.10950;%%
\bibitem [{\citenamefont {Bern}\ \emph {et~al.}(2018)\citenamefont {Bern},
  \citenamefont {Carrasco}, \citenamefont {Chen}, \citenamefont {Edison},
  \citenamefont {Johansson}, \citenamefont {Parra-Martinez}, \citenamefont
  {Roiban},\ and\ \citenamefont {Zeng}}]{Bern:2018jmv}%
  \BibitemOpen
  \bibfield  {author} {\bibinfo {author} {\bibfnamefont {Zvi}\ \bibnamefont
  {Bern}}, \bibinfo {author} {\bibfnamefont {John~Joseph}\ \bibnamefont
  {Carrasco}}, \bibinfo {author} {\bibfnamefont {Wei-Ming}\ \bibnamefont
  {Chen}}, \bibinfo {author} {\bibfnamefont {Alex}\ \bibnamefont {Edison}},
  \bibinfo {author} {\bibfnamefont {Henrik}\ \bibnamefont {Johansson}},
  \bibinfo {author} {\bibfnamefont {Julio}\ \bibnamefont {Parra-Martinez}},
  \bibinfo {author} {\bibfnamefont {Radu}\ \bibnamefont {Roiban}}, \ and\
  \bibinfo {author} {\bibfnamefont {Mao}\ \bibnamefont {Zeng}},\ }\bibfield
  {title} {\enquote {\bibinfo {title} {{Ultraviolet Properties of $\mathcal N =
  8$ Supergravity at Five Loops}},}\ }\href {\doibase
  10.1103/PhysRevD.98.086021} {\bibfield  {journal} {\bibinfo  {journal} {Phys.
  Rev.}\ }\textbf {\bibinfo {volume} {D98}},\ \bibinfo {pages} {086021}
  (\bibinfo {year} {2018})},\ \Eprint {http://arxiv.org/abs/1804.09311}
  {arXiv:1804.09311 [hep-th]} \BibitemShut {NoStop}%
%%CITATION = ARXIV:1804.09311;%%
\bibitem [{\citenamefont {Monteiro}\ \emph {et~al.}(2014)\citenamefont
  {Monteiro}, \citenamefont {O'Connell},\ and\ \citenamefont
  {White}}]{Monteiro:2014cda}%
  \BibitemOpen
  \bibfield  {author} {\bibinfo {author} {\bibfnamefont {Ricardo}\ \bibnamefont
  {Monteiro}}, \bibinfo {author} {\bibfnamefont {Donal}\ \bibnamefont
  {O'Connell}}, \ and\ \bibinfo {author} {\bibfnamefont {Chris~D.}\
  \bibnamefont {White}},\ }\bibfield  {title} {\enquote {\bibinfo {title}
  {{Black holes and the double copy}},}\ }\href {\doibase
  10.1007/JHEP12(2014)056} {\bibfield  {journal} {\bibinfo  {journal} {JHEP}\
  }\textbf {\bibinfo {volume} {12}},\ \bibinfo {pages} {056} (\bibinfo {year}
  {2014})},\ \Eprint {http://arxiv.org/abs/1410.0239} {arXiv:1410.0239
  [hep-th]} \BibitemShut {NoStop}%
%%CITATION = ARXIV:1410.0239;%%
\bibitem [{\citenamefont {Luna}\ \emph {et~al.}(2015)\citenamefont {Luna},
  \citenamefont {Monteiro}, \citenamefont {O'Connell},\ and\ \citenamefont
  {White}}]{Luna:2015paa}%
  \BibitemOpen
  \bibfield  {author} {\bibinfo {author} {\bibfnamefont {Andrés}\ \bibnamefont
  {Luna}}, \bibinfo {author} {\bibfnamefont {Ricardo}\ \bibnamefont
  {Monteiro}}, \bibinfo {author} {\bibfnamefont {Donal}\ \bibnamefont
  {O'Connell}}, \ and\ \bibinfo {author} {\bibfnamefont {Chris~D.}\
  \bibnamefont {White}},\ }\bibfield  {title} {\enquote {\bibinfo {title} {{The
  classical double copy for Taub–NUT spacetime}},}\ }\href {\doibase
  10.1016/j.physletb.2015.09.021} {\bibfield  {journal} {\bibinfo  {journal}
  {Phys. Lett.}\ }\textbf {\bibinfo {volume} {B750}},\ \bibinfo {pages}
  {272--277} (\bibinfo {year} {2015})},\ \Eprint
  {http://arxiv.org/abs/1507.01869} {arXiv:1507.01869 [hep-th]} \BibitemShut
  {NoStop}%
%%CITATION = ARXIV:1507.01869;%%
\bibitem [{\citenamefont {Luna}\ \emph {et~al.}(2017)\citenamefont {Luna},
  \citenamefont {Monteiro}, \citenamefont {Nicholson}, \citenamefont {Ochirov},
  \citenamefont {O'Connell}, \citenamefont {Westerberg},\ and\ \citenamefont
  {White}}]{Luna:2016hge}%
  \BibitemOpen
  \bibfield  {author} {\bibinfo {author} {\bibfnamefont {Andr{\'e}s}\
  \bibnamefont {Luna}}, \bibinfo {author} {\bibfnamefont {Ricardo}\
  \bibnamefont {Monteiro}}, \bibinfo {author} {\bibfnamefont {Isobel}\
  \bibnamefont {Nicholson}}, \bibinfo {author} {\bibfnamefont {Alexander}\
  \bibnamefont {Ochirov}}, \bibinfo {author} {\bibfnamefont {Donal}\
  \bibnamefont {O'Connell}}, \bibinfo {author} {\bibfnamefont {Niclas}\
  \bibnamefont {Westerberg}}, \ and\ \bibinfo {author} {\bibfnamefont
  {Chris~D.}\ \bibnamefont {White}},\ }\bibfield  {title} {\enquote {\bibinfo
  {title} {{Perturbative spacetimes from Yang-Mills theory}},}\ }\href
  {\doibase 10.1007/JHEP04(2017)069} {\bibfield  {journal} {\bibinfo  {journal}
  {JHEP}\ }\textbf {\bibinfo {volume} {04}},\ \bibinfo {pages} {069} (\bibinfo
  {year} {2017})},\ \Eprint {http://arxiv.org/abs/1611.07508} {arXiv:1611.07508
  [hep-th]} \BibitemShut {NoStop}%
%%CITATION = ARXIV:1611.07508;%%
\bibitem [{\citenamefont {Luna}\ \emph {et~al.}(2016)\citenamefont {Luna},
  \citenamefont {Monteiro}, \citenamefont {Nicholson}, \citenamefont
  {O'Connell},\ and\ \citenamefont {White}}]{Luna:2016due}%
  \BibitemOpen
  \bibfield  {author} {\bibinfo {author} {\bibfnamefont {Andrés}\ \bibnamefont
  {Luna}}, \bibinfo {author} {\bibfnamefont {Ricardo}\ \bibnamefont
  {Monteiro}}, \bibinfo {author} {\bibfnamefont {Isobel}\ \bibnamefont
  {Nicholson}}, \bibinfo {author} {\bibfnamefont {Donal}\ \bibnamefont
  {O'Connell}}, \ and\ \bibinfo {author} {\bibfnamefont {Chris~D.}\
  \bibnamefont {White}},\ }\bibfield  {title} {\enquote {\bibinfo {title} {{The
  double copy: Bremsstrahlung and accelerating black holes}},}\ }\href
  {\doibase 10.1007/JHEP06(2016)023} {\bibfield  {journal} {\bibinfo  {journal}
  {JHEP}\ }\textbf {\bibinfo {volume} {06}},\ \bibinfo {pages} {023} (\bibinfo
  {year} {2016})},\ \Eprint {http://arxiv.org/abs/1603.05737} {arXiv:1603.05737
  [hep-th]} \BibitemShut {NoStop}%
%%CITATION = ARXIV:1603.05737;%%
\bibitem [{\citenamefont {Luna}\ \emph {et~al.}(2018)\citenamefont {Luna},
  \citenamefont {Nicholson}, \citenamefont {O'Connell},\ and\ \citenamefont
  {White}}]{Luna:2017dtq}%
  \BibitemOpen
  \bibfield  {author} {\bibinfo {author} {\bibfnamefont {Andr{\'e}s}\
  \bibnamefont {Luna}}, \bibinfo {author} {\bibfnamefont {Isobel}\ \bibnamefont
  {Nicholson}}, \bibinfo {author} {\bibfnamefont {Donal}\ \bibnamefont
  {O'Connell}}, \ and\ \bibinfo {author} {\bibfnamefont {Chris~D.}\
  \bibnamefont {White}},\ }\bibfield  {title} {\enquote {\bibinfo {title}
  {{Inelastic Black Hole Scattering from Charged Scalar Amplitudes}},}\ }\href
  {\doibase 10.1007/JHEP03(2018)044} {\bibfield  {journal} {\bibinfo  {journal}
  {JHEP}\ }\textbf {\bibinfo {volume} {03}},\ \bibinfo {pages} {044} (\bibinfo
  {year} {2018})},\ \Eprint {http://arxiv.org/abs/1711.03901} {arXiv:1711.03901
  [hep-th]} \BibitemShut {NoStop}%
%%CITATION = ARXIV:1711.03901;%%
\bibitem [{\citenamefont {Goldberger}\ and\ \citenamefont
  {Ridgway}(2017)}]{Goldberger:2016iau}%
  \BibitemOpen
  \bibfield  {author} {\bibinfo {author} {\bibfnamefont {Walter~D.}\
  \bibnamefont {Goldberger}}\ and\ \bibinfo {author} {\bibfnamefont
  {Alexander~K.}\ \bibnamefont {Ridgway}},\ }\bibfield  {title} {\enquote
  {\bibinfo {title} {{Radiation and the classical double copy for color
  charges}},}\ }\href {\doibase 10.1103/PhysRevD.95.125010} {\bibfield
  {journal} {\bibinfo  {journal} {Phys. Rev.}\ }\textbf {\bibinfo {volume}
  {D95}},\ \bibinfo {pages} {125010} (\bibinfo {year} {2017})},\ \Eprint
  {http://arxiv.org/abs/1611.03493} {arXiv:1611.03493 [hep-th]} \BibitemShut
  {NoStop}%
%%CITATION = ARXIV:1611.03493;%%
\bibitem [{\citenamefont {Goldberger}\ and\ \citenamefont
  {Ridgway}(2018)}]{Goldberger:2017vcg}%
  \BibitemOpen
  \bibfield  {author} {\bibinfo {author} {\bibfnamefont {Walter~D.}\
  \bibnamefont {Goldberger}}\ and\ \bibinfo {author} {\bibfnamefont
  {Alexander~K.}\ \bibnamefont {Ridgway}},\ }\bibfield  {title} {\enquote
  {\bibinfo {title} {{Bound states and the classical double copy}},}\ }\href
  {\doibase 10.1103/PhysRevD.97.085019} {\bibfield  {journal} {\bibinfo
  {journal} {Phys. Rev.}\ }\textbf {\bibinfo {volume} {D97}},\ \bibinfo {pages}
  {085019} (\bibinfo {year} {2018})},\ \Eprint
  {http://arxiv.org/abs/1711.09493} {arXiv:1711.09493 [hep-th]} \BibitemShut
  {NoStop}%
%%CITATION = ARXIV:1711.09493;%%
\bibitem [{\citenamefont {Goldberger}\ \emph {et~al.}(2018)\citenamefont
  {Goldberger}, \citenamefont {Li},\ and\ \citenamefont
  {Prabhu}}]{Goldberger:2017ogt}%
  \BibitemOpen
  \bibfield  {author} {\bibinfo {author} {\bibfnamefont {Walter~D.}\
  \bibnamefont {Goldberger}}, \bibinfo {author} {\bibfnamefont {Jingping}\
  \bibnamefont {Li}}, \ and\ \bibinfo {author} {\bibfnamefont {Siddharth~G.}\
  \bibnamefont {Prabhu}},\ }\bibfield  {title} {\enquote {\bibinfo {title}
  {{Spinning particles, axion radiation, and the classical double copy}},}\
  }\href {\doibase 10.1103/PhysRevD.97.105018} {\bibfield  {journal} {\bibinfo
  {journal} {Phys. Rev.}\ }\textbf {\bibinfo {volume} {D97}},\ \bibinfo {pages}
  {105018} (\bibinfo {year} {2018})},\ \Eprint
  {http://arxiv.org/abs/1712.09250} {arXiv:1712.09250 [hep-th]} \BibitemShut
  {NoStop}%
%%CITATION = ARXIV:1712.09250;%%
\bibitem [{\citenamefont {Goldberger}\ \emph {et~al.}(2017)\citenamefont
  {Goldberger}, \citenamefont {Prabhu},\ and\ \citenamefont
  {Thompson}}]{Goldberger:2017frp}%
  \BibitemOpen
  \bibfield  {author} {\bibinfo {author} {\bibfnamefont {Walter~D.}\
  \bibnamefont {Goldberger}}, \bibinfo {author} {\bibfnamefont {Siddharth~G.}\
  \bibnamefont {Prabhu}}, \ and\ \bibinfo {author} {\bibfnamefont
  {Jedidiah~O.}\ \bibnamefont {Thompson}},\ }\bibfield  {title} {\enquote
  {\bibinfo {title} {{Classical gluon and graviton radiation from the
  bi-adjoint scalar double copy}},}\ }\href {\doibase
  10.1103/PhysRevD.96.065009} {\bibfield  {journal} {\bibinfo  {journal} {Phys.
  Rev.}\ }\textbf {\bibinfo {volume} {D96}},\ \bibinfo {pages} {065009}
  (\bibinfo {year} {2017})},\ \Eprint {http://arxiv.org/abs/1705.09263}
  {arXiv:1705.09263 [hep-th]} \BibitemShut {NoStop}%
%%CITATION = ARXIV:1705.09263;%%
\bibitem [{\citenamefont {Chester}(2018)}]{Chester:2017vcz}%
  \BibitemOpen
  \bibfield  {author} {\bibinfo {author} {\bibfnamefont {David}\ \bibnamefont
  {Chester}},\ }\bibfield  {title} {\enquote {\bibinfo {title} {{Radiative
  double copy for Einstein-Yang-Mills theory}},}\ }\href {\doibase
  10.1103/PhysRevD.97.084025} {\bibfield  {journal} {\bibinfo  {journal} {Phys.
  Rev.}\ }\textbf {\bibinfo {volume} {D97}},\ \bibinfo {pages} {084025}
  (\bibinfo {year} {2018})},\ \Eprint {http://arxiv.org/abs/1712.08684}
  {arXiv:1712.08684 [hep-th]} \BibitemShut {NoStop}%
%%CITATION = ARXIV:1712.08684;%%
\bibitem [{\citenamefont {Shen}(2018)}]{Shen:2018ebu}%
  \BibitemOpen
  \bibfield  {author} {\bibinfo {author} {\bibfnamefont {Chia-Hsien}\
  \bibnamefont {Shen}},\ }\bibfield  {title} {\enquote {\bibinfo {title}
  {{Gravitational Radiation from Color-Kinematics Duality}},}\ }\href@noop {}
  {\  (\bibinfo {year} {2018})},\ \Eprint {http://arxiv.org/abs/1806.07388}
  {arXiv:1806.07388 [hep-th]} \BibitemShut {NoStop}%
%%CITATION = ARXIV:1806.07388;%%
\bibitem [{\citenamefont {Plefka}\ \emph {et~al.}(2018)\citenamefont {Plefka},
  \citenamefont {Steinhoff},\ and\ \citenamefont
  {Wormsbecher}}]{Plefka:2018dpa}%
  \BibitemOpen
  \bibfield  {author} {\bibinfo {author} {\bibfnamefont {Jan}\ \bibnamefont
  {Plefka}}, \bibinfo {author} {\bibfnamefont {Jan}\ \bibnamefont {Steinhoff}},
  \ and\ \bibinfo {author} {\bibfnamefont {Wadim}\ \bibnamefont
  {Wormsbecher}},\ }\bibfield  {title} {\enquote {\bibinfo {title} {{Effective
  action of dilaton gravity as the classical double copy of Yang-Mills
  theory}},}\ }\href@noop {} {\  (\bibinfo {year} {2018})},\ \Eprint
  {http://arxiv.org/abs/1807.09859} {arXiv:1807.09859 [hep-th]} \BibitemShut
  {NoStop}%
%%CITATION = ARXIV:1807.09859;%%
\bibitem [{\citenamefont {Buonanno}\ and\ \citenamefont
  {Damour}(1999)}]{Buonanno:1998gg}%
  \BibitemOpen
  \bibfield  {author} {\bibinfo {author} {\bibfnamefont {A.}~\bibnamefont
  {Buonanno}}\ and\ \bibinfo {author} {\bibfnamefont {T.}~\bibnamefont
  {Damour}},\ }\bibfield  {title} {\enquote {\bibinfo {title} {{Effective
  one-body approach to general relativistic two-body dynamics}},}\ }\href
  {\doibase 10.1103/PhysRevD.59.084006} {\bibfield  {journal} {\bibinfo
  {journal} {Phys. Rev.}\ }\textbf {\bibinfo {volume} {D59}},\ \bibinfo {pages}
  {084006} (\bibinfo {year} {1999})},\ \Eprint
  {http://arxiv.org/abs/gr-qc/9811091} {arXiv:gr-qc/9811091 [gr-qc]}
  \BibitemShut {NoStop}%
%%CITATION = GR-QC/9811091;%%
\bibitem [{\citenamefont {Buonanno}\ and\ \citenamefont
  {Damour}(2000)}]{Buonanno:2000ef}%
  \BibitemOpen
  \bibfield  {author} {\bibinfo {author} {\bibfnamefont {Alessandra}\
  \bibnamefont {Buonanno}}\ and\ \bibinfo {author} {\bibfnamefont {Thibault}\
  \bibnamefont {Damour}},\ }\bibfield  {title} {\enquote {\bibinfo {title}
  {{Transition from inspiral to plunge in binary black hole coalescences}},}\
  }\href {\doibase 10.1103/PhysRevD.62.064015} {\bibfield  {journal} {\bibinfo
  {journal} {Phys. Rev.}\ }\textbf {\bibinfo {volume} {D62}},\ \bibinfo {pages}
  {064015} (\bibinfo {year} {2000})},\ \Eprint
  {http://arxiv.org/abs/gr-qc/0001013} {arXiv:gr-qc/0001013 [gr-qc]}
  \BibitemShut {NoStop}%
%%CITATION = GR-QC/0001013;%%
\bibitem [{\citenamefont {Damour}(2001)}]{Damour:2001tu}%
  \BibitemOpen
  \bibfield  {author} {\bibinfo {author} {\bibfnamefont {Thibault}\
  \bibnamefont {Damour}},\ }\bibfield  {title} {\enquote {\bibinfo {title}
  {{Coalescence of two spinning black holes: an effective one-body
  approach}},}\ }\href {\doibase 10.1103/PhysRevD.64.124013} {\bibfield
  {journal} {\bibinfo  {journal} {Phys. Rev.}\ }\textbf {\bibinfo {volume}
  {D64}},\ \bibinfo {pages} {124013} (\bibinfo {year} {2001})},\ \Eprint
  {http://arxiv.org/abs/gr-qc/0103018} {arXiv:gr-qc/0103018 [gr-qc]}
  \BibitemShut {NoStop}%
%%CITATION = GR-QC/0103018;%%
\bibitem [{\citenamefont {Damour}\ \emph
  {et~al.}(2008{\natexlab{a}})\citenamefont {Damour}, \citenamefont
  {Jaranowski},\ and\ \citenamefont {Sch{\"a}fer}}]{Damour:2008qf}%
  \BibitemOpen
  \bibfield  {author} {\bibinfo {author} {\bibfnamefont {Thibault}\
  \bibnamefont {Damour}}, \bibinfo {author} {\bibfnamefont {Piotr}\
  \bibnamefont {Jaranowski}}, \ and\ \bibinfo {author} {\bibfnamefont
  {Gerhard}\ \bibnamefont {Sch{\"a}fer}},\ }\bibfield  {title} {\enquote
  {\bibinfo {title} {{Effective one body approach to the dynamics of two
  spinning black holes with next-to-leading order spin-orbit coupling}},}\
  }\href {\doibase 10.1103/PhysRevD.78.024009} {\bibfield  {journal} {\bibinfo
  {journal} {Phys. Rev.}\ }\textbf {\bibinfo {volume} {D78}},\ \bibinfo {pages}
  {024009} (\bibinfo {year} {2008}{\natexlab{a}})},\ \Eprint
  {http://arxiv.org/abs/0803.0915} {arXiv:0803.0915 [gr-qc]} \BibitemShut
  {NoStop}%
%%CITATION = ARXIV:0803.0915;%%
\bibitem [{\citenamefont {Barausse}\ and\ \citenamefont
  {Buonanno}(2010)}]{Barausse:2009xi}%
  \BibitemOpen
  \bibfield  {author} {\bibinfo {author} {\bibfnamefont {Enrico}\ \bibnamefont
  {Barausse}}\ and\ \bibinfo {author} {\bibfnamefont {Alessandra}\ \bibnamefont
  {Buonanno}},\ }\bibfield  {title} {\enquote {\bibinfo {title} {{An Improved
  effective-one-body Hamiltonian for spinning black-hole binaries}},}\ }\href
  {\doibase 10.1103/PhysRevD.81.084024} {\bibfield  {journal} {\bibinfo
  {journal} {Phys. Rev.}\ }\textbf {\bibinfo {volume} {D81}},\ \bibinfo {pages}
  {084024} (\bibinfo {year} {2010})},\ \Eprint {http://arxiv.org/abs/0912.3517}
  {arXiv:0912.3517 [gr-qc]} \BibitemShut {NoStop}%
%%CITATION = ARXIV:0912.3517;%%
\bibitem [{\citenamefont {Mathisson}(1937)}]{Mathisson:1937}%
  \BibitemOpen
  \bibfield  {author} {\bibinfo {author} {\bibfnamefont {Myron}\ \bibnamefont
  {Mathisson}},\ }\bibfield  {title} {\enquote {\bibinfo {title} {Neue mechanik
  materieller systeme},}\ }\href@noop {} {\bibfield  {journal} {\bibinfo
  {journal} {Acta Physica Polonica}\ }\textbf {\bibinfo {volume} {6}},\
  \bibinfo {pages} {163--200} (\bibinfo {year} {1937})}\BibitemShut {NoStop}%
\bibitem [{\citenamefont {Mathisson}(2010)}]{Mathisson:2010}%
  \BibitemOpen
  \bibfield  {author} {\bibinfo {author} {\bibfnamefont {Myron}\ \bibnamefont
  {Mathisson}},\ }\bibfield  {title} {\enquote {\bibinfo {title} {Republication
  of: {N}ew mechanics of material systems},}\ }\href {\doibase
  10.1007/s10714-010-0939-y} {\bibfield  {journal} {\bibinfo  {journal} {Gen.
  Relativ. Gravit.}\ }\textbf {\bibinfo {volume} {42}},\ \bibinfo {pages}
  {1011--1048} (\bibinfo {year} {2010})}\BibitemShut {NoStop}%
\bibitem [{\citenamefont {Dixon}(1979)}]{Dixon:1979}%
  \BibitemOpen
  \bibfield  {author} {\bibinfo {author} {\bibfnamefont {W.~Graham}\
  \bibnamefont {Dixon}},\ }\bibfield  {title} {\enquote {\bibinfo {title}
  {Extended bodies in general relativity: {T}heir description and motion},}\
  }in\ \href@noop {} {\emph {\bibinfo {booktitle} {Proceedings of the
  International School of Physics Enrico Fermi LXVII}}},\ \bibinfo {editor}
  {edited by\ \bibinfo {editor} {\bibfnamefont {J{\"u}rgen}\ \bibnamefont
  {Ehlers}}}\ (\bibinfo  {publisher} {North Holland},\ \bibinfo {address}
  {Amsterdam},\ \bibinfo {year} {1979})\ pp.\ \bibinfo {pages}
  {156--219}\BibitemShut {NoStop}%
\bibitem [{\citenamefont {Dixon}(2015)}]{Dixon:2015vxa}%
  \BibitemOpen
  \bibfield  {author} {\bibinfo {author} {\bibfnamefont {W.~G.}\ \bibnamefont
  {Dixon}},\ }\bibfield  {title} {\enquote {\bibinfo {title} {{The New
  Mechanics of Myron Mathisson and Its Subsequent Development}},}\ }\bibfield
  {booktitle} {\emph {\bibinfo {booktitle} {{Proceedings, 524th
  WE-Heraeus-Seminar: Equations of Motion in Relativistic Gravity (EOM
  2013)}}},\ }\href {\doibase 10.1007/978-3-319-18335-0_1} {\bibfield
  {journal} {\bibinfo  {journal} {Fund. Theor. Phys.}\ }\textbf {\bibinfo
  {volume} {179}},\ \bibinfo {pages} {1--66} (\bibinfo {year}
  {2015})}\BibitemShut {NoStop}%
%%CITATION = INSPIRE-1381011;%%
\bibitem [{\citenamefont {Hansen}(1974)}]{Hansen:1974zz}%
  \BibitemOpen
  \bibfield  {author} {\bibinfo {author} {\bibfnamefont {R.~O.}\ \bibnamefont
  {Hansen}},\ }\bibfield  {title} {\enquote {\bibinfo {title} {{Multipole
  moments of stationary space-times}},}\ }\href {\doibase 10.1063/1.1666501}
  {\bibfield  {journal} {\bibinfo  {journal} {J. Math. Phys.}\ }\textbf
  {\bibinfo {volume} {15}},\ \bibinfo {pages} {46--52} (\bibinfo {year}
  {1974})}\BibitemShut {NoStop}%
%%CITATION = JMAPA,15,46;%%
\bibitem [{\citenamefont {Vines}\ and\ \citenamefont
  {Steinhoff}(2018)}]{Vines:2016qwa}%
  \BibitemOpen
  \bibfield  {author} {\bibinfo {author} {\bibfnamefont {Justin}\ \bibnamefont
  {Vines}}\ and\ \bibinfo {author} {\bibfnamefont {Jan}\ \bibnamefont
  {Steinhoff}},\ }\bibfield  {title} {\enquote {\bibinfo {title}
  {{Spin-multipole effects in binary black holes and the test-body limit}},}\
  }\href {\doibase 10.1103/PhysRevD.97.064010} {\bibfield  {journal} {\bibinfo
  {journal} {Phys. Rev.}\ }\textbf {\bibinfo {volume} {D97}},\ \bibinfo {pages}
  {064010} (\bibinfo {year} {2018})},\ \Eprint
  {http://arxiv.org/abs/1606.08832} {arXiv:1606.08832 [gr-qc]} \BibitemShut
  {NoStop}%
%%CITATION = ARXIV:1606.08832;%%
\bibitem [{\citenamefont {Siemonsen}\ \emph {et~al.}(2018)\citenamefont
  {Siemonsen}, \citenamefont {Steinhoff},\ and\ \citenamefont
  {Vines}}]{Siemonsen:2017yux}%
  \BibitemOpen
  \bibfield  {author} {\bibinfo {author} {\bibfnamefont {Nils}\ \bibnamefont
  {Siemonsen}}, \bibinfo {author} {\bibfnamefont {Jan}\ \bibnamefont
  {Steinhoff}}, \ and\ \bibinfo {author} {\bibfnamefont {Justin}\ \bibnamefont
  {Vines}},\ }\bibfield  {title} {\enquote {\bibinfo {title} {{Gravitational
  waves from spinning binary black holes at the leading post-Newtonian orders
  at all orders in spin}},}\ }\href {\doibase 10.1103/PhysRevD.97.124046}
  {\bibfield  {journal} {\bibinfo  {journal} {Phys. Rev.}\ }\textbf {\bibinfo
  {volume} {D97}},\ \bibinfo {pages} {124046} (\bibinfo {year} {2018})},\
  \Eprint {http://arxiv.org/abs/1712.08603} {arXiv:1712.08603 [gr-qc]}
  \BibitemShut {NoStop}%
%%CITATION = ARXIV:1712.08603;%%
\bibitem [{\citenamefont {Bini}\ and\ \citenamefont
  {Damour}(2017{\natexlab{b}})}]{Bini:2017wfr}%
  \BibitemOpen
  \bibfield  {author} {\bibinfo {author} {\bibfnamefont {Donato}\ \bibnamefont
  {Bini}}\ and\ \bibinfo {author} {\bibfnamefont {Thibault}\ \bibnamefont
  {Damour}},\ }\bibfield  {title} {\enquote {\bibinfo {title} {{Gravitational
  scattering of two black holes at the fourth post-Newtonian approximation}},}\
  }\href {\doibase 10.1103/PhysRevD.96.064021} {\bibfield  {journal} {\bibinfo
  {journal} {Phys. Rev.}\ }\textbf {\bibinfo {volume} {D96}},\ \bibinfo {pages}
  {064021} (\bibinfo {year} {2017}{\natexlab{b}})},\ \Eprint
  {http://arxiv.org/abs/1706.06877} {arXiv:1706.06877 [gr-qc]} \BibitemShut
  {NoStop}%
%%CITATION = ARXIV:1706.06877;%%
\bibitem [{\citenamefont {Damour}\ \emph {et~al.}(2014)\citenamefont {Damour},
  \citenamefont {Jaranowski},\ and\ \citenamefont
  {Sch{\"a}fer}}]{Damour:2014jta}%
  \BibitemOpen
  \bibfield  {author} {\bibinfo {author} {\bibfnamefont {Thibault}\
  \bibnamefont {Damour}}, \bibinfo {author} {\bibfnamefont {Piotr}\
  \bibnamefont {Jaranowski}}, \ and\ \bibinfo {author} {\bibfnamefont
  {Gerhard}\ \bibnamefont {Sch{\"a}fer}},\ }\bibfield  {title} {\enquote
  {\bibinfo {title} {{Nonlocal-in-time action for the fourth post-Newtonian
  conservative dynamics of two-body systems}},}\ }\href {\doibase
  10.1103/PhysRevD.89.064058} {\bibfield  {journal} {\bibinfo  {journal} {Phys.
  Rev.}\ }\textbf {\bibinfo {volume} {D89}},\ \bibinfo {pages} {064058}
  (\bibinfo {year} {2014})},\ \Eprint {http://arxiv.org/abs/1401.4548}
  {arXiv:1401.4548 [gr-qc]} \BibitemShut {NoStop}%
%%CITATION = ARXIV:1401.4548;%%
\bibitem [{\citenamefont {Cachazo}\ and\ \citenamefont
  {Guevara}(2017)}]{Cachazo:2017jef}%
  \BibitemOpen
  \bibfield  {author} {\bibinfo {author} {\bibfnamefont {Freddy}\ \bibnamefont
  {Cachazo}}\ and\ \bibinfo {author} {\bibfnamefont {Alfredo}\ \bibnamefont
  {Guevara}},\ }\bibfield  {title} {\enquote {\bibinfo {title} {{Leading
  Singularities and Classical Gravitational Scattering}},}\ }\href@noop {} {\
  (\bibinfo {year} {2017})},\ \Eprint {http://arxiv.org/abs/1705.10262}
  {arXiv:1705.10262 [hep-th]} \BibitemShut {NoStop}%
%%CITATION = ARXIV:1705.10262;%%
\bibitem [{\citenamefont {Tulczyjew}(1959)}]{Tulczyjew:1959}%
  \BibitemOpen
  \bibfield  {author} {\bibinfo {author} {\bibfnamefont {W{\l}odzimierz~M.}\
  \bibnamefont {Tulczyjew}},\ }\bibfield  {title} {\enquote {\bibinfo {title}
  {{M}otion of multipole particles in general relativity theory},}\ }\href@noop
  {} {\bibfield  {journal} {\bibinfo  {journal} {Acta Phys. Pol.}\ }\textbf
  {\bibinfo {volume} {18}},\ \bibinfo {pages} {393--409} (\bibinfo {year}
  {1959})}\BibitemShut {NoStop}%
\bibitem [{\citenamefont {Schattner}(1979)}]{Schattner:1979vp}%
  \BibitemOpen
  \bibfield  {author} {\bibinfo {author} {\bibfnamefont {R.}~\bibnamefont
  {Schattner}},\ }\bibfield  {title} {\enquote {\bibinfo {title} {The
  center-of-mass in general relativity},}\ }\href {\doibase 10.1007/BF00760221}
  {\bibfield  {journal} {\bibinfo  {journal} {Gen. Rel. Grav.}\ }\textbf
  {\bibinfo {volume} {10}},\ \bibinfo {pages} {377--393} (\bibinfo {year}
  {1979})}\BibitemShut {NoStop}%
%%CITATION = GRGVA,10,377;%%
\bibitem [{\citenamefont {Jaranowski}\ and\ \citenamefont
  {Sch{\"a}fer}(1998)}]{Jaranowski:1997ky}%
  \BibitemOpen
  \bibfield  {author} {\bibinfo {author} {\bibfnamefont {Piotr}\ \bibnamefont
  {Jaranowski}}\ and\ \bibinfo {author} {\bibfnamefont {Gerhard}\ \bibnamefont
  {Sch{\"a}fer}},\ }\bibfield  {title} {\enquote {\bibinfo {title} {{Third
  post-Newtonian higher order ADM Hamilton dynamics for two-body point mass
  systems}},}\ }\href {\doibase 10.1103/PhysRevD.57.7274,
  10.1103/PhysRevD.63.029902} {\bibfield  {journal} {\bibinfo  {journal} {Phys.
  Rev.}\ }\textbf {\bibinfo {volume} {D57}},\ \bibinfo {pages} {7274--7291}
  (\bibinfo {year} {1998})},\ \bibinfo {note} {[Erratum: Phys.
  Rev.D63,029902(2001)]},\ \Eprint {http://arxiv.org/abs/gr-qc/9712075}
  {arXiv:gr-qc/9712075 [gr-qc]} \BibitemShut {NoStop}%
%%CITATION = GR-QC/9712075;%%
\bibitem [{\citenamefont {Jaranowski}\ and\ \citenamefont
  {Sch{\"a}fer}(1999)}]{Jaranowski:1999ye}%
  \BibitemOpen
  \bibfield  {author} {\bibinfo {author} {\bibfnamefont {Piotr}\ \bibnamefont
  {Jaranowski}}\ and\ \bibinfo {author} {\bibfnamefont {Gerhard}\ \bibnamefont
  {Sch{\"a}fer}},\ }\bibfield  {title} {\enquote {\bibinfo {title} {{The Binary
  black hole problem at the third post-Newtonian approximation in the orbital
  motion: Static part}},}\ }\href {\doibase 10.1103/PhysRevD.60.124003}
  {\bibfield  {journal} {\bibinfo  {journal} {Phys. Rev.}\ }\textbf {\bibinfo
  {volume} {D60}},\ \bibinfo {pages} {124003} (\bibinfo {year} {1999})},\
  \Eprint {http://arxiv.org/abs/gr-qc/9906092} {arXiv:gr-qc/9906092 [gr-qc]}
  \BibitemShut {NoStop}%
%%CITATION = GR-QC/9906092;%%
\bibitem [{\citenamefont {Damour}\ \emph
  {et~al.}(2000{\natexlab{a}})\citenamefont {Damour}, \citenamefont
  {Jaranowski},\ and\ \citenamefont {Sch{\"a}fer}}]{Damour:2000kk}%
  \BibitemOpen
  \bibfield  {author} {\bibinfo {author} {\bibfnamefont {Thibault}\
  \bibnamefont {Damour}}, \bibinfo {author} {\bibfnamefont {Piotr}\
  \bibnamefont {Jaranowski}}, \ and\ \bibinfo {author} {\bibfnamefont
  {Gerhard}\ \bibnamefont {Sch{\"a}fer}},\ }\bibfield  {title} {\enquote
  {\bibinfo {title} {{Poincare invariance in the ADM Hamiltonian approach to
  the general relativistic two-body problem}},}\ }\href {\doibase
  10.1103/PhysRevD.63.029903, 10.1103/PhysRevD.62.021501} {\bibfield  {journal}
  {\bibinfo  {journal} {Phys. Rev.}\ }\textbf {\bibinfo {volume} {D62}},\
  \bibinfo {pages} {021501} (\bibinfo {year} {2000}{\natexlab{a}})},\ \bibinfo
  {note} {[Erratum: Phys. Rev.D63,029903(2001)]},\ \Eprint
  {http://arxiv.org/abs/gr-qc/0003051} {arXiv:gr-qc/0003051 [gr-qc]}
  \BibitemShut {NoStop}%
%%CITATION = GR-QC/0003051;%%
\bibitem [{\citenamefont {Jaranowski}\ and\ \citenamefont
  {Sch{\"a}fer}(2000)}]{Jaranowski:1999qd}%
  \BibitemOpen
  \bibfield  {author} {\bibinfo {author} {\bibfnamefont {Piotr}\ \bibnamefont
  {Jaranowski}}\ and\ \bibinfo {author} {\bibfnamefont {Gerhard}\ \bibnamefont
  {Sch{\"a}fer}},\ }\bibfield  {title} {\enquote {\bibinfo {title} {{The Binary
  black hole dynamics at the third post-Newtonian order in the orbital
  motion}},}\ }\bibfield  {booktitle} {\emph {\bibinfo {booktitle}
  {{Gravitation. Proceedings, International European Conference, 27th Session
  of the Journees Relativistes, Weimar, Germany, September 12-17, 1999}}},\
  }\href {\doibase
  10.1002/(SICI)1521-3889(200005)9:3/5<378::AID-ANDP378>3.0.CO;2-M} {\bibfield
  {journal} {\bibinfo  {journal} {Annalen Phys.}\ }\textbf {\bibinfo {volume}
  {9}},\ \bibinfo {pages} {378--383} (\bibinfo {year} {2000})},\ \Eprint
  {http://arxiv.org/abs/gr-qc/0003054} {arXiv:gr-qc/0003054 [gr-qc]}
  \BibitemShut {NoStop}%
%%CITATION = GR-QC/0003054;%%
\bibitem [{\citenamefont {Blanchet}\ and\ \citenamefont
  {Faye}(2000{\natexlab{a}})}]{Blanchet:2000nv}%
  \BibitemOpen
  \bibfield  {author} {\bibinfo {author} {\bibfnamefont {Luc}\ \bibnamefont
  {Blanchet}}\ and\ \bibinfo {author} {\bibfnamefont {Guillaume}\ \bibnamefont
  {Faye}},\ }\bibfield  {title} {\enquote {\bibinfo {title} {{Equations of
  motion of point particle binaries at the third post-Newtonian order}},}\
  }\href {\doibase 10.1016/S0375-9601(00)00360-1} {\bibfield  {journal}
  {\bibinfo  {journal} {Phys. Lett.}\ }\textbf {\bibinfo {volume} {A271}},\
  \bibinfo {pages} {58} (\bibinfo {year} {2000}{\natexlab{a}})},\ \Eprint
  {http://arxiv.org/abs/gr-qc/0004009} {arXiv:gr-qc/0004009 [gr-qc]}
  \BibitemShut {NoStop}%
%%CITATION = GR-QC/0004009;%%
\bibitem [{\citenamefont {Blanchet}\ and\ \citenamefont
  {Faye}(2001{\natexlab{a}})}]{Blanchet:2000ub}%
  \BibitemOpen
  \bibfield  {author} {\bibinfo {author} {\bibfnamefont {Luc}\ \bibnamefont
  {Blanchet}}\ and\ \bibinfo {author} {\bibfnamefont {Guillaume}\ \bibnamefont
  {Faye}},\ }\bibfield  {title} {\enquote {\bibinfo {title} {{General
  relativistic dynamics of compact binaries at the third post-Newtonian
  order}},}\ }\href {\doibase 10.1103/PhysRevD.63.062005} {\bibfield  {journal}
  {\bibinfo  {journal} {Phys. Rev.}\ }\textbf {\bibinfo {volume} {D63}},\
  \bibinfo {pages} {062005} (\bibinfo {year} {2001}{\natexlab{a}})},\ \Eprint
  {http://arxiv.org/abs/gr-qc/0007051} {arXiv:gr-qc/0007051 [gr-qc]}
  \BibitemShut {NoStop}%
%%CITATION = GR-QC/0007051;%%
\bibitem [{\citenamefont {Blanchet}\ and\ \citenamefont
  {Faye}(2000{\natexlab{b}})}]{Blanchet:2000nu}%
  \BibitemOpen
  \bibfield  {author} {\bibinfo {author} {\bibfnamefont {Luc}\ \bibnamefont
  {Blanchet}}\ and\ \bibinfo {author} {\bibfnamefont {Guillaume}\ \bibnamefont
  {Faye}},\ }\bibfield  {title} {\enquote {\bibinfo {title} {{Hadamard
  regularization}},}\ }\href {\doibase 10.1063/1.1308506} {\bibfield  {journal}
  {\bibinfo  {journal} {J. Math. Phys.}\ }\textbf {\bibinfo {volume} {41}},\
  \bibinfo {pages} {7675--7714} (\bibinfo {year} {2000}{\natexlab{b}})},\
  \Eprint {http://arxiv.org/abs/gr-qc/0004008} {arXiv:gr-qc/0004008 [gr-qc]}
  \BibitemShut {NoStop}%
%%CITATION = GR-QC/0004008;%%
\bibitem [{\citenamefont {Blanchet}\ and\ \citenamefont
  {Faye}(2001{\natexlab{b}})}]{Blanchet:2000cw}%
  \BibitemOpen
  \bibfield  {author} {\bibinfo {author} {\bibfnamefont {Luc}\ \bibnamefont
  {Blanchet}}\ and\ \bibinfo {author} {\bibfnamefont {Guillaume}\ \bibnamefont
  {Faye}},\ }\bibfield  {title} {\enquote {\bibinfo {title} {{Lorentzian
  regularization and the problem of point - like particles in general
  relativity}},}\ }\href {\doibase 10.1063/1.1384864} {\bibfield  {journal}
  {\bibinfo  {journal} {J. Math. Phys.}\ }\textbf {\bibinfo {volume} {42}},\
  \bibinfo {pages} {4391--4418} (\bibinfo {year} {2001}{\natexlab{b}})},\
  \Eprint {http://arxiv.org/abs/gr-qc/0006100} {arXiv:gr-qc/0006100 [gr-qc]}
  \BibitemShut {NoStop}%
%%CITATION = GR-QC/0006100;%%
\bibitem [{\citenamefont {Damour}\ \emph
  {et~al.}(2001{\natexlab{a}})\citenamefont {Damour}, \citenamefont
  {Jaranowski},\ and\ \citenamefont {Sch{\"a}fer}}]{Damour:2000ni}%
  \BibitemOpen
  \bibfield  {author} {\bibinfo {author} {\bibfnamefont {Thibault}\
  \bibnamefont {Damour}}, \bibinfo {author} {\bibfnamefont {Piotr}\
  \bibnamefont {Jaranowski}}, \ and\ \bibinfo {author} {\bibfnamefont
  {Gerhard}\ \bibnamefont {Sch{\"a}fer}},\ }\bibfield  {title} {\enquote
  {\bibinfo {title} {{Equivalence between the ADM-Hamiltonian and the harmonic
  coordinates approaches to the third post-Newtonian dynamics of compact
  binaries}},}\ }\href {\doibase 10.1103/PhysRevD.63.044021,
  10.1103/PhysRevD.66.029901} {\bibfield  {journal} {\bibinfo  {journal} {Phys.
  Rev.}\ }\textbf {\bibinfo {volume} {D63}},\ \bibinfo {pages} {044021}
  (\bibinfo {year} {2001}{\natexlab{a}})},\ \bibinfo {note} {[Erratum: Phys.
  Rev.D66,029901(2002)]},\ \Eprint {http://arxiv.org/abs/gr-qc/0010040}
  {arXiv:gr-qc/0010040 [gr-qc]} \BibitemShut {NoStop}%
%%CITATION = GR-QC/0010040;%%
\bibitem [{\citenamefont {de~Andrade}\ \emph {et~al.}(2001)\citenamefont
  {de~Andrade}, \citenamefont {Blanchet},\ and\ \citenamefont
  {Faye}}]{deAndrade:2000gf}%
  \BibitemOpen
  \bibfield  {author} {\bibinfo {author} {\bibfnamefont {Vanessa~C.}\
  \bibnamefont {de~Andrade}}, \bibinfo {author} {\bibfnamefont {Luc}\
  \bibnamefont {Blanchet}}, \ and\ \bibinfo {author} {\bibfnamefont
  {Guillaume}\ \bibnamefont {Faye}},\ }\bibfield  {title} {\enquote {\bibinfo
  {title} {{Third post-Newtonian dynamics of compact binaries: Noetherian
  conserved quantities and equivalence between the harmonic coordinate and ADM
  Hamiltonian formalisms}},}\ }\href {\doibase 10.1088/0264-9381/18/5/301}
  {\bibfield  {journal} {\bibinfo  {journal} {Class. Quant. Grav.}\ }\textbf
  {\bibinfo {volume} {18}},\ \bibinfo {pages} {753--778} (\bibinfo {year}
  {2001})},\ \Eprint {http://arxiv.org/abs/gr-qc/0011063} {arXiv:gr-qc/0011063
  [gr-qc]} \BibitemShut {NoStop}%
%%CITATION = GR-QC/0011063;%%
\bibitem [{\citenamefont {Damour}\ \emph
  {et~al.}(2001{\natexlab{b}})\citenamefont {Damour}, \citenamefont
  {Jaranowski},\ and\ \citenamefont {Sch{\"a}fer}}]{Damour:2001bu}%
  \BibitemOpen
  \bibfield  {author} {\bibinfo {author} {\bibfnamefont {Thibault}\
  \bibnamefont {Damour}}, \bibinfo {author} {\bibfnamefont {Piotr}\
  \bibnamefont {Jaranowski}}, \ and\ \bibinfo {author} {\bibfnamefont
  {Gerhard}\ \bibnamefont {Sch{\"a}fer}},\ }\bibfield  {title} {\enquote
  {\bibinfo {title} {{Dimensional regularization of the gravitational
  interaction of point masses}},}\ }\href {\doibase
  10.1016/S0370-2693(01)00642-6} {\bibfield  {journal} {\bibinfo  {journal}
  {Phys. Lett.}\ }\textbf {\bibinfo {volume} {B513}},\ \bibinfo {pages}
  {147--155} (\bibinfo {year} {2001}{\natexlab{b}})},\ \Eprint
  {http://arxiv.org/abs/gr-qc/0105038} {arXiv:gr-qc/0105038 [gr-qc]}
  \BibitemShut {NoStop}%
%%CITATION = GR-QC/0105038;%%
\bibitem [{\citenamefont {Blanchet}\ \emph {et~al.}(2004)\citenamefont
  {Blanchet}, \citenamefont {Damour},\ and\ \citenamefont
  {Esposito-Farese}}]{Blanchet:2003gy}%
  \BibitemOpen
  \bibfield  {author} {\bibinfo {author} {\bibfnamefont {Luc}\ \bibnamefont
  {Blanchet}}, \bibinfo {author} {\bibfnamefont {Thibault}\ \bibnamefont
  {Damour}}, \ and\ \bibinfo {author} {\bibfnamefont {Gilles}\ \bibnamefont
  {Esposito-Farese}},\ }\bibfield  {title} {\enquote {\bibinfo {title}
  {{Dimensional regularization of the third post-Newtonian dynamics of point
  particles in harmonic coordinates}},}\ }\href {\doibase
  10.1103/PhysRevD.69.124007} {\bibfield  {journal} {\bibinfo  {journal} {Phys.
  Rev.}\ }\textbf {\bibinfo {volume} {D69}},\ \bibinfo {pages} {124007}
  (\bibinfo {year} {2004})},\ \Eprint {http://arxiv.org/abs/gr-qc/0311052}
  {arXiv:gr-qc/0311052 [gr-qc]} \BibitemShut {NoStop}%
%%CITATION = GR-QC/0311052;%%
\bibitem [{\citenamefont {Blanchet}\ and\ \citenamefont
  {Iyer}(2003)}]{Blanchet:2002mb}%
  \BibitemOpen
  \bibfield  {author} {\bibinfo {author} {\bibfnamefont {Luc}\ \bibnamefont
  {Blanchet}}\ and\ \bibinfo {author} {\bibfnamefont {Bala~R.}\ \bibnamefont
  {Iyer}},\ }\bibfield  {title} {\enquote {\bibinfo {title} {{Third
  post-Newtonian dynamics of compact binaries: Equations of motion in the
  center-of-mass frame}},}\ }\href {\doibase 10.1088/0264-9381/20/4/309}
  {\bibfield  {journal} {\bibinfo  {journal} {Class. Quant. Grav.}\ }\textbf
  {\bibinfo {volume} {20}},\ \bibinfo {pages} {755} (\bibinfo {year} {2003})},\
  \Eprint {http://arxiv.org/abs/gr-qc/0209089} {arXiv:gr-qc/0209089 [gr-qc]}
  \BibitemShut {NoStop}%
%%CITATION = GR-QC/0209089;%%
\bibitem [{\citenamefont {Itoh}\ and\ \citenamefont
  {Futamase}(2003)}]{Itoh:2003fy}%
  \BibitemOpen
  \bibfield  {author} {\bibinfo {author} {\bibfnamefont {Yousuke}\ \bibnamefont
  {Itoh}}\ and\ \bibinfo {author} {\bibfnamefont {Toshifumi}\ \bibnamefont
  {Futamase}},\ }\bibfield  {title} {\enquote {\bibinfo {title} {{New
  derivation of a third post-Newtonian equation of motion for relativistic
  compact binaries without ambiguity}},}\ }\href {\doibase
  10.1103/PhysRevD.68.121501} {\bibfield  {journal} {\bibinfo  {journal} {Phys.
  Rev.}\ }\textbf {\bibinfo {volume} {D68}},\ \bibinfo {pages} {121501}
  (\bibinfo {year} {2003})},\ \Eprint {http://arxiv.org/abs/gr-qc/0310028}
  {arXiv:gr-qc/0310028 [gr-qc]} \BibitemShut {NoStop}%
%%CITATION = GR-QC/0310028;%%
\bibitem [{\citenamefont {Itoh}(2004)}]{Itoh:2003fz}%
  \BibitemOpen
  \bibfield  {author} {\bibinfo {author} {\bibfnamefont {Yousuke}\ \bibnamefont
  {Itoh}},\ }\bibfield  {title} {\enquote {\bibinfo {title} {{Equation of
  motion for relativistic compact binaries with the strong field point particle
  limit: Third post-Newtonian order}},}\ }\href {\doibase
  10.1103/PhysRevD.69.064018} {\bibfield  {journal} {\bibinfo  {journal} {Phys.
  Rev.}\ }\textbf {\bibinfo {volume} {D69}},\ \bibinfo {pages} {064018}
  (\bibinfo {year} {2004})},\ \Eprint {http://arxiv.org/abs/gr-qc/0310029}
  {arXiv:gr-qc/0310029 [gr-qc]} \BibitemShut {NoStop}%
%%CITATION = GR-QC/0310029;%%
\bibitem [{\citenamefont {Foffa}\ and\ \citenamefont
  {Sturani}(2011)}]{Foffa:2011ub}%
  \BibitemOpen
  \bibfield  {author} {\bibinfo {author} {\bibfnamefont {Stefano}\ \bibnamefont
  {Foffa}}\ and\ \bibinfo {author} {\bibfnamefont {Riccardo}\ \bibnamefont
  {Sturani}},\ }\bibfield  {title} {\enquote {\bibinfo {title} {{Effective
  field theory calculation of conservative binary dynamics at third
  post-Newtonian order}},}\ }\href {\doibase 10.1103/PhysRevD.84.044031}
  {\bibfield  {journal} {\bibinfo  {journal} {Phys. Rev.}\ }\textbf {\bibinfo
  {volume} {D84}},\ \bibinfo {pages} {044031} (\bibinfo {year} {2011})},\
  \Eprint {http://arxiv.org/abs/1104.1122} {arXiv:1104.1122 [gr-qc]}
  \BibitemShut {NoStop}%
%%CITATION = ARXIV:1104.1122;%%
\bibitem [{\citenamefont {Barker}\ and\ \citenamefont
  {O'Connell}(1970)}]{Barker:1970zr}%
  \BibitemOpen
  \bibfield  {author} {\bibinfo {author} {\bibfnamefont {B.~M.}\ \bibnamefont
  {Barker}}\ and\ \bibinfo {author} {\bibfnamefont {R.~F.}\ \bibnamefont
  {O'Connell}},\ }\bibfield  {title} {\enquote {\bibinfo {title} {{Derivation
  of the equations of motion of a gyroscope from the quantum theory of
  gravitation}},}\ }\href {\doibase 10.1103/PhysRevD.2.1428} {\bibfield
  {journal} {\bibinfo  {journal} {Phys. Rev.}\ }\textbf {\bibinfo {volume}
  {D2}},\ \bibinfo {pages} {1428--1435} (\bibinfo {year} {1970})}\BibitemShut
  {NoStop}%
%%CITATION = PHRVA,D2,1428;%%
\bibitem [{\citenamefont {Barker}\ and\ \citenamefont
  {O'Connell}(1975)}]{Barker:1975ae}%
  \BibitemOpen
  \bibfield  {author} {\bibinfo {author} {\bibfnamefont {B.~M.}\ \bibnamefont
  {Barker}}\ and\ \bibinfo {author} {\bibfnamefont {R.~F.}\ \bibnamefont
  {O'Connell}},\ }\bibfield  {title} {\enquote {\bibinfo {title}
  {{Gravitational Two-Body Problem with Arbitrary Masses, Spins, and Quadrupole
  Moments}},}\ }\href {\doibase 10.1103/PhysRevD.12.329} {\bibfield  {journal}
  {\bibinfo  {journal} {Phys. Rev.}\ }\textbf {\bibinfo {volume} {D12}},\
  \bibinfo {pages} {329--335} (\bibinfo {year} {1975})}\BibitemShut {NoStop}%
%%CITATION = PHRVA,D12,329;%%
\bibitem [{\citenamefont {D'Eath}(1975)}]{D'Eath:1975vw}%
  \BibitemOpen
  \bibfield  {author} {\bibinfo {author} {\bibfnamefont {Peter~D.}\
  \bibnamefont {D'Eath}},\ }\bibfield  {title} {\enquote {\bibinfo {title}
  {{Interaction of two black holes in the slow-motion limit}},}\ }\href
  {\doibase 10.1103/PhysRevD.12.2183} {\bibfield  {journal} {\bibinfo
  {journal} {Phys. Rev.}\ }\textbf {\bibinfo {volume} {D12}},\ \bibinfo {pages}
  {2183--2199} (\bibinfo {year} {1975})}\BibitemShut {NoStop}%
%%CITATION = PHRVA,D12,2183;%%
\bibitem [{\citenamefont {{Barker}}\ and\ \citenamefont
  {{Oconnell}}(1979)}]{Barker:1979}%
  \BibitemOpen
  \bibfield  {author} {\bibinfo {author} {\bibfnamefont {B.~M.}\ \bibnamefont
  {{Barker}}}\ and\ \bibinfo {author} {\bibfnamefont {R.~F.}\ \bibnamefont
  {{Oconnell}}},\ }\bibfield  {title} {\enquote {\bibinfo {title} {{The
  gravitational interaction: spin, rotation, and quantum effects - a
  review.}}}\ }\href {\doibase 10.1007/BF00756587} {\bibfield  {journal}
  {\bibinfo  {journal} {General Relativity and Gravitation}\ }\textbf {\bibinfo
  {volume} {11}},\ \bibinfo {pages} {149--175} (\bibinfo {year}
  {1979})}\BibitemShut {NoStop}%
\bibitem [{\citenamefont {Damour}(1982)}]{Damour:1982}%
  \BibitemOpen
  \bibfield  {author} {\bibinfo {author} {\bibfnamefont {Thibault}\
  \bibnamefont {Damour}},\ }\bibfield  {title} {\enquote {\bibinfo {title}
  {{Probl{\'e}me des deux corps et freinage de rayonnement en relativit{\'e}
  g{\'e}n{\'e}rale}},}\ }\href@noop {} {\bibfield  {journal} {\bibinfo
  {journal} {C. R. Acad. Sci. Paris S{\'e}r. II}\ }\textbf {\bibinfo {volume}
  {294}},\ \bibinfo {pages} {1355--1357} (\bibinfo {year} {1982})}\BibitemShut
  {NoStop}%
\bibitem [{\citenamefont {Thorne}\ and\ \citenamefont
  {Hartle}(1984)}]{Thorne:1984mz}%
  \BibitemOpen
  \bibfield  {author} {\bibinfo {author} {\bibfnamefont {Kip~S.}\ \bibnamefont
  {Thorne}}\ and\ \bibinfo {author} {\bibfnamefont {James~B.}\ \bibnamefont
  {Hartle}},\ }\bibfield  {title} {\enquote {\bibinfo {title} {{Laws of motion
  and precession for black holes and other bodies}},}\ }\href {\doibase
  10.1103/PhysRevD.31.1815} {\bibfield  {journal} {\bibinfo  {journal} {Phys.
  Rev.}\ }\textbf {\bibinfo {volume} {D31}},\ \bibinfo {pages} {1815--1837}
  (\bibinfo {year} {1984})}\BibitemShut {NoStop}%
%%CITATION = PHRVA,D31,1815;%%
\bibitem [{\citenamefont {Damour}\ and\ \citenamefont
  {Sch{\"a}fer}(1988)}]{Damour:1988mr}%
  \BibitemOpen
  \bibfield  {author} {\bibinfo {author} {\bibfnamefont {T.}~\bibnamefont
  {Damour}}\ and\ \bibinfo {author} {\bibfnamefont {Gerhard}\ \bibnamefont
  {Sch{\"a}fer}},\ }\bibfield  {title} {\enquote {\bibinfo {title} {{Higher
  Order Relativistic Periastron Advances and Binary Pulsars}},}\ }\href
  {\doibase 10.1007/BF02828697} {\bibfield  {journal} {\bibinfo  {journal}
  {Nuovo Cim.}\ }\textbf {\bibinfo {volume} {B101}},\ \bibinfo {pages} {127}
  (\bibinfo {year} {1988})}\BibitemShut {NoStop}%
%%CITATION = NUCIA,B101,127;%%
\bibitem [{\citenamefont {Tagoshi}\ \emph {et~al.}(2001)\citenamefont
  {Tagoshi}, \citenamefont {Ohashi},\ and\ \citenamefont
  {Owen}}]{Tagoshi:2000zg}%
  \BibitemOpen
  \bibfield  {author} {\bibinfo {author} {\bibfnamefont {Hideyuki}\
  \bibnamefont {Tagoshi}}, \bibinfo {author} {\bibfnamefont {Akira}\
  \bibnamefont {Ohashi}}, \ and\ \bibinfo {author} {\bibfnamefont
  {Benjamin~J.}\ \bibnamefont {Owen}},\ }\bibfield  {title} {\enquote {\bibinfo
  {title} {{Gravitational field and equations of motion of spinning compact
  binaries to 2.5 post-Newtonian order}},}\ }\href {\doibase
  10.1103/PhysRevD.63.044006} {\bibfield  {journal} {\bibinfo  {journal} {Phys.
  Rev.}\ }\textbf {\bibinfo {volume} {D63}},\ \bibinfo {pages} {044006}
  (\bibinfo {year} {2001})},\ \Eprint {http://arxiv.org/abs/gr-qc/0010014}
  {arXiv:gr-qc/0010014 [gr-qc]} \BibitemShut {NoStop}%
%%CITATION = GR-QC/0010014;%%
\bibitem [{\citenamefont {Faye}\ \emph {et~al.}(2006)\citenamefont {Faye},
  \citenamefont {Blanchet},\ and\ \citenamefont {Buonanno}}]{Faye:2006gx}%
  \BibitemOpen
  \bibfield  {author} {\bibinfo {author} {\bibfnamefont {Guillaume}\
  \bibnamefont {Faye}}, \bibinfo {author} {\bibfnamefont {Luc}\ \bibnamefont
  {Blanchet}}, \ and\ \bibinfo {author} {\bibfnamefont {Alessandra}\
  \bibnamefont {Buonanno}},\ }\bibfield  {title} {\enquote {\bibinfo {title}
  {{Higher-order spin effects in the dynamics of compact binaries. I. Equations
  of motion}},}\ }\href {\doibase 10.1103/PhysRevD.74.104033} {\bibfield
  {journal} {\bibinfo  {journal} {Phys. Rev.}\ }\textbf {\bibinfo {volume}
  {D74}},\ \bibinfo {pages} {104033} (\bibinfo {year} {2006})},\ \Eprint
  {http://arxiv.org/abs/gr-qc/0605139} {arXiv:gr-qc/0605139 [gr-qc]}
  \BibitemShut {NoStop}%
%%CITATION = GR-QC/0605139;%%
\bibitem [{\citenamefont {Damour}\ \emph
  {et~al.}(2008{\natexlab{b}})\citenamefont {Damour}, \citenamefont
  {Jaranowski},\ and\ \citenamefont {Sch{\"a}fer}}]{Damour:2007nc}%
  \BibitemOpen
  \bibfield  {author} {\bibinfo {author} {\bibfnamefont {Thibault}\
  \bibnamefont {Damour}}, \bibinfo {author} {\bibfnamefont {Piotr}\
  \bibnamefont {Jaranowski}}, \ and\ \bibinfo {author} {\bibfnamefont
  {Gerhard}\ \bibnamefont {Sch{\"a}fer}},\ }\bibfield  {title} {\enquote
  {\bibinfo {title} {{Hamiltonian of two spinning compact bodies with
  next-to-leading order gravitational spin-orbit coupling}},}\ }\href {\doibase
  10.1103/PhysRevD.77.064032} {\bibfield  {journal} {\bibinfo  {journal} {Phys.
  Rev.}\ }\textbf {\bibinfo {volume} {D77}},\ \bibinfo {pages} {064032}
  (\bibinfo {year} {2008}{\natexlab{b}})},\ \Eprint
  {http://arxiv.org/abs/0711.1048} {arXiv:0711.1048 [gr-qc]} \BibitemShut
  {NoStop}%
%%CITATION = ARXIV:0711.1048;%%
\bibitem [{\citenamefont {Steinhoff}\ \emph
  {et~al.}(2008{\natexlab{a}})\citenamefont {Steinhoff}, \citenamefont
  {Sch{\"a}fer},\ and\ \citenamefont {Hergt}}]{Steinhoff:2008zr}%
  \BibitemOpen
  \bibfield  {author} {\bibinfo {author} {\bibfnamefont {Jan}\ \bibnamefont
  {Steinhoff}}, \bibinfo {author} {\bibfnamefont {Gerhard}\ \bibnamefont
  {Sch{\"a}fer}}, \ and\ \bibinfo {author} {\bibfnamefont {Steven}\
  \bibnamefont {Hergt}},\ }\bibfield  {title} {\enquote {\bibinfo {title} {{ADM
  canonical formalism for gravitating spinning objects}},}\ }\href {\doibase
  10.1103/PhysRevD.77.104018} {\bibfield  {journal} {\bibinfo  {journal} {Phys.
  Rev.}\ }\textbf {\bibinfo {volume} {D77}},\ \bibinfo {pages} {104018}
  (\bibinfo {year} {2008}{\natexlab{a}})},\ \Eprint
  {http://arxiv.org/abs/0805.3136} {arXiv:0805.3136 [gr-qc]} \BibitemShut
  {NoStop}%
%%CITATION = ARXIV:0805.3136;%%
\bibitem [{\citenamefont {Perrodin}(2010)}]{Perrodin:2010dy}%
  \BibitemOpen
  \bibfield  {author} {\bibinfo {author} {\bibfnamefont {Delphine~L.}\
  \bibnamefont {Perrodin}},\ }\bibfield  {title} {\enquote {\bibinfo {title}
  {{Subleading Spin-Orbit Correction to the Newtonian Potential in Effective
  Field Theory Formalism}},}\ }in\ \href {\doibase 10.1142/9789814374552_0041}
  {\emph {\bibinfo {booktitle} {{On recent developments in theoretical and
  experimental general relativity, astrophysics and relativistic field
  theories. Proceedings, 12th Marcel Grossmann Meeting on General Relativity,
  Paris, France, July 12-18, 2009. Vol. 1-3}}}}\ (\bibinfo {year} {2010})\ pp.\
  \bibinfo {pages} {725--727},\ \Eprint {http://arxiv.org/abs/1005.0634}
  {arXiv:1005.0634 [gr-qc]} \BibitemShut {NoStop}%
%%CITATION = ARXIV:1005.0634;%%
\bibitem [{\citenamefont {Porto}(2010)}]{Porto:2010tr}%
  \BibitemOpen
  \bibfield  {author} {\bibinfo {author} {\bibfnamefont {Rafael~A.}\
  \bibnamefont {Porto}},\ }\bibfield  {title} {\enquote {\bibinfo {title}
  {{Next to leading order spin-orbit effects in the motion of inspiralling
  compact binaries}},}\ }\href {\doibase 10.1088/0264-9381/27/20/205001}
  {\bibfield  {journal} {\bibinfo  {journal} {Class. Quant. Grav.}\ }\textbf
  {\bibinfo {volume} {27}},\ \bibinfo {pages} {205001} (\bibinfo {year}
  {2010})},\ \Eprint {http://arxiv.org/abs/1005.5730} {arXiv:1005.5730 [gr-qc]}
  \BibitemShut {NoStop}%
%%CITATION = ARXIV:1005.5730;%%
\bibitem [{\citenamefont {Levi}(2010{\natexlab{a}})}]{Levi:2010zu}%
  \BibitemOpen
  \bibfield  {author} {\bibinfo {author} {\bibfnamefont {Michele}\ \bibnamefont
  {Levi}},\ }\bibfield  {title} {\enquote {\bibinfo {title} {{Next to Leading
  Order gravitational Spin-Orbit coupling in an Effective Field Theory
  approach}},}\ }\href {\doibase 10.1103/PhysRevD.82.104004} {\bibfield
  {journal} {\bibinfo  {journal} {Phys. Rev.}\ }\textbf {\bibinfo {volume}
  {D82}},\ \bibinfo {pages} {104004} (\bibinfo {year} {2010}{\natexlab{a}})},\
  \Eprint {http://arxiv.org/abs/1006.4139} {arXiv:1006.4139 [gr-qc]}
  \BibitemShut {NoStop}%
%%CITATION = ARXIV:1006.4139;%%
\bibitem [{\citenamefont {Hartung}\ and\ \citenamefont
  {Steinhoff}(2011{\natexlab{a}})}]{Hartung:2011te}%
  \BibitemOpen
  \bibfield  {author} {\bibinfo {author} {\bibfnamefont {Johannes}\
  \bibnamefont {Hartung}}\ and\ \bibinfo {author} {\bibfnamefont {Jan}\
  \bibnamefont {Steinhoff}},\ }\bibfield  {title} {\enquote {\bibinfo {title}
  {{Next-to-next-to-leading order post-Newtonian spin-orbit Hamiltonian for
  self-gravitating binaries}},}\ }\href {\doibase 10.1002/andp.201100094}
  {\bibfield  {journal} {\bibinfo  {journal} {Annalen Phys.}\ }\textbf
  {\bibinfo {volume} {523}},\ \bibinfo {pages} {783--790} (\bibinfo {year}
  {2011}{\natexlab{a}})},\ \Eprint {http://arxiv.org/abs/1104.3079}
  {arXiv:1104.3079 [gr-qc]} \BibitemShut {NoStop}%
%%CITATION = ARXIV:1104.3079;%%
\bibitem [{\citenamefont {Hartung}\ \emph {et~al.}(2013)\citenamefont
  {Hartung}, \citenamefont {Steinhoff},\ and\ \citenamefont
  {Sch{\"a}fer}}]{Hartung:2013dza}%
  \BibitemOpen
  \bibfield  {author} {\bibinfo {author} {\bibfnamefont {Johannes}\
  \bibnamefont {Hartung}}, \bibinfo {author} {\bibfnamefont {Jan}\ \bibnamefont
  {Steinhoff}}, \ and\ \bibinfo {author} {\bibfnamefont {Gerhard}\ \bibnamefont
  {Sch{\"a}fer}},\ }\bibfield  {title} {\enquote {\bibinfo {title}
  {{Next-to-next-to-leading order post-Newtonian linear-in-spin binary
  Hamiltonians}},}\ }\href {\doibase 10.1002/andp.201200271} {\bibfield
  {journal} {\bibinfo  {journal} {Annalen Phys.}\ }\textbf {\bibinfo {volume}
  {525}},\ \bibinfo {pages} {359--394} (\bibinfo {year} {2013})},\ \Eprint
  {http://arxiv.org/abs/1302.6723} {arXiv:1302.6723 [gr-qc]} \BibitemShut
  {NoStop}%
%%CITATION = ARXIV:1302.6723;%%
\bibitem [{\citenamefont {Marsat}\ \emph {et~al.}(2013)\citenamefont {Marsat},
  \citenamefont {Bohe}, \citenamefont {Faye},\ and\ \citenamefont
  {Blanchet}}]{Marsat:2012fn}%
  \BibitemOpen
  \bibfield  {author} {\bibinfo {author} {\bibfnamefont {Sylvain}\ \bibnamefont
  {Marsat}}, \bibinfo {author} {\bibfnamefont {Alejandro}\ \bibnamefont
  {Bohe}}, \bibinfo {author} {\bibfnamefont {Guillaume}\ \bibnamefont {Faye}},
  \ and\ \bibinfo {author} {\bibfnamefont {Luc}\ \bibnamefont {Blanchet}},\
  }\bibfield  {title} {\enquote {\bibinfo {title} {{Next-to-next-to-leading
  order spin-orbit effects in the equations of motion of compact binary
  systems}},}\ }\href {\doibase 10.1088/0264-9381/30/5/055007} {\bibfield
  {journal} {\bibinfo  {journal} {Class. Quant. Grav.}\ }\textbf {\bibinfo
  {volume} {30}},\ \bibinfo {pages} {055007} (\bibinfo {year} {2013})},\
  \Eprint {http://arxiv.org/abs/1210.4143} {arXiv:1210.4143 [gr-qc]}
  \BibitemShut {NoStop}%
%%CITATION = ARXIV:1210.4143;%%
\bibitem [{\citenamefont {Bohe}\ \emph {et~al.}(2013)\citenamefont {Bohe},
  \citenamefont {Marsat}, \citenamefont {Faye},\ and\ \citenamefont
  {Blanchet}}]{Bohe:2012mr}%
  \BibitemOpen
  \bibfield  {author} {\bibinfo {author} {\bibfnamefont {Alejandro}\
  \bibnamefont {Bohe}}, \bibinfo {author} {\bibfnamefont {Sylvain}\
  \bibnamefont {Marsat}}, \bibinfo {author} {\bibfnamefont {Guillaume}\
  \bibnamefont {Faye}}, \ and\ \bibinfo {author} {\bibfnamefont {Luc}\
  \bibnamefont {Blanchet}},\ }\bibfield  {title} {\enquote {\bibinfo {title}
  {{Next-to-next-to-leading order spin-orbit effects in the near-zone metric
  and precession equations of compact binaries}},}\ }\href {\doibase
  10.1088/0264-9381/30/7/075017} {\bibfield  {journal} {\bibinfo  {journal}
  {Class. Quant. Grav.}\ }\textbf {\bibinfo {volume} {30}},\ \bibinfo {pages}
  {075017} (\bibinfo {year} {2013})},\ \Eprint {http://arxiv.org/abs/1212.5520}
  {arXiv:1212.5520} \BibitemShut {NoStop}%
%%CITATION = ARXIV:1212.5520;%%
\bibitem [{\citenamefont {Levi}\ and\ \citenamefont
  {Steinhoff}(2016{\natexlab{a}})}]{Levi:2015uxa}%
  \BibitemOpen
  \bibfield  {author} {\bibinfo {author} {\bibfnamefont {Michele}\ \bibnamefont
  {Levi}}\ and\ \bibinfo {author} {\bibfnamefont {Jan}\ \bibnamefont
  {Steinhoff}},\ }\bibfield  {title} {\enquote {\bibinfo {title}
  {{Next-to-next-to-leading order gravitational spin-orbit coupling via the
  effective field theory for spinning objects in the post-Newtonian scheme}},}\
  }\href {\doibase 10.1088/1475-7516/2016/01/011} {\bibfield  {journal}
  {\bibinfo  {journal} {JCAP}\ }\textbf {\bibinfo {volume} {1601}},\ \bibinfo
  {pages} {011} (\bibinfo {year} {2016}{\natexlab{a}})},\ \Eprint
  {http://arxiv.org/abs/1506.05056} {arXiv:1506.05056 [gr-qc]} \BibitemShut
  {NoStop}%
%%CITATION = ARXIV:1506.05056;%%
\bibitem [{\citenamefont {Levi}\ and\ \citenamefont
  {Steinhoff}(2015)}]{Levi:2015msa}%
  \BibitemOpen
  \bibfield  {author} {\bibinfo {author} {\bibfnamefont {Michele}\ \bibnamefont
  {Levi}}\ and\ \bibinfo {author} {\bibfnamefont {Jan}\ \bibnamefont
  {Steinhoff}},\ }\bibfield  {title} {\enquote {\bibinfo {title} {{Spinning
  gravitating objects in the effective field theory in the post-Newtonian
  scheme}},}\ }\href {\doibase 10.1007/JHEP09(2015)219} {\bibfield  {journal}
  {\bibinfo  {journal} {JHEP}\ }\textbf {\bibinfo {volume} {09}},\ \bibinfo
  {pages} {219} (\bibinfo {year} {2015})},\ \Eprint
  {http://arxiv.org/abs/1501.04956} {arXiv:1501.04956 [gr-qc]} \BibitemShut
  {NoStop}%
%%CITATION = ARXIV:1501.04956;%%
\bibitem [{\citenamefont {Poisson}(1998)}]{Poisson:1997ha}%
  \BibitemOpen
  \bibfield  {author} {\bibinfo {author} {\bibfnamefont {Eric}\ \bibnamefont
  {Poisson}},\ }\bibfield  {title} {\enquote {\bibinfo {title} {{Gravitational
  waves from inspiraling compact binaries: The Quadrupole moment term}},}\
  }\href {\doibase 10.1103/PhysRevD.57.5287} {\bibfield  {journal} {\bibinfo
  {journal} {Phys. Rev.}\ }\textbf {\bibinfo {volume} {D57}},\ \bibinfo {pages}
  {5287--5290} (\bibinfo {year} {1998})},\ \Eprint
  {http://arxiv.org/abs/gr-qc/9709032} {arXiv:gr-qc/9709032 [gr-qc]}
  \BibitemShut {NoStop}%
%%CITATION = GR-QC/9709032;%%
\bibitem [{\citenamefont {Racine}(2008)}]{Racine:2008qv}%
  \BibitemOpen
  \bibfield  {author} {\bibinfo {author} {\bibfnamefont {{\'E}tienne}\
  \bibnamefont {Racine}},\ }\bibfield  {title} {\enquote {\bibinfo {title}
  {{Analysis of spin precession in binary black hole systems including
  quadrupole-monopole interaction}},}\ }\href {\doibase
  10.1103/PhysRevD.78.044021} {\bibfield  {journal} {\bibinfo  {journal} {Phys.
  Rev.}\ }\textbf {\bibinfo {volume} {D78}},\ \bibinfo {pages} {044021}
  (\bibinfo {year} {2008})},\ \Eprint {http://arxiv.org/abs/0803.1820}
  {arXiv:0803.1820 [gr-qc]} \BibitemShut {NoStop}%
%%CITATION = ARXIV:0803.1820;%%
\bibitem [{\citenamefont {Steinhoff}\ \emph
  {et~al.}(2008{\natexlab{b}})\citenamefont {Steinhoff}, \citenamefont
  {Hergt},\ and\ \citenamefont {Sch{\"a}fer}}]{Steinhoff:2007mb}%
  \BibitemOpen
  \bibfield  {author} {\bibinfo {author} {\bibfnamefont {Jan}\ \bibnamefont
  {Steinhoff}}, \bibinfo {author} {\bibfnamefont {Steven}\ \bibnamefont
  {Hergt}}, \ and\ \bibinfo {author} {\bibfnamefont {Gerhard}\ \bibnamefont
  {Sch{\"a}fer}},\ }\bibfield  {title} {\enquote {\bibinfo {title} {{On the
  next-to-leading order gravitational spin(1)-spin(2) dynamics}},}\ }\href
  {\doibase 10.1103/PhysRevD.77.081501} {\bibfield  {journal} {\bibinfo
  {journal} {Phys. Rev.}\ }\textbf {\bibinfo {volume} {D77}},\ \bibinfo {pages}
  {081501} (\bibinfo {year} {2008}{\natexlab{b}})},\ \Eprint
  {http://arxiv.org/abs/0712.1716} {arXiv:0712.1716 [gr-qc]} \BibitemShut
  {NoStop}%
%%CITATION = ARXIV:0712.1716;%%
\bibitem [{\citenamefont {Porto}\ and\ \citenamefont
  {Rothstein}(2008{\natexlab{a}})}]{Porto:2008tb}%
  \BibitemOpen
  \bibfield  {author} {\bibinfo {author} {\bibfnamefont {Rafael~A.}\
  \bibnamefont {Porto}}\ and\ \bibinfo {author} {\bibfnamefont {Ira~Z.}\
  \bibnamefont {Rothstein}},\ }\bibfield  {title} {\enquote {\bibinfo {title}
  {{Spin(1)Spin(2) Effects in the Motion of Inspiralling Compact Binaries at
  Third Order in the Post-Newtonian Expansion}},}\ }\href {\doibase
  10.1103/PhysRevD.78.044012, 10.1103/PhysRevD.81.029904} {\bibfield  {journal}
  {\bibinfo  {journal} {Phys. Rev.}\ }\textbf {\bibinfo {volume} {D78}},\
  \bibinfo {pages} {044012} (\bibinfo {year} {2008}{\natexlab{a}})},\ \bibinfo
  {note} {[Erratum: Phys. Rev.D81,029904(2010)]},\ \Eprint
  {http://arxiv.org/abs/0802.0720} {arXiv:0802.0720 [gr-qc]} \BibitemShut
  {NoStop}%
%%CITATION = ARXIV:0802.0720;%%
\bibitem [{\citenamefont {Porto}\ and\ \citenamefont
  {Rothstein}(2006)}]{Porto:2006bt}%
  \BibitemOpen
  \bibfield  {author} {\bibinfo {author} {\bibfnamefont {Rafael~A.}\
  \bibnamefont {Porto}}\ and\ \bibinfo {author} {\bibfnamefont {Ira~Z.}\
  \bibnamefont {Rothstein}},\ }\bibfield  {title} {\enquote {\bibinfo {title}
  {Calculation of the first nonlinear contribution to the general-relativistic
  spin-spin interaction for binary systems},}\ }\href {\doibase
  10.1103/PhysRevLett.97.021101} {\bibfield  {journal} {\bibinfo  {journal}
  {Phys. Rev. Lett.}\ }\textbf {\bibinfo {volume} {97}},\ \bibinfo {pages}
  {021101} (\bibinfo {year} {2006})},\ \Eprint
  {http://arxiv.org/abs/gr-qc/0604099} {arXiv:gr-qc/0604099 [gr-qc]}
  \BibitemShut {NoStop}%
%%CITATION = GR-QC/0604099;%%
\bibitem [{\citenamefont {Levi}(2010{\natexlab{b}})}]{Levi:2008nh}%
  \BibitemOpen
  \bibfield  {author} {\bibinfo {author} {\bibfnamefont {Michele}\ \bibnamefont
  {Levi}},\ }\bibfield  {title} {\enquote {\bibinfo {title} {{Next to Leading
  Order gravitational Spin1-Spin2 coupling with Kaluza-Klein reduction}},}\
  }\href {\doibase 10.1103/PhysRevD.82.064029} {\bibfield  {journal} {\bibinfo
  {journal} {Phys. Rev.}\ }\textbf {\bibinfo {volume} {D82}},\ \bibinfo {pages}
  {064029} (\bibinfo {year} {2010}{\natexlab{b}})},\ \Eprint
  {http://arxiv.org/abs/0802.1508} {arXiv:0802.1508 [gr-qc]} \BibitemShut
  {NoStop}%
%%CITATION = ARXIV:0802.1508;%%
\bibitem [{\citenamefont {Porto}\ and\ \citenamefont
  {Rothstein}(2008{\natexlab{b}})}]{Porto:2008jj}%
  \BibitemOpen
  \bibfield  {author} {\bibinfo {author} {\bibfnamefont {Rafael~A}\
  \bibnamefont {Porto}}\ and\ \bibinfo {author} {\bibfnamefont {Ira~Z.}\
  \bibnamefont {Rothstein}},\ }\bibfield  {title} {\enquote {\bibinfo {title}
  {{Next to Leading Order Spin(1)Spin(1) Effects in the Motion of Inspiralling
  Compact Binaries}},}\ }\bibfield  {booktitle} {\emph {\bibinfo {booktitle}
  {{Workshop on Effective Field Theory Techniques in Gravitational Wave Physics
  Pittsburgh, Pennsylvania, July 23-25, 2007}}},\ }\href {\doibase
  10.1103/PhysRevD.81.029905, 10.1103/PhysRevD.78.044013} {\bibfield  {journal}
  {\bibinfo  {journal} {Phys. Rev.}\ }\textbf {\bibinfo {volume} {D78}},\
  \bibinfo {pages} {044013} (\bibinfo {year} {2008}{\natexlab{b}})},\ \bibinfo
  {note} {[Erratum: Phys. Rev.D81,029905(2010)]},\ \Eprint
  {http://arxiv.org/abs/0804.0260} {arXiv:0804.0260 [gr-qc]} \BibitemShut
  {NoStop}%
%%CITATION = ARXIV:0804.0260;%%
\bibitem [{\citenamefont {Steinhoff}\ \emph
  {et~al.}(2008{\natexlab{c}})\citenamefont {Steinhoff}, \citenamefont
  {Hergt},\ and\ \citenamefont {Sch{\"a}fer}}]{Steinhoff:2008ji}%
  \BibitemOpen
  \bibfield  {author} {\bibinfo {author} {\bibfnamefont {Jan}\ \bibnamefont
  {Steinhoff}}, \bibinfo {author} {\bibfnamefont {Steven}\ \bibnamefont
  {Hergt}}, \ and\ \bibinfo {author} {\bibfnamefont {Gerhard}\ \bibnamefont
  {Sch{\"a}fer}},\ }\bibfield  {title} {\enquote {\bibinfo {title}
  {{Spin-squared Hamiltonian of next-to-leading order gravitational
  interaction}},}\ }\href {\doibase 10.1103/PhysRevD.78.101503} {\bibfield
  {journal} {\bibinfo  {journal} {Phys. Rev.}\ }\textbf {\bibinfo {volume}
  {D78}},\ \bibinfo {pages} {101503} (\bibinfo {year} {2008}{\natexlab{c}})},\
  \Eprint {http://arxiv.org/abs/0809.2200} {arXiv:0809.2200 [gr-qc]}
  \BibitemShut {NoStop}%
%%CITATION = ARXIV:0809.2200;%%
\bibitem [{\citenamefont {Hergt}\ \emph {et~al.}(2010)\citenamefont {Hergt},
  \citenamefont {Steinhoff},\ and\ \citenamefont {Sch{\"a}fer}}]{Hergt:2010pa}%
  \BibitemOpen
  \bibfield  {author} {\bibinfo {author} {\bibfnamefont {Steven}\ \bibnamefont
  {Hergt}}, \bibinfo {author} {\bibfnamefont {Jan}\ \bibnamefont {Steinhoff}},
  \ and\ \bibinfo {author} {\bibfnamefont {Gerhard}\ \bibnamefont
  {Sch{\"a}fer}},\ }\bibfield  {title} {\enquote {\bibinfo {title} {{Reduced
  Hamiltonian for next-to-leading order Spin-Squared Dynamics of General
  Compact Binaries}},}\ }\href {\doibase 10.1088/0264-9381/27/13/135007}
  {\bibfield  {journal} {\bibinfo  {journal} {Class. Quant. Grav.}\ }\textbf
  {\bibinfo {volume} {27}},\ \bibinfo {pages} {135007} (\bibinfo {year}
  {2010})},\ \Eprint {http://arxiv.org/abs/1002.2093} {arXiv:1002.2093 [gr-qc]}
  \BibitemShut {NoStop}%
%%CITATION = ARXIV:1002.2093;%%
\bibitem [{\citenamefont {Hergt}\ \emph {et~al.}(2012)\citenamefont {Hergt},
  \citenamefont {Steinhoff},\ and\ \citenamefont {Sch{\"a}fer}}]{Hergt:2011ik}%
  \BibitemOpen
  \bibfield  {author} {\bibinfo {author} {\bibfnamefont {Steven}\ \bibnamefont
  {Hergt}}, \bibinfo {author} {\bibfnamefont {Jan}\ \bibnamefont {Steinhoff}},
  \ and\ \bibinfo {author} {\bibfnamefont {Gerhard}\ \bibnamefont
  {Sch{\"a}fer}},\ }\bibfield  {title} {\enquote {\bibinfo {title}
  {{Elimination of the spin supplementary condition in the effective field
  theory approach to the post-Newtonian approximation}},}\ }\href {\doibase
  10.1016/j.aop.2012.02.006} {\bibfield  {journal} {\bibinfo  {journal} {Annals
  Phys.}\ }\textbf {\bibinfo {volume} {327}},\ \bibinfo {pages} {1494--1537}
  (\bibinfo {year} {2012})},\ \Eprint {http://arxiv.org/abs/1110.2094}
  {arXiv:1110.2094 [gr-qc]} \BibitemShut {NoStop}%
%%CITATION = ARXIV:1110.2094;%%
\bibitem [{\citenamefont {Boh\'e}\ \emph {et~al.}(2015)\citenamefont {Boh\'e},
  \citenamefont {Faye}, \citenamefont {Marsat},\ and\ \citenamefont
  {Porter}}]{Bohe:2015ana}%
  \BibitemOpen
  \bibfield  {author} {\bibinfo {author} {\bibfnamefont {Alejandro}\
  \bibnamefont {Boh\'e}}, \bibinfo {author} {\bibfnamefont {Guillaume}\
  \bibnamefont {Faye}}, \bibinfo {author} {\bibfnamefont {Sylvain}\
  \bibnamefont {Marsat}}, \ and\ \bibinfo {author} {\bibfnamefont {Edward~K.}\
  \bibnamefont {Porter}},\ }\bibfield  {title} {\enquote {\bibinfo {title}
  {{Quadratic-in-spin effects in the orbital dynamics and gravitational-wave
  energy flux of compact binaries at the 3PN order}},}\ }\href {\doibase
  10.1088/0264-9381/32/19/195010} {\bibfield  {journal} {\bibinfo  {journal}
  {Class. Quant. Grav.}\ }\textbf {\bibinfo {volume} {32}},\ \bibinfo {pages}
  {195010} (\bibinfo {year} {2015})},\ \Eprint
  {http://arxiv.org/abs/1501.01529} {arXiv:1501.01529 [gr-qc]} \BibitemShut
  {NoStop}%
%%CITATION = ARXIV:1501.01529;%%
\bibitem [{\citenamefont {Hartung}\ and\ \citenamefont
  {Steinhoff}(2011{\natexlab{b}})}]{Hartung:2011ea}%
  \BibitemOpen
  \bibfield  {author} {\bibinfo {author} {\bibfnamefont {Johannes}\
  \bibnamefont {Hartung}}\ and\ \bibinfo {author} {\bibfnamefont {Jan}\
  \bibnamefont {Steinhoff}},\ }\bibfield  {title} {\enquote {\bibinfo {title}
  {{Next-to-next-to-leading order post-Newtonian spin(1)-spin(2) Hamiltonian
  for self-gravitating binaries}},}\ }\href {\doibase 10.1002/andp.201100163}
  {\bibfield  {journal} {\bibinfo  {journal} {Annalen Phys.}\ }\textbf
  {\bibinfo {volume} {523}},\ \bibinfo {pages} {919--924} (\bibinfo {year}
  {2011}{\natexlab{b}})},\ \Eprint {http://arxiv.org/abs/1107.4294}
  {arXiv:1107.4294 [gr-qc]} \BibitemShut {NoStop}%
%%CITATION = ARXIV:1107.4294;%%
\bibitem [{\citenamefont {Levi}(2012)}]{Levi:2011eq}%
  \BibitemOpen
  \bibfield  {author} {\bibinfo {author} {\bibfnamefont {Michele}\ \bibnamefont
  {Levi}},\ }\bibfield  {title} {\enquote {\bibinfo {title} {{Binary dynamics
  from spin1-spin2 coupling at fourth post-Newtonian order}},}\ }\href
  {\doibase 10.1103/PhysRevD.85.064043} {\bibfield  {journal} {\bibinfo
  {journal} {Phys. Rev.}\ }\textbf {\bibinfo {volume} {D85}},\ \bibinfo {pages}
  {064043} (\bibinfo {year} {2012})},\ \Eprint {http://arxiv.org/abs/1107.4322}
  {arXiv:1107.4322 [gr-qc]} \BibitemShut {NoStop}%
%%CITATION = ARXIV:1107.4322;%%
\bibitem [{\citenamefont {Levi}\ and\ \citenamefont
  {Steinhoff}(2014)}]{Levi:2014sba}%
  \BibitemOpen
  \bibfield  {author} {\bibinfo {author} {\bibfnamefont {Michele}\ \bibnamefont
  {Levi}}\ and\ \bibinfo {author} {\bibfnamefont {Jan}\ \bibnamefont
  {Steinhoff}},\ }\bibfield  {title} {\enquote {\bibinfo {title} {{Equivalence
  of ADM Hamiltonian and Effective Field Theory approaches at
  next-to-next-to-leading order spin1-spin2 coupling of binary inspirals}},}\
  }\href {\doibase 10.1088/1475-7516/2014/12/003} {\bibfield  {journal}
  {\bibinfo  {journal} {JCAP}\ }\textbf {\bibinfo {volume} {1412}},\ \bibinfo
  {pages} {003} (\bibinfo {year} {2014})},\ \Eprint
  {http://arxiv.org/abs/1408.5762} {arXiv:1408.5762 [gr-qc]} \BibitemShut
  {NoStop}%
%%CITATION = ARXIV:1408.5762;%%
\bibitem [{\citenamefont {Levi}\ and\ \citenamefont
  {Steinhoff}(2016{\natexlab{b}})}]{Levi:2015ixa}%
  \BibitemOpen
  \bibfield  {author} {\bibinfo {author} {\bibfnamefont {Michele}\ \bibnamefont
  {Levi}}\ and\ \bibinfo {author} {\bibfnamefont {Jan}\ \bibnamefont
  {Steinhoff}},\ }\bibfield  {title} {\enquote {\bibinfo {title}
  {{Next-to-next-to-leading order gravitational spin-squared potential via the
  effective field theory for spinning objects in the post-Newtonian scheme}},}\
  }\href {\doibase 10.1088/1475-7516/2016/01/008} {\bibfield  {journal}
  {\bibinfo  {journal} {JCAP}\ }\textbf {\bibinfo {volume} {1601}},\ \bibinfo
  {pages} {008} (\bibinfo {year} {2016}{\natexlab{b}})},\ \Eprint
  {http://arxiv.org/abs/1506.05794} {arXiv:1506.05794 [gr-qc]} \BibitemShut
  {NoStop}%
%%CITATION = ARXIV:1506.05794;%%
\bibitem [{\citenamefont {Damour}\ \emph
  {et~al.}(2000{\natexlab{b}})\citenamefont {Damour}, \citenamefont
  {Jaranowski},\ and\ \citenamefont {Sch{\"a}fer}}]{Damour:2000we}%
  \BibitemOpen
  \bibfield  {author} {\bibinfo {author} {\bibfnamefont {Thibault}\
  \bibnamefont {Damour}}, \bibinfo {author} {\bibfnamefont {Piotr}\
  \bibnamefont {Jaranowski}}, \ and\ \bibinfo {author} {\bibfnamefont
  {Gerhard}\ \bibnamefont {Sch{\"a}fer}},\ }\bibfield  {title} {\enquote
  {\bibinfo {title} {{On the determination of the last stable orbit for
  circular general relativistic binaries at the third post-Newtonian
  approximation}},}\ }\href {\doibase 10.1103/PhysRevD.62.084011} {\bibfield
  {journal} {\bibinfo  {journal} {Phys. Rev.}\ }\textbf {\bibinfo {volume}
  {D62}},\ \bibinfo {pages} {084011} (\bibinfo {year} {2000}{\natexlab{b}})},\
  \Eprint {http://arxiv.org/abs/gr-qc/0005034} {arXiv:gr-qc/0005034 [gr-qc]}
  \BibitemShut {NoStop}%
%%CITATION = GR-QC/0005034;%%
\bibitem [{\citenamefont {Levi}\ and\ \citenamefont
  {Steinhoff}(2016{\natexlab{c}})}]{Levi:2016ofk}%
  \BibitemOpen
  \bibfield  {author} {\bibinfo {author} {\bibfnamefont {Michele}\ \bibnamefont
  {Levi}}\ and\ \bibinfo {author} {\bibfnamefont {Jan}\ \bibnamefont
  {Steinhoff}},\ }\bibfield  {title} {\enquote {\bibinfo {title} {{Complete
  conservative dynamics for inspiralling compact binaries with spins at fourth
  post-Newtonian order}},}\ }\href@noop {} {\  (\bibinfo {year}
  {2016}{\natexlab{c}})},\ \Eprint {http://arxiv.org/abs/1607.04252}
  {arXiv:1607.04252 [gr-qc]} \BibitemShut {NoStop}%
%%CITATION = ARXIV:1607.04252;%%
\bibitem [{\citenamefont {Barausse}\ \emph {et~al.}(2009)\citenamefont
  {Barausse}, \citenamefont {Racine},\ and\ \citenamefont
  {Buonanno}}]{Barausse:2009aa}%
  \BibitemOpen
  \bibfield  {author} {\bibinfo {author} {\bibfnamefont {Enrico}\ \bibnamefont
  {Barausse}}, \bibinfo {author} {\bibfnamefont {Etienne}\ \bibnamefont
  {Racine}}, \ and\ \bibinfo {author} {\bibfnamefont {Alessandra}\ \bibnamefont
  {Buonanno}},\ }\bibfield  {title} {\enquote {\bibinfo {title} {{Hamiltonian
  of a spinning test-particle in curved spacetime}},}\ }\href {\doibase
  10.1103/PhysRevD.85.069904, 10.1103/PhysRevD.80.104025} {\bibfield  {journal}
  {\bibinfo  {journal} {Phys. Rev.}\ }\textbf {\bibinfo {volume} {D80}},\
  \bibinfo {pages} {104025} (\bibinfo {year} {2009})},\ \bibinfo {note}
  {[Erratum: Phys. Rev.D85,069904(2012)]},\ \Eprint
  {http://arxiv.org/abs/0907.4745} {arXiv:0907.4745 [gr-qc]} \BibitemShut
  {NoStop}%
%%CITATION = ARXIV:0907.4745;%%
\bibitem [{\citenamefont {Vines}\ \emph {et~al.}(2016)\citenamefont {Vines},
  \citenamefont {Kunst}, \citenamefont {Steinhoff},\ and\ \citenamefont
  {Hinderer}}]{Vines:2016unv}%
  \BibitemOpen
  \bibfield  {author} {\bibinfo {author} {\bibfnamefont {Justin}\ \bibnamefont
  {Vines}}, \bibinfo {author} {\bibfnamefont {Daniela}\ \bibnamefont {Kunst}},
  \bibinfo {author} {\bibfnamefont {Jan}\ \bibnamefont {Steinhoff}}, \ and\
  \bibinfo {author} {\bibfnamefont {Tanja}\ \bibnamefont {Hinderer}},\
  }\bibfield  {title} {\enquote {\bibinfo {title} {{Canonical Hamiltonian for
  an extended test body in curved spacetime: To quadratic order in spin}},}\
  }\href {\doibase 10.1103/PhysRevD.93.103008} {\bibfield  {journal} {\bibinfo
  {journal} {Phys. Rev.}\ }\textbf {\bibinfo {volume} {D93}},\ \bibinfo {pages}
  {103008} (\bibinfo {year} {2016})},\ \Eprint
  {http://arxiv.org/abs/1601.07529} {arXiv:1601.07529 [gr-qc]} \BibitemShut
  {NoStop}%
%%CITATION = ARXIV:1601.07529;%%
\bibitem [{\citenamefont {Pryce}(1935)}]{Pryce:1935ibt}%
  \BibitemOpen
  \bibfield  {author} {\bibinfo {author} {\bibfnamefont {M.~H.~L.}\
  \bibnamefont {Pryce}},\ }\bibfield  {title} {\enquote {\bibinfo {title}
  {{Commuting co-ordinates in the new field theory}},}\ }\href {\doibase
  10.1098/rspa.1935.0094} {\bibfield  {journal} {\bibinfo  {journal} {Proc.
  Roy. Soc. Lond.}\ }\textbf {\bibinfo {volume} {A150}},\ \bibinfo {pages}
  {166--172} (\bibinfo {year} {1935})}\BibitemShut {NoStop}%
%%CITATION = PRSLA,A150,166;%%
\bibitem [{\citenamefont {Pryce}(1948)}]{Pryce:1948pf}%
  \BibitemOpen
  \bibfield  {author} {\bibinfo {author} {\bibfnamefont {M.~H.~L.}\
  \bibnamefont {Pryce}},\ }\bibfield  {title} {\enquote {\bibinfo {title} {{The
  Mass center in the restricted theory of relativity and its connection with
  the quantum theory of elementary particles}},}\ }\href {\doibase
  10.1098/rspa.1948.0103} {\bibfield  {journal} {\bibinfo  {journal} {Proc.
  Roy. Soc. Lond.}\ }\textbf {\bibinfo {volume} {A195}},\ \bibinfo {pages}
  {62--81} (\bibinfo {year} {1948})}\BibitemShut {NoStop}%
%%CITATION = PRSLA,A195,62;%%
\bibitem [{\citenamefont {Newton}\ and\ \citenamefont
  {Wigner}(1949)}]{Newton:1949cq}%
  \BibitemOpen
  \bibfield  {author} {\bibinfo {author} {\bibfnamefont {T.~D.}\ \bibnamefont
  {Newton}}\ and\ \bibinfo {author} {\bibfnamefont {Eugene~P.}\ \bibnamefont
  {Wigner}},\ }\bibfield  {title} {\enquote {\bibinfo {title} {{Localized
  States for Elementary Systems}},}\ }\href {\doibase
  10.1103/RevModPhys.21.400} {\bibfield  {journal} {\bibinfo  {journal} {Rev.
  Mod. Phys.}\ }\textbf {\bibinfo {volume} {21}},\ \bibinfo {pages} {400--406}
  (\bibinfo {year} {1949})}\BibitemShut {NoStop}%
%%CITATION = RMPHA,21,400;%%
\bibitem [{\citenamefont {Papapetrou}(1951)}]{Papapetrou:1951pa}%
  \BibitemOpen
  \bibfield  {author} {\bibinfo {author} {\bibfnamefont {Achille}\ \bibnamefont
  {Papapetrou}},\ }\bibfield  {title} {\enquote {\bibinfo {title} {{Spinning
  test particles in general relativity. 1.}}}\ }\href {\doibase
  10.1098/rspa.1951.0200} {\bibfield  {journal} {\bibinfo  {journal} {Proc.
  Roy. Soc. Lond.}\ }\textbf {\bibinfo {volume} {A209}},\ \bibinfo {pages}
  {248--258} (\bibinfo {year} {1951})}\BibitemShut {NoStop}%
%%CITATION = PRSLA,A209,248;%%
\bibitem [{\citenamefont {{Ehlers}}\ and\ \citenamefont
  {{Rudolph}}(1977)}]{Ehlers:1977}%
  \BibitemOpen
  \bibfield  {author} {\bibinfo {author} {\bibfnamefont {J{\"u}rgen}\
  \bibnamefont {{Ehlers}}}\ and\ \bibinfo {author} {\bibfnamefont {Ekkart}\
  \bibnamefont {{Rudolph}}},\ }\bibfield  {title} {\enquote {\bibinfo {title}
  {{Dynamics of extended bodies in general relativity center-of-mass
  description and quasirigidity}},}\ }\href {\doibase 10.1007/BF00763547}
  {\bibfield  {journal} {\bibinfo  {journal} {General Relativity and
  Gravitation}\ }\textbf {\bibinfo {volume} {8}},\ \bibinfo {pages} {197--217}
  (\bibinfo {year} {1977})}\BibitemShut {NoStop}%
\bibitem [{\citenamefont {Harte}(2012)}]{Harte:2011ku}%
  \BibitemOpen
  \bibfield  {author} {\bibinfo {author} {\bibfnamefont {Abraham~I.}\
  \bibnamefont {Harte}},\ }\bibfield  {title} {\enquote {\bibinfo {title}
  {{Mechanics of extended masses in general relativity}},}\ }\href {\doibase
  10.1088/0264-9381/29/5/055012} {\bibfield  {journal} {\bibinfo  {journal}
  {Class. Quant. Grav.}\ }\textbf {\bibinfo {volume} {29}},\ \bibinfo {pages}
  {055012} (\bibinfo {year} {2012})},\ \Eprint {http://arxiv.org/abs/1103.0543}
  {arXiv:1103.0543 [gr-qc]} \BibitemShut {NoStop}%
%%CITATION = ARXIV:1103.0543;%%
\bibitem [{\citenamefont {Harte}(2015)}]{Harte:2014wya}%
  \BibitemOpen
  \bibfield  {author} {\bibinfo {author} {\bibfnamefont {Abraham~I.}\
  \bibnamefont {Harte}},\ }\bibfield  {title} {\enquote {\bibinfo {title}
  {{Motion in classical field theories and the foundations of the self-force
  problem}},}\ }\bibfield  {booktitle} {\emph {\bibinfo {booktitle}
  {{Proceedings, 524th WE-Heraeus-Seminar: Equations of Motion in Relativistic
  Gravity (EOM 2013): Bad Honnef, Germany, February 17-23, 2013}}},\ }\href
  {\doibase 10.1007/978-3-319-18335-0_12} {\bibfield  {journal} {\bibinfo
  {journal} {Fund. Theor. Phys.}\ }\textbf {\bibinfo {volume} {179}},\ \bibinfo
  {pages} {327--398} (\bibinfo {year} {2015})},\ \Eprint
  {http://arxiv.org/abs/1405.5077} {arXiv:1405.5077 [gr-qc]} \BibitemShut
  {NoStop}%
%%CITATION = ARXIV:1405.5077;%%
\bibitem [{\citenamefont {Steinhoff}(2011)}]{Steinhoff:2010zz}%
  \BibitemOpen
  \bibfield  {author} {\bibinfo {author} {\bibfnamefont {Jan}\ \bibnamefont
  {Steinhoff}},\ }\bibfield  {title} {\enquote {\bibinfo {title} {{Canonical
  formulation of spin in general relativity}},}\ }\href {\doibase
  10.1002/andp.201000178} {\bibfield  {journal} {\bibinfo  {journal} {Annalen
  Phys.}\ }\textbf {\bibinfo {volume} {523}},\ \bibinfo {pages} {296--353}
  (\bibinfo {year} {2011})},\ \Eprint {http://arxiv.org/abs/1106.4203}
  {arXiv:1106.4203 [gr-qc]} \BibitemShut {NoStop}%
%%CITATION = ARXIV:1106.4203;%%
\bibitem [{\citenamefont {Steinhoff}\ and\ \citenamefont
  {Puetzfeld}(2012)}]{Steinhoff:2012rw}%
  \BibitemOpen
  \bibfield  {author} {\bibinfo {author} {\bibfnamefont {Jan}\ \bibnamefont
  {Steinhoff}}\ and\ \bibinfo {author} {\bibfnamefont {Dirk}\ \bibnamefont
  {Puetzfeld}},\ }\bibfield  {title} {\enquote {\bibinfo {title} {{Influence of
  internal structure on the motion of test bodies in extreme mass ratio
  situations}},}\ }\href {\doibase 10.1103/PhysRevD.86.044033} {\bibfield
  {journal} {\bibinfo  {journal} {Phys. Rev.}\ }\textbf {\bibinfo {volume}
  {D86}},\ \bibinfo {pages} {044033} (\bibinfo {year} {2012})},\ \Eprint
  {http://arxiv.org/abs/1205.3926} {arXiv:1205.3926 [gr-qc]} \BibitemShut
  {NoStop}%
%%CITATION = ARXIV:1205.3926;%%
\bibitem [{\citenamefont {Buonanno}\ \emph {et~al.}(2013)\citenamefont
  {Buonanno}, \citenamefont {Faye},\ and\ \citenamefont
  {Hinderer}}]{Buonanno:2012rv}%
  \BibitemOpen
  \bibfield  {author} {\bibinfo {author} {\bibfnamefont {Alessandra}\
  \bibnamefont {Buonanno}}, \bibinfo {author} {\bibfnamefont {Guillaume}\
  \bibnamefont {Faye}}, \ and\ \bibinfo {author} {\bibfnamefont {Tanja}\
  \bibnamefont {Hinderer}},\ }\bibfield  {title} {\enquote {\bibinfo {title}
  {{Spin effects on gravitational waves from inspiraling compact binaries at
  second post-Newtonian order}},}\ }\href {\doibase 10.1103/PhysRevD.87.044009}
  {\bibfield  {journal} {\bibinfo  {journal} {Phys. Rev.}\ }\textbf {\bibinfo
  {volume} {D87}},\ \bibinfo {pages} {044009} (\bibinfo {year} {2013})},\
  \Eprint {http://arxiv.org/abs/1209.6349} {arXiv:1209.6349 [gr-qc]}
  \BibitemShut {NoStop}%
%%CITATION = ARXIV:1209.6349;%%
\bibitem [{\citenamefont {Bini}\ and\ \citenamefont
  {Geralico}(2013)}]{Bini:2013nw}%
  \BibitemOpen
  \bibfield  {author} {\bibinfo {author} {\bibfnamefont {Donato}\ \bibnamefont
  {Bini}}\ and\ \bibinfo {author} {\bibfnamefont {Andrea}\ \bibnamefont
  {Geralico}},\ }\bibfield  {title} {\enquote {\bibinfo {title} {{Dynamics of
  quadrupolar bodies in a Schwarzschild spacetime}},}\ }\href {\doibase
  10.1103/PhysRevD.87.024028} {\bibfield  {journal} {\bibinfo  {journal} {Phys.
  Rev.}\ }\textbf {\bibinfo {volume} {D87}},\ \bibinfo {pages} {024028}
  (\bibinfo {year} {2013})},\ \Eprint {http://arxiv.org/abs/1408.5261}
  {arXiv:1408.5261 [gr-qc]} \BibitemShut {NoStop}%
%%CITATION = ARXIV:1408.5261;%%
\bibitem [{\citenamefont {Bini}\ and\ \citenamefont
  {Geralico}(2014)}]{Bini:2013uwa}%
  \BibitemOpen
  \bibfield  {author} {\bibinfo {author} {\bibfnamefont {Donato}\ \bibnamefont
  {Bini}}\ and\ \bibinfo {author} {\bibfnamefont {Andrea}\ \bibnamefont
  {Geralico}},\ }\bibfield  {title} {\enquote {\bibinfo {title} {{Deviation of
  quadrupolar bodies from geodesic motion in a Kerr spacetime}},}\ }\href
  {\doibase 10.1103/PhysRevD.89.044013} {\bibfield  {journal} {\bibinfo
  {journal} {Phys. Rev.}\ }\textbf {\bibinfo {volume} {D89}},\ \bibinfo {pages}
  {044013} (\bibinfo {year} {2014})},\ \Eprint {http://arxiv.org/abs/1311.7512}
  {arXiv:1311.7512 [gr-qc]} \BibitemShut {NoStop}%
%%CITATION = ARXIV:1311.7512;%%
\bibitem [{\citenamefont {Marsat}(2015)}]{Marsat:2014xea}%
  \BibitemOpen
  \bibfield  {author} {\bibinfo {author} {\bibfnamefont {Sylvain}\ \bibnamefont
  {Marsat}},\ }\bibfield  {title} {\enquote {\bibinfo {title} {{Cubic order
  spin effects in the dynamics and gravitational wave energy flux of compact
  object binaries}},}\ }\href {\doibase 10.1088/0264-9381/32/8/085008}
  {\bibfield  {journal} {\bibinfo  {journal} {Class. Quant. Grav.}\ }\textbf
  {\bibinfo {volume} {32}},\ \bibinfo {pages} {085008} (\bibinfo {year}
  {2015})},\ \Eprint {http://arxiv.org/abs/1411.4118} {arXiv:1411.4118 [gr-qc]}
  \BibitemShut {NoStop}%
%%CITATION = ARXIV:1411.4118;%%
\bibitem [{\citenamefont {Bini}\ \emph {et~al.}(2017)\citenamefont {Bini},
  \citenamefont {Geralico},\ and\ \citenamefont {Vines}}]{Bini:2017pee}%
  \BibitemOpen
  \bibfield  {author} {\bibinfo {author} {\bibfnamefont {Donato}\ \bibnamefont
  {Bini}}, \bibinfo {author} {\bibfnamefont {Andrea}\ \bibnamefont {Geralico}},
  \ and\ \bibinfo {author} {\bibfnamefont {Justin}\ \bibnamefont {Vines}},\
  }\bibfield  {title} {\enquote {\bibinfo {title} {{Hyperbolic scattering of
  spinning particles by a Kerr black hole}},}\ }\href {\doibase
  10.1103/PhysRevD.96.084044} {\bibfield  {journal} {\bibinfo  {journal} {Phys.
  Rev.}\ }\textbf {\bibinfo {volume} {D96}},\ \bibinfo {pages} {084044}
  (\bibinfo {year} {2017})},\ \Eprint {http://arxiv.org/abs/1707.09814}
  {arXiv:1707.09814 [gr-qc]} \BibitemShut {NoStop}%
%%CITATION = ARXIV:1707.09814;%%
\end{thebibliography}%

% run: cp ?.bbl ?_refs.tex    (done in make script)
%\input{paper_refs}

\end{document}